\documentclass[aps,prb,amsmath,amssymb,twocolumn,superscriptaddress,floatfix]{revtex4-2}
\usepackage{graphicx}
\usepackage{hyperref}
\usepackage{bm}
\usepackage{booktabs}
\usepackage[dvipsnames]{xcolor}
\hypersetup{pdftitle={Antlerite},pdfauthor={Anton},pdfsubject={Frustrated magnetism},pdfdisplaydoctitle}

\def\antH{Cu$_3$SO$_4$(OH)$_4$}
\def\antD{Cu$_3$SO$_4$(OD)$_4$}
\def\TN{\ensuremath{T_\text{N}}}

\begin{document}

\title{Incommensurate and multiple-{\bfseries\itshape q} magnetic misfit order in the frustrated quantum-spin-ladder material antlerite, \antH}

\author{Anton A.\ Kulbakov}
\author{Elaheh Sadrollahi}
\author{Florian Rasch}
\affiliation{Institut f\"ur Festk\"orper- und Materialphysik, Technische Universit\"at Dresden, 01069 Dresden, Germany}

\author{Maxim Avdeev}
\affiliation{Australian Nuclear Science and Technology Organisation, Lucas Heights, NSW 2234, Australia}
\affiliation{School of Chemistry, The University of Sydney, Sydney 2006, Australia}

\author{Sebastian Ga\ss}
\author{Laura Teresa {Corredor Bohorquez}}
\affiliation{Institute for Solid State Research, Leibniz IFW Dresden, 01069 Dresden, Germany}
\author{Anja U.\ B.\ Wolter}
\affiliation{Institute for Solid State Research, Leibniz IFW Dresden, 01069 Dresden, Germany}
\affiliation{W\"urzburg-Dresden Cluster of Excellence on Complexity and Topology in Quantum Matter\,--\,ct.qmat, Technische Universit\"at Dresden, 01069 Dresden, Germany}

\author{Manuel Feig}
\author{Roman Gumeniuk}
\affiliation{Institut f\"ur Experimentelle Physik, TU Bergakademie Freiberg, 09596 Freiberg, Germany}

\author{Hagen Poddig}
\affiliation{Anorganische Chemie II, Technische Universit\"at Dresden, 01069 Dresden, Germany}
\author{Markus St{\"o}tzer}
\affiliation{Anorganische Chemie I, Technische Universit\"at Dresden, 01069 Dresden, Germany}

\author{F.\ Jochen Litterst}
\affiliation{Institut f{\"u}r Physik der Kondensierten Materie, Technische Universit{\"a}t Braunschweig, 38106 Braunschweig, Germany}

\author{In\'es Puente-Orench}
\affiliation{Instituto de Nanociencia y Materiales de Arag{\'o}n (INMA), CSIC-Universidad de Zaragoza, Zaragoza 50009, Spain}
\affiliation{Institut Laue-Langevin, 71 Avenue des Martyrs, CS 20156, CEDEX 9, 38042 Grenoble, France}
\author{Andrew Wildes}
\affiliation{Institut Laue-Langevin, 71 Avenue des Martyrs, CS 20156, CEDEX 9, 38042 Grenoble, France}

\author{Eugen Weschke}
\affiliation{Helmholtz-Zentrum Berlin f\"ur Materialien und Energie, BESSY II, 12489 Berlin, Germany}

\author{Jochen Geck}
\author{Dmytro S.\ Inosov}
\email{dmytro.inosov@tu-dresden.de}
\affiliation{Institut f\"ur Festk\"orper- und Materialphysik, Technische Universit\"at Dresden, 01069 Dresden, Germany}
\affiliation{W\"urzburg-Dresden Cluster of Excellence on Complexity and Topology in Quantum Matter -- ct.qmat, Technische Universit\"at Dresden, 01069 Dresden, Germany}

\author{Darren C.\ Peets}
\email{darren.peets@tu-dresden.de}
\affiliation{Institut f\"ur Festk\"orper- und Materialphysik, Technische Universit\"at Dresden, 01069 Dresden, Germany}

\begin{abstract}

In frustrated magnetic systems, the competition amongst interactions can introduce extremely high degeneracy and prevent the system from readily selecting a unique ground state.  In such cases, the magnetic order is often exquisitely sensitive to the balance among the interactions, allowing tuning among novel magnetically ordered phases. In antlerite, \antH, Cu$^{2+}$ ($S=1/2$) quantum spins populate three-leg zigzag ladders in a highly frustrated quasi-one-dimensional structural motif.  We demonstrate that at zero applied field, in addition to its recently reported low-temperature phase of coupled ferromagnetic and antiferromagnetic spin chains, this mineral hosts an incommensurate helical+cycloidal state, an idle-spin state, and a multiple-$q$ phase which is the magnetic analog of misfit crystal structures.  The antiferromagnetic order on the central leg is reentrant.  The high tunability of the magnetism in antlerite makes it a particularly promising platform for pursuing exotic magnetic order.

\end{abstract}

\maketitle

\section{Introduction}

When attempting to expand our understanding of magnetism and reveal exotic new magnetic phases or the influence of less-studied magnetic interactions, it is necessary to prevent the exchange interactions from producing a more-conventional magnetic ground state.  This can be accomplished by introducing strong frustration --- the competition among interactions --- most commonly by arranging the spins in a geometry that pits different interactions against each other\,\cite{Ramirez1994,Batista2016,Schmidt2017}.  
This impedes the spin system selecting a unique global ground state, leading to a wide variety of physical phenomena in which fluctuations, quantum mechanical effects, and fine details of the spin-spin interactions can be crucial\,\cite{Lacroix2011}.  Another approach is to investigate low-dimensional structures, where the reduced number of magnetic interactions makes it more difficult to stabilize magnetic order.  

To observe quantum mechanical effects, the spin must be small, preferably $S=1/2$.  The stability of the $3d^9$ electronic configuration of Cu$^{2+}$ makes it probably the most accessible magnetic ion for quantum magnetism, while its low spin-orbit coupling makes it a nearly pure-spin moment.  Cu$^{2+}$ is best known in complex oxides, typically in octahedral coordination.  These materials predominantly form square magnetic sublattices, for instance in the cuprate superconductors\,\cite{Park1995}.  Frustration can arise in such lattices from a competition between nearest- and next-nearest-neighbor interactions, but not through geometric constraints.  Low-dimensional copper ladder compounds are well-established and are known to exhibit incommensurate magnetic order, among other interesting physics\,\cite{Smaalen1999}, but geometric frustration does not play a significant role in these materials since their copper sublattices are typically bipartite.

\begin{figure}[htb]
  \includegraphics[width=\columnwidth]{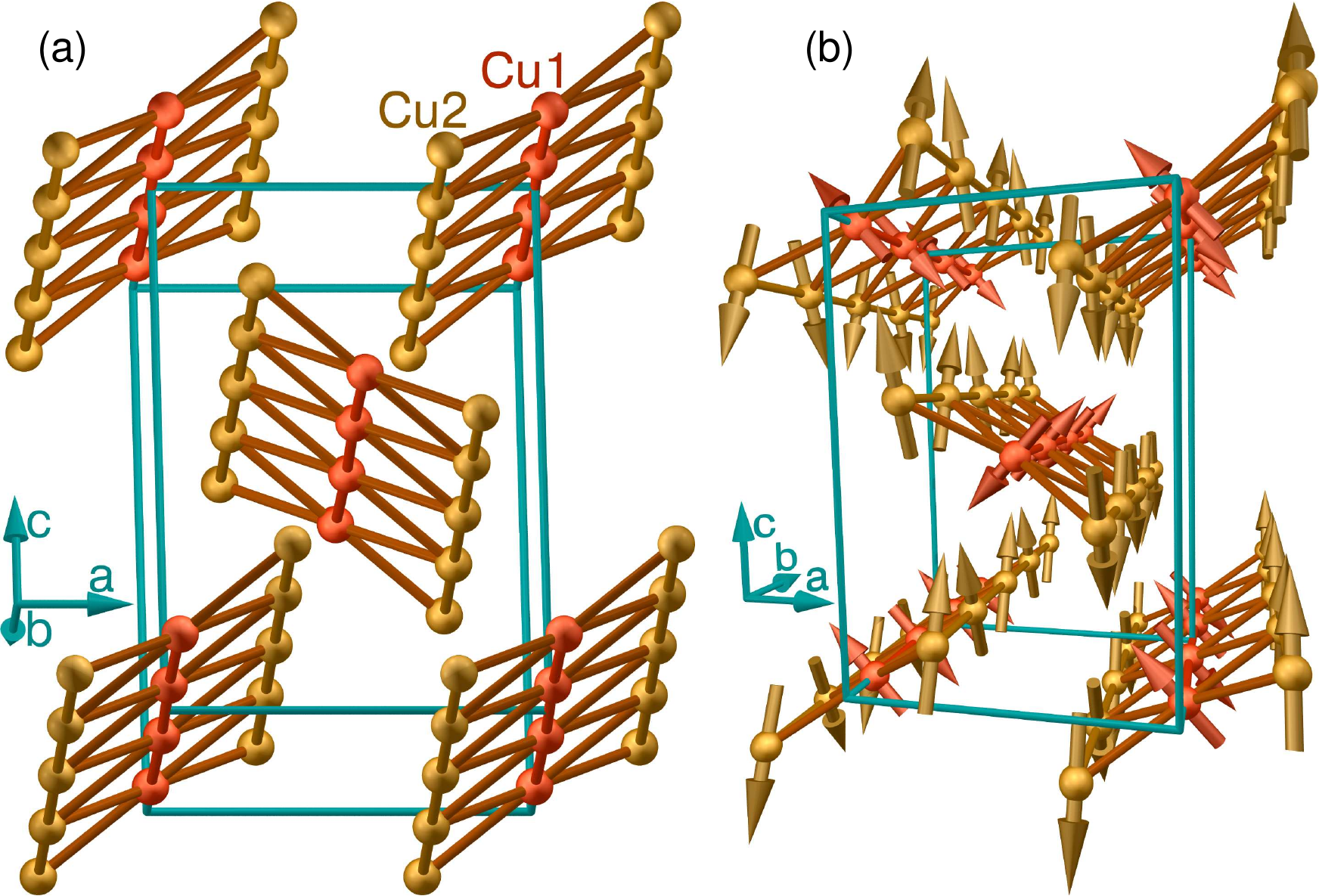}
  \caption{\label{fig:struct}(a) Copper sublattice in antlerite. (b) Low-temperature magnetically ordered phase at 2.2\,K, based on Ref.~\cite{Kulbakov2022a}.}
\end{figure}

Copper-based minerals are nevertheless a rich source of novel frustrated lattices, often built up of distorted Cu$^{2+}$ triangles\,\cite{Inosov2018}, and have proven a rich platform for novel physics.  Examples include the candidate quantum spin-liquid state in herbertsmithite\,\cite{Shores2005,Norman2016}; enormous effective moments in atacamite Cu$_2$Cl(OH)$_3$\,\cite{Heinze2021}; and spinon-magnon interaction in botallackite Cu$_2$(OH)$_3$Br\,\cite{Zhang2020bot}.  The natural mineral antlerite considered in this work is a three-leg-ladder compound with a magnetic sublattice comprising Cu$^{2+}$ ions arranged in distorted triangles\,\cite{Hawthorne1989}, realizing a zigzag ladder --- its copper sublattice is depicted in Fig.~\ref{fig:struct}(a).  This structure has been studied far less than the conventional ladder cuprates, and may be expected to reveal new physics.  

Neutron diffraction on antlerite initially indicated that only the outer legs of the ladder possess an ordered moment, whereas the central leg exhibited ``idle-spin'' behavior\,\cite{Vilminot2003,Hara2011}.  While such a picture would be very interesting, follow-up studies seriously questioned this result.  First, density-functional-theory (DFT) calculations showed strong antiferromagnetic coupling along the central leg of the ladder, which would be expected to lead to order\,\cite{Koo2012}. In this work, most relevant exchange interactions were considered, including Dzyaloshinskii-Moriya terms, and a number of possible spin configurations were suggested. Then, Fujii {\itshape et al}.\,\cite{Fujii2013} performed specific-heat characterization and proton NMR measurements on antlerite single crystals in magnetic field. They found a vastly more complex magnetic phase diagram than previously imagined, which hosts four distinct magnetic phases in zero field and at least five other field-induced phases for $B\parallel c$ alone. Their NMR results could only be explained if the system had at least four distinct magnetic sites, at odds with the idle-spin picture.  Finally, the low-temperature magnetic state was determined by neutron powder diffraction and DFT calculations to comprise mutually antialigned ferromagnetic outer legs and an antiferromagnetic inner leg, with significant canting\,\cite{Kulbakov2022a}, as depicted in Fig.~\ref{fig:struct}(b).  The ordered moment on the outer legs of the ladder was nearly the full moment expected for Cu$^{2+}$, while the moment on the central leg was $\sim$20\%\ lower at 2.2\,K.  This suggests that quantum fluctuations play a relatively minor role in the ferromagnetic legs of the ladder.  DFT calculations were only able to converge on this state once updated atomic positions were available, and even then indicated other states at similar energy, with both the inner and outer legs of the ladder independently on the verge of a phase transition.  This suggests a rich and tunable phase diagram with parameters such as field, strain, pressure, or chemical substitution, as the published $H$--$T$ phase diagrams already hint at\,\cite{Hara2011,Fujii2013}.  

Here, we verify and identify the remaining low-field magnetic phases, revealing incommensurate and multiple-$q$ order, as well as an idle-spin state at intermediate temperatures which closely resembles the previously proposed ground state\,\cite{Vilminot2003}.  

\section{Experimental Details}

\paragraph{Sample preparation.}Synthetic antlerite was prepared by hydrothermal synthesis under autogenous pressure at 180\,$^\circ$C in a Teflon-lined stainless steel autoclave.  CuSO$_4\cdot 5$H$_2$O (Alfa Aesar, 99\%) and Cu(OH)$_2$ (Alfa Aesar, 94\%) in a typical molar ratio of 1:2 were well mixed in distilled water inside the liner, which was then sealed inside the autoclave.  Deuterated samples for neutron scattering were prepared using D$_2$O (Acros Organics, 99.8\% D) in place of water.  The mixtures were heated at 90\,$^\circ$C/h to 180\,$^\circ$C in a convection drying oven, held at that temperature to react, cooled at 50--65\,$^\circ$C/h to 50\,$^\circ$C, then cooled freely to room temperature.  Reaction times of 1--3 days were used for powder, 5 days for typical (100--200\,$\mu$m) single crystals, or up to several months when larger crystals were desired.  Examples of several larger crystals are shown on millimetre-ruled graph paper in the inset to Fig.~\ref{fig:cP}(b).

\paragraph{Physical properties.}Magnetization was measured using a Quantum Design Magnetic Property Measurement System MPMS3, through a SQUID-detected vibrating-sample magnetometer (VSM), in both zero-field-cooled-warming (ZFC) and field-cooled-warming (FC) conditions. The 0.215-mg single crystal was mounted to a quartz bar sample holder using Duosan glue, the contribution from which was measured separately and subtracted.  The extremely fine point spacing chosen to clearly resolve the transitions led to increased noise in the derivative d$(M/H)$/d$T$, so the derivatives shown in this paper incorporate a 6-point moving average, which leads to slight broadening.

Low-temperature specific heat measurements were performed on a 4.75-mg mosaic of $\sim$20 crystals using a Quantum Design Physical Property Measurement System (PPMS DynaCool) equipped with a $^3$He refrigerator.  Contributions from the sample holder and grease were subtracted.  Multiple data points were collected at each temperature and averaged; it was necessary to discard the first data point at most temperatures due to incomplete thermal stabilization.  The ac specific heat was measured on a small single crystal in a Cryogenic Limited Cryogen-Free Measurement System using its ac calorimetry probe.  Measurements were performed at 16\,Hz, using an ac power of 9.5\,$\mu$W.  Decreasing the power to 2.3\,$\mu$W made the peaks sharper at the cost of an increased noise level; since we use these data primarily for higher temperatures, the higher-power data are shown.  In both experiments, the sample was mounted to the measurement puck using a thin layer of Apiezon N grease.

\paragraph{Diffraction.}To investigate the thermal evolution of the magnetic structure, high-intensity neutron powder diffraction was performed on a $\sim$5\,g sample at the WOMBAT diffractometer at ANSTO\,\cite{Wombat}, in Sydney, Australia, from 14 to 135.5$^\circ$ in steps of 0.125$^\circ$, using 2.41-\AA\ neutrons and count times of 30\,min, in temperature steps of 0.1\,K through the transitions.  Further high-intensity diffraction data were collected on a $\sim$7\,g powder sample at the D1B\,\cite{D1B} and D20\,\cite{D20} diffraction beamlines at the Institut Laue-Langevin (ILL), in Grenoble, France.  Count times were 30--60\,min to visualize the transitions, and $\sim$20\,h for data required for refinement of the highest-temperature phase.  A scan through the transitions with 0.1-K temperature steps was performed before all higher-resolution measurements to validate the thermometry. At D1B, data were collected from 1 to 129$^\circ$ in steps of 0.1$^\circ$, and at D20 from 0.17 to 151.37$^\circ$ in steps of 0.05$^\circ$. Neutrons with calibrated wavelengths $\lambda = 2.529$\,\AA\ (D1B) or 2.412\,\AA\ (D20) were selected with highly oriented pyrolytic graphite (HOPG) [002] monochromators. On both beamlines, the contribution of the instrument to the peak broadening was determined from the refinement of a Na$_2$Ca$_3$Al$_2$F$_{14}$ standard sample, while the wavelength was refined using a Si standard. Parasitic diffraction peaks arising from the sample environment were eliminated using radial oscillating collimators.  Additional powder neutron diffraction with polarization analysis was performed on the D7 beamline\,\cite{D7_2} at the ILL, using a wavelength of 4.8707\,\AA.  Powder diffraction data were Rietveld-refined in {\sc FullProf} by the full-matrix least-squares method\,\cite{FullProf}, using the scattering factors from Ref.~\onlinecite{Sears1992}.  

\paragraph{Spectroscopy.}X-ray absorption spectroscopy was measured at room temperature at beamline UE46 PGM-1 at BESSY-II, in Berlin\,\cite{UE46}.  A counter voltage was applied to prevent photoelectrons from entering the detector.  Data were collected in total-fluorescence-yield mode to ensure bulk sensitivity, and were normalized to the incoming beam current.  Further normalization was based on the average signals in the 909.9--917.9 and 952.5--953.1\,eV energy ranges, well below and above the L$_2$ and L$_3$ edges.  

Optical absorption spectra in the ultraviolet (UV), visible (vis) and near-infrared (NIR) range from 200 to 900\,nm were collected at room temperature in a Varian Cary 4000 spectrophotometer with a scan rate of 300\,nm/min, a step of 0.5\,nm, a spectral bandwidth of 2\,nm, and averaging time of 0.1\,s.  15\,mg of antlerite crystal was ground together with 200\,mg of BaSO$_4$ to avoid excessive absorption.  

\paragraph{Muon Spin Rotation.}Muon spin rotation and relaxation ($\mu$SR) experiments were performed on \antD\ powder at the Swiss Muon Source (S$\mu$S) at the Paul Scherrer Institut (PSI), Switzerland, using the nearly 100\%\ spin-polarized positive-muon beam at the GPS instrument\,\cite{GPS}. Spectra were measured at temperatures down to 1.5\,K in zero field (ZF), weak longitudinal field (wLF, 2\,mT field applied along the initial muon spin direction) and weak transverse field modes (wTF, 5\,mT field applied perpendicular to the initial muon spin direction). The sample was wrapped in an aluminized mylar foil and mounted within the beam spot between the tines of an ultrapure copper fork. Thermal contact was achieved {\itshape via} helium exchange gas. Analysis of the data was performed in the time domain by the least-squares method using the program {\sc Musrfit}\,\cite{musrfit}. 

\section{X-ray Absorption Spectroscopy}

\begin{figure}[t!]
  \includegraphics[width=\columnwidth]{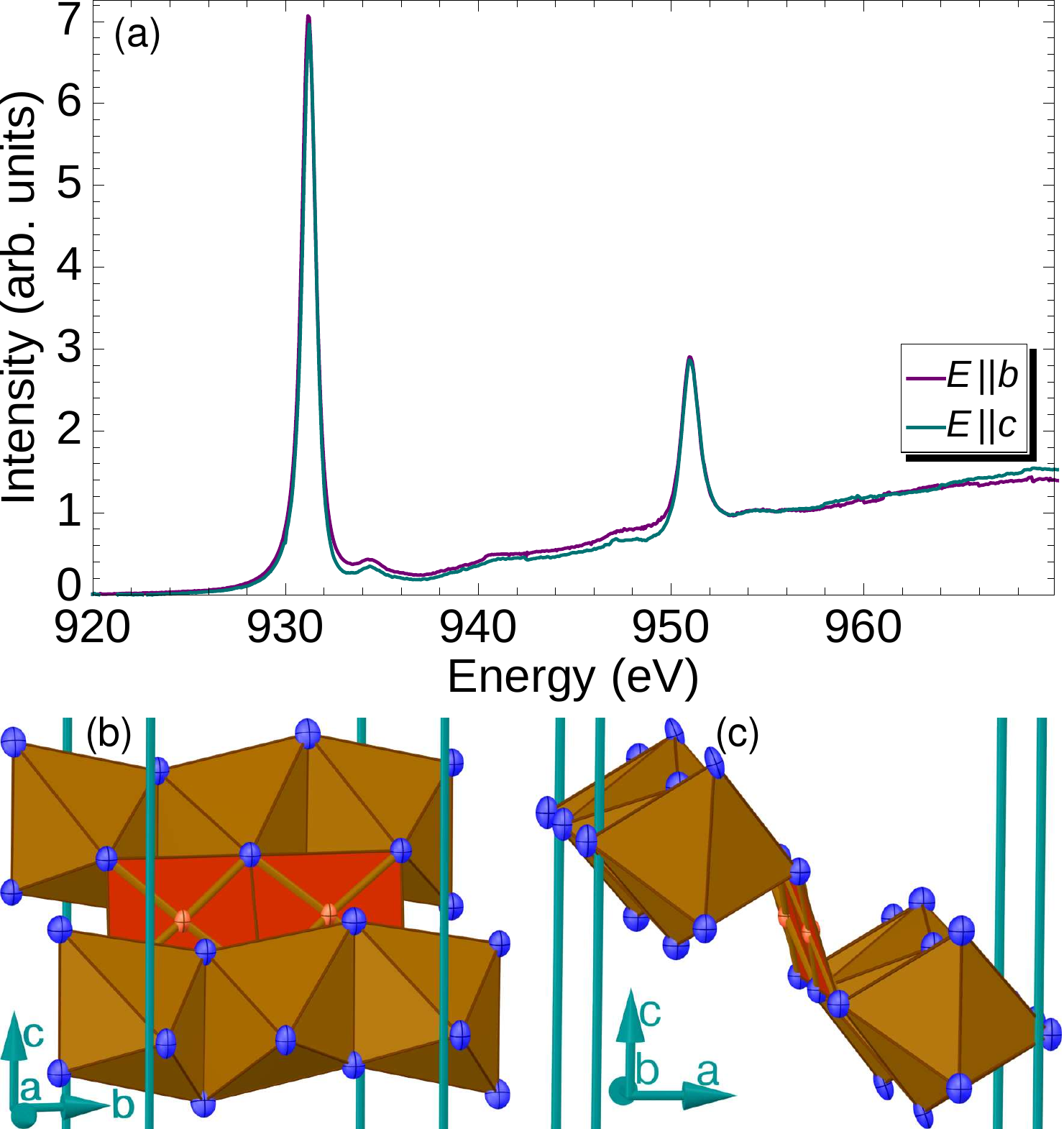}
  \caption{\label{XAS}(a) X-ray absorption spectra of antlerite at the Cu L$_2$/L$_3$ edges at room temperature, taken in total-fluorescence-yield mode for $\vec E\parallel\hat b$ and $\hat c$.  (b,c) Views of the first coordination sphere for the Cu sites on one ladder.}
\end{figure}

We begin by verifying the oxidation state of Cu in \antH.  Cu L-edge x-ray absorption spectra collected in total-fluorescence-yield mode for beam polarizations along $\hat b$ and $\hat c$ are shown in Fig.~\ref{XAS}(a).  The strong peak at 931.2\,eV is the characteristic L$_3$ line of $3d^9$ Cu$^{2+}$, while the weak hump at 934.3\,eV suggests a small $3d^{10}$ contribution; these energies and the Cu$^{2+}$ peak shape are in good agreement with those found in Ref.~\onlinecite{Grioni1992} for CuO and Cu$_2$O.  This indicates that copper atoms in antlerite exist predominantly in the expected localized $3d^9$ (Cu$^{2+}$) configuration, while the $3d^{10}$ contribution may indicate a small excess population of protons, perhaps due to some OH$^-$ ligands actually being H$_2$O. This was not detectable in our structure refinements.  A $3d^{10}$ contribution is also observed in materials with strongly-correlated electrons, where some hole density is transferred to the ligands to reduce on-site repulsion at the Cu site, and in some CuO samples\,\cite{Grioni1989}.  The peak heights are slightly lower than expected compared to the jump across the absorption edge, an indication of strong self absorption which we have not corrected for.  The splitting between the L$_3$ and L$_2$ lines is 19.8\,eV, consistent with reports on other Cu$^{2+}$ materials\,\cite{Jiang2013}.  This is a measure of the spin-orbit coupling in the $2p$ orbitals\,\cite{Chen1997}.

No significant difference was observed between polarizations parallel and perpendicular to the ladders.  As can be seen from the CuO$_x$ polyhedra for a single ladder in Figs.~\ref{XAS}(b) and \ref{XAS}(c), the local coordinate system has very different orientations on the two copper sites as well as on adjacent sites on the outer legs of the ladder, so x-ray absorption is not able to determine exactly in which orbitals the holes in antlerite reside.

\section{Phase Transitions}

\begin{figure}[t!]
  \includegraphics[width=\columnwidth]{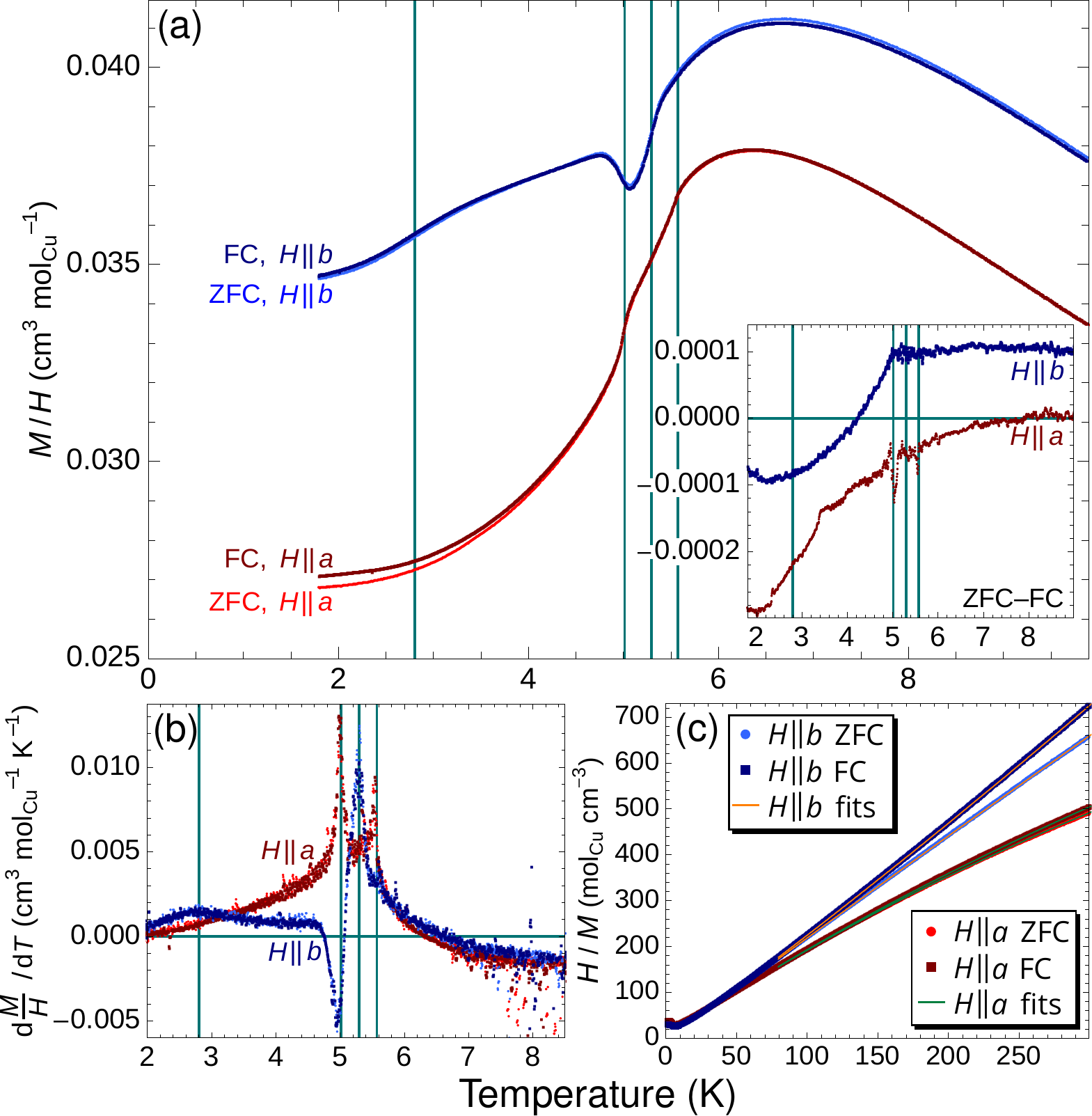}
  \caption{\label{fig:M}(a) Normalized magnetization $M/H$ data for antlerite, in fields $H\parallel a$ and $b$ of 1000\,Oe.  The inset plots the difference between ZFC and FC data for each axis, showing a sharp onset at the 5.0-K transition for $H\parallel b$.  Vertical lines represent the transition temperatures extracted from the specific heat data in Fig.~\ref{fig:cP}.  (b) Derivatives of the magnetization curves in (a), identifying the transitions.  (c) Curie-Weiss fits to all four inverse normalized magnetization datasets above 80\,K.}
\end{figure}

We now turn to the phase transitions at low field. Magnetization data collected under zero-field-cooled- and field-cooled-warming conditions are shown in Fig.~\ref{fig:M}(a) for 1000\,Oe fields parallel to the $a$ and $b$ axes.  Two transitions are visible for fields along $a$, at 5.00 and 5.55\,K, while for fields along $b$ transitions appear at 4.97 and 5.30\,K and are accompanied by a broad transition centered near 2.9\,K.  The transitions around 5.0--5.5\,K are clearer in the derivative plot shown in Fig.~\ref{fig:M}(b).  For fields along $b$ the transition at 4.97\,K is a ferromagnetic-like increase in magnetization on cooling, while all other observed transitions manifest as an antiferromagnetic-like decrease.  A slight splitting can be observed between the field-cooled and zero-field-cooled data, with a clear onset at 5.0\,K for $b$-axis fields as seen in the difference between ZFC and FC data, plotted in the inset to Fig.~\ref{fig:M}(a), indicating either the polarization or freezing of some spin degree of freedom.

The unambiguous detection of three distinct, sharp magnetic ordering transitions around 5\,K with an additional transition at lower temperature confirms the result of Fujii {\itshape et al}.~\cite{Fujii2013} based on comprehensive $^1$H-NMR and specific heat data, and indicates that their rather complex $H$--$T$ phase diagram is most likely correct.  The visibility of the two highest-temperature transitions only for fields along $a$ and $b$, respectively, but not for both orientations, is unusual, and may explain why other groups did not observe all transitions.  

The first transition encountered upon cooling, at 5.55\,K, evidently concerns spin components perpendicular to $b$, the direction along which the three-leg ladders run, and is antiferromagnetic.  The second transition, at 5.30\,K, is more surprising since it evidently concerns antiferromagnetic moments perpendicular to $a$, but neither the plane of any three-leg spin ladder nor its normal lies along $a$, so $a$ might not be expected to be a special direction for spin order.  The third transition around 5.0\,K occurs at temperatures that are within the uncertainty for the two field orientations, and is assumed to be a single transition.  This is antiferromagnetic along $a$, but the small increase in magnetization and slight difference between ZFC and FC data below this temperature for fields along $b$ would appear to suggest a transition involving a canting angle toward the $b$ axis. We will show that this is actually the result of a partial loss of long-range order as we enter the gap between two phases in a reentrant system.  Finally, around 2.9\,K there is a loss of susceptibility for $H\parallel b$, which we will show is the same reentrant magnetism.

The paramagnetic state was characterized by an extended Curie-Weiss law which considers a small temperature-independent paramagnetic contribution $\chi_0$ as in Ref.~\onlinecite{Hara2011}:  $\chi(T) = \chi_0 + C/(T-\Theta_\text{CW})$. This constant offset reproduces the curvature seen in the inverse normalized magnetization in Fig.~\ref{fig:M}(c). Due to the more pronounced anisotropy in the paramagnetic state reported below 50\,K, the temperature region from 80 to 300\,K was fit as in Ref.~\onlinecite{Hara2011}. Our extended Curie-Weiss fits yield an effective moment $\mu_\text{eff} = 1.96\pm0.04$\,$\mu_\text{B}$ and $\Theta_\text{CW} = 0.3\pm0.2$\,K for fields along $a$, and $\mu_\text{eff} = 1.87\pm0.04$\,$\mu_\text{B}$ and $\Theta_\text{CW} = 4\pm3$\,K for fields along $b$.  



The effective moments $\mu_\text{eff}$ are similar to those in Ref.~\onlinecite{Hara2011} and significantly above the spin-only value of 1.73\,$\mu_\text{B}$ expected for $S=1/2$ Cu$^{2+}$, indicating an orbital contribution. This results in a $g$ factor slightly larger than 2, as also reported based on ESR results\,\cite{Okubo2009}, and supports the scenario of non-negligible spin-orbit coupling proposed previously\,\cite{Hara2011,Vilminot2003}. The small and slightly positive Curie-Weiss temperatures indicate a balance between antiferromagnetic and ferromagnetic exchange pathways, with a slight edge for ferromagnetism.  DFT calculations and the low-temperature magnetic structure indeed suggest ferromagnetic interactions on the two outer legs of the ladder and antiferromagnetic exchange on the central leg\,\cite{Kulbakov2022a}.  

\begin{figure}[tb]
  \includegraphics[width=\columnwidth]{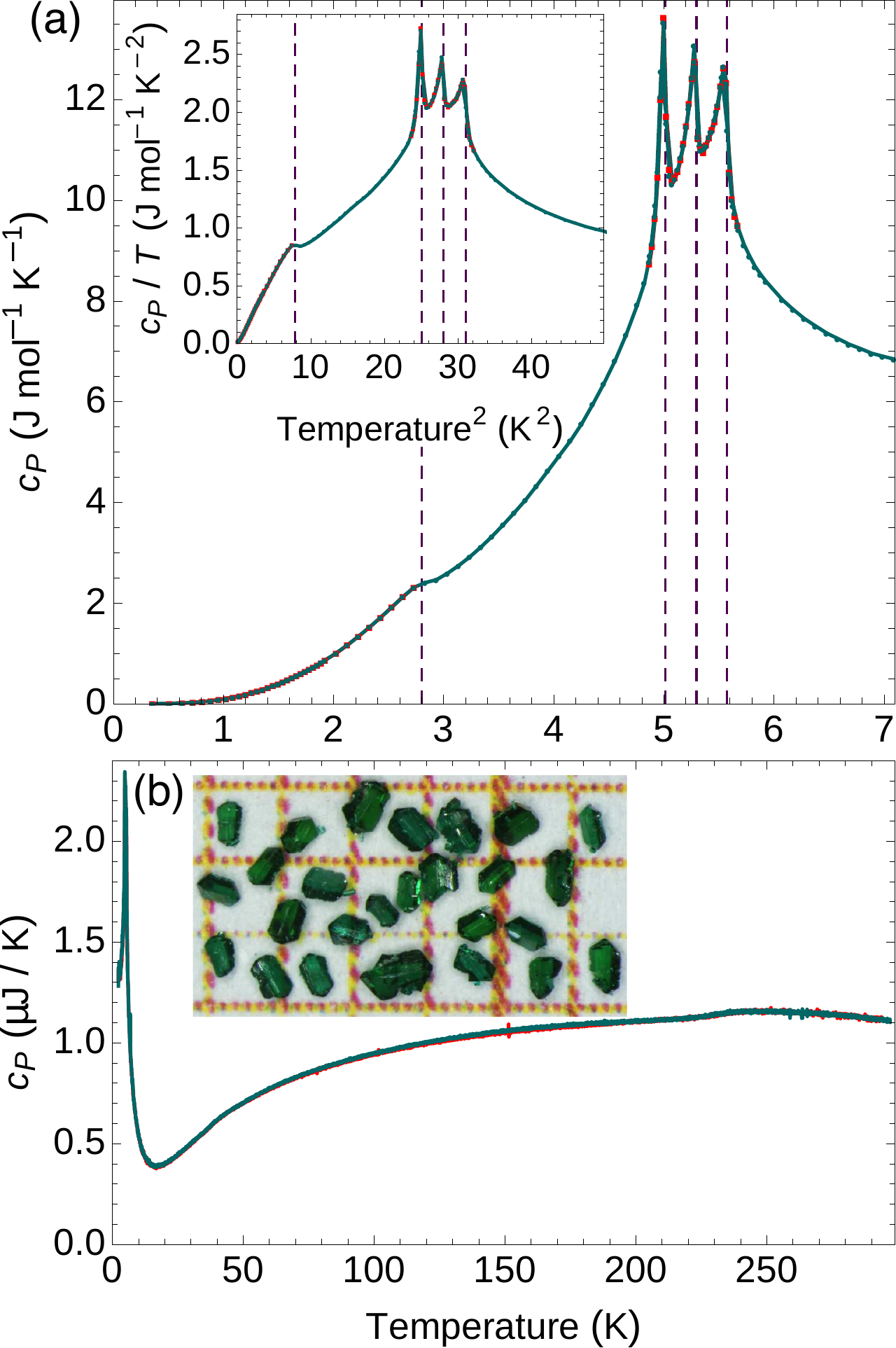}
  \caption{\label{fig:cP}(a) Zero-field low-temperature specific heat data on antlerite, showing four clear transitions.  Red points were measured on warming and turquoise on cooling.  The inset plots $c_P/T$ vs $T^2$.  (b) ac Calorimetry of antlerite to higher temperatures.  The inset shows several single crystals on mm-ruled graph paper.}
\end{figure}

Our low-temperature specific heat data, shown in Fig.~\ref{fig:cP}(a), indicate sharp phase transitions at 2.80, 5.01, 5.30, and 5.57\,K, defined using entropy-conserving constructions around each jump; measurements through the higher-temperature peaks on warming and cooling detected no hysteresis.  The transition temperatures are consistent with Ref.~\onlinecite{Fujii2013} and our magnetization data above --- this agreement is emphasized by the vertical lines in Fig.~\ref{fig:M}, which mark the transition temperatures extracted from the specific heat data.  The upper three transitions were also clearly distinguished in our ac specific heat data, shown to higher temperatures in Fig.~\ref{fig:cP}(b), which did not extend to low enough temperature to clearly resolve the 2.8-K transition.  The inset to Fig.~\ref{fig:cP}(a), which replots the low-temperature data as $c_P/T$ vs $T^2$, shows that the 2.8-K transition is associated with a collapse in entropy.  Below 1.6\,K the specific heat data vary as $T^{2.5}$, with a slight upward deviation below 0.5\,K that may be associated with nuclear magnetism.  As previously reported\,\cite{Kulbakov2022a}, no further phase transitions are observed down to 0.35\,K.

Figure~\ref{fig:cP}(b) shows ac specific heat data to higher temperatures;  
it was not possible to put these data in absolute molar units.  A weak, broad hump around 230\,K is associated with the glass transition of Apiezon N grease.  The magnetic entropy persists up to at least 20\,K and conceivably as high as 40\,K, several times higher than any magnetic transition.  This likely arises from the low dimensionality, as well as frustration among the spins in each ladder.  

\section{Magnetic Phases}
\hyphenation{wombat}

\begin{figure*}[htb]
  \includegraphics[width=0.75\textwidth]{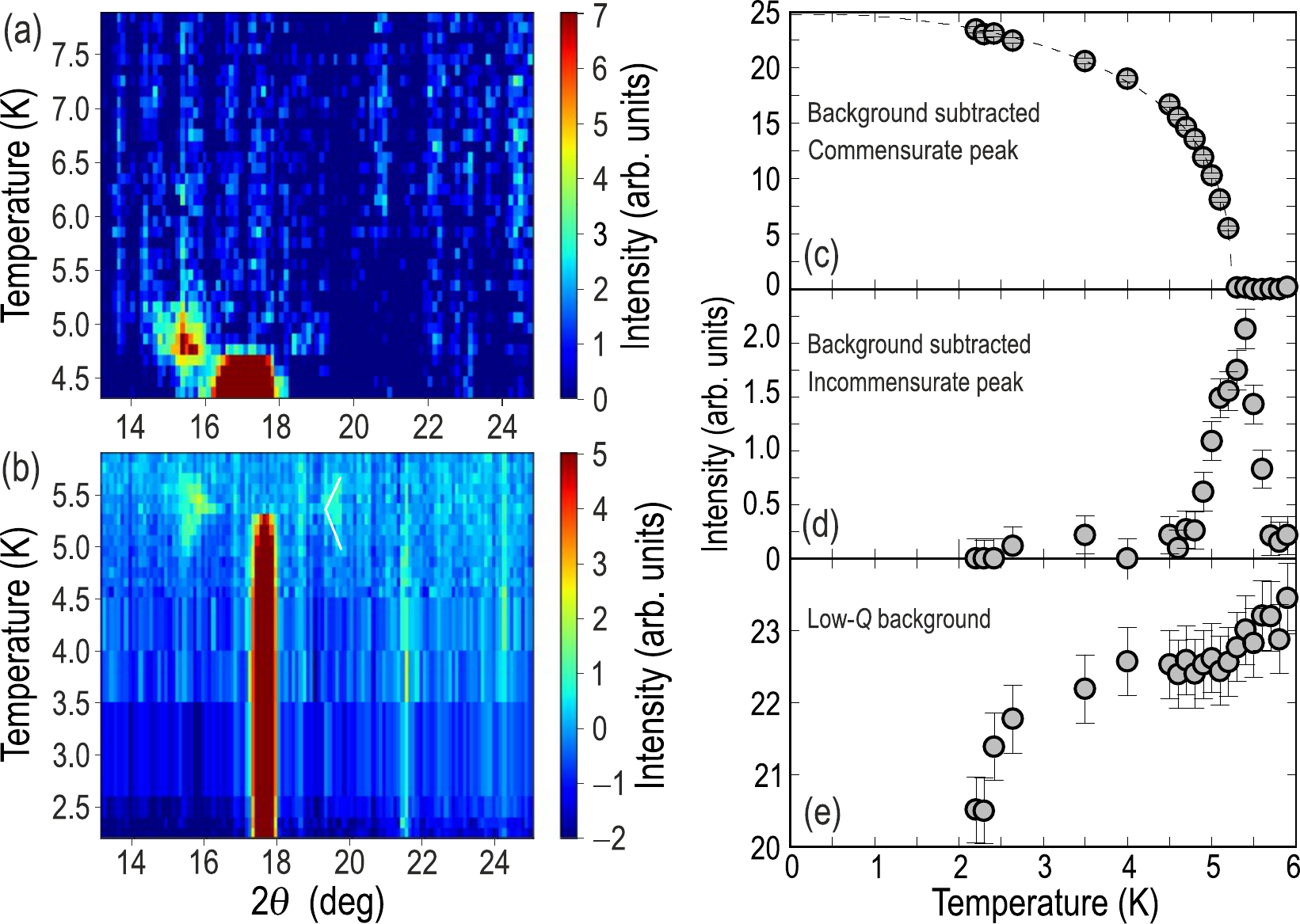}
  \caption{\label{fig:Mickey}Temperature-dependent magnetic peak intensity, after subtraction of a high-temperature background.  (a,b) Magnetic intensity color maps (after subtraction of background measured above \TN) from (a) WOMBAT, ANSTO and (b) D1B, ILL, highlighting the coupling between the incommensurate and commensurate magnetism. The strong magnetic peak between 16$^\circ$ and 18$^\circ$ is (100).  The white line in (b) is a guide to the eye.  (c) Peak intensity of (100), the strongest commensurate magnetic reflection in (b), as a function of temperature.  (d) Temperature dependence of the peak intensity in the stronger incommensurate reflection in (b).  (e) Integrated intensity in the background at low angles (11--12$^\circ$) of the background-unsubtracted data presented in (b), as a function of temperature.}
\end{figure*}

The magnetic order in the lowest-temperature phase has already been reported\,\cite{Kulbakov2022a} [see also Fig.~\ref{fig:struct}(b)], so we concentrate here on the higher-temperature phases. A temperature-dependent map of magnetic intensity through the sharp transitions at 5.0--5.6\,K near the strongest magnetic reflection was obtained from WOMBAT (ANSTO) data by subtracting the intensity in the paramagnetic phase at 7.9\,K.  The result is shown in Fig.~\ref{fig:Mickey}(a).  It is immediately clear that the highest-temperature magnetically ordered phase is incommensurate --- the (1\,0\,0) reflection splits into (1$\pm\delta$\,0\,0), where $\delta$ varies between $\sim 0.07$ and 0.12. The noninteger $\delta$ appears to vary continuously without locking into a rational value, indicating that this phase is not commensurate. The $\sim$0.5\,K temperature shift in these data is most likely associated with a difference in calibration of temperature sensors used in the measurements.

Similar data collected on the D1B diffractometer at the ILL are shown in Fig.~\ref{fig:Mickey}(b), where the intensity in the paramagnetic phase at 5.9\,K has been subtracted. Of particular note, the incommensurate and commensurate peaks {\slshape coexist} for $\sim$0.5\,K.  This is not phase separation or coincidental coexistence, for instance from different regions of the sample or separate Cu ladders.  The incommensurate peaks form an hourglass, pinching toward the point at which the commensurate peak is extinguished, which implies that these wavevectors are strongly coupled.  A similar pinching is observed in structural Bragg peaks in the incommensurate charge-density-wave phase of 2H-TaSe$_2$\,\cite{Leininger2011}, and is a fingerprint of multiple-$q$ order.  The phase diagram of Fujii {\itshape et al}.~\cite{Fujii2013} implies that this multiple-$q$ phase exists in only a very narrow window for fields along $b$, but it may be their ``$\beta$1'' phase for fields along $c$, in which case it would occupy a far broader swath of the phase diagram.  In an earlier $H$--$T$ phase diagram based on magnetization measurements\,\cite{Hara2011}, a similar bubble was also observed for fields along $a$, extending down to quite low temperatures.

The temperature dependence of the strongest magnetic reflection is shown in Fig.~\ref{fig:Mickey}(c).  The integrated peak intensity extracted from a Gaussian fit decreases toward the transition in an order-parameter-like fashion, as might be expected.  This peak vanishes at the middle transition, which according to magnetization and specific heat data is at 5.30\,K and represents antiferromagnetic order largely perpendicular to $a$.  The most obvious candidate from the low-temperature state would be the outer (Cu2) legs of the ladder --- the spins on these legs lie perpendicular to $a$ at low temperature, with opposite legs antialigned\,\cite{Kulbakov2022a}. We may thus expect to find commensurate ordering of the Cu2 sites below 5.30\,K.  

The integrated intensity in the incommensurate peaks, shown in Fig.~\ref{fig:Mickey}(d) based on a three-Gaussian fit, is peaked immediately above the temperature at which the commensurate peak is extinguished.  The magnetization data suggest a spin orientation perpendicular to $b$ in the incommensurate phase, consistent with a ripple along $a$ as suggested by the peaks at (1$\pm\delta$\,0\,0) and suggesting a cycloidal state.  Since all spins in the lowest-temperature state are perpendicular or nearly perpendicular to $b$, it is difficult to speculate based on the magnetization data alone which spins participate in the incommensurate phase, although it may be natural to expect the Cu1 site to play a key role if we associate the commensurate peak primarily with Cu2.

\begin{figure*}[htb]
\includegraphics[width=\textwidth]{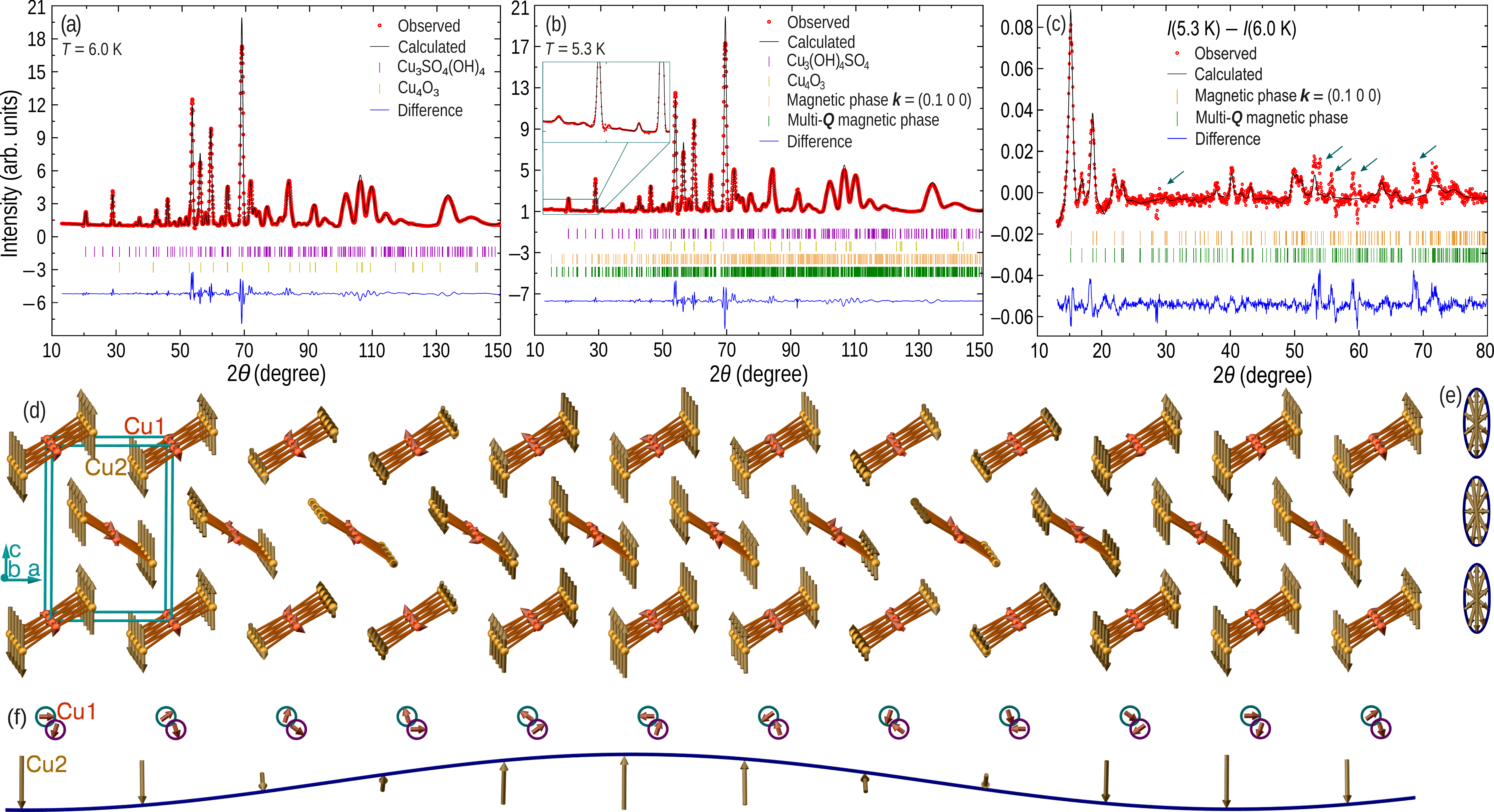}
\caption{\label{IC}Refinement of the incommensurate phase.  Data from D20 are shown at (a) 6.0\,K in the paramagnetic state and (b) 5.3\,K where the incommensurate phase begins to yield to multiple-$q$ order;  a refinement of their difference is plotted in (c), and the new magnetic peaks are highlighted in an inset in (b).  Artifacts due to imperfect subtraction of strong nuclear peaks are marked with arrows.  The refined incommensurate order is shown in (d), and (e) shows a view along $a$ of the elliptical helical order on the corner (upper and lower) and body-centered (middle) Cu2 ladders. The order pattern on selected Cu1 and Cu2 sites is shown in (f), using the same axes as in (d).}
\end{figure*}

Another feature in Fig.~\ref{fig:Mickey}(b) may not be immediately obvious --- below roughly 3\,K, the background magnetic intensity values in this colormap are strongly negative.  The presence of a diffuse component in the subtracted 5.9\,K data would be expected, since we see indications in the specific heat of short-range order up to $\sim$40\,K.  However, such a diffuse component should condense into the magnetic Bragg peaks upon entering the long-range-ordered state, presumably around 5.0-5.6\,K.  The temperature dependence of the diffuse background does indeed show a decrease in that temperature range, but it exhibits a much stronger collapse below 3\,K, as seen in Fig.~\ref{fig:Mickey}(e).  The order-parameter-like behavior seen in Fig.~\ref{fig:Mickey}(c) shows no evidence of this intensity condensing into the (100) peak.  

\section{Magnetic Structures}

Neutron diffraction was also able to fully identify the zero-field magnetic phases. To obtain the statistics required to refine the higher-temperature incommensurate structure based on its much-weaker peaks, we collected additional diffraction data on the D20 neutron diffractometer at the ILL. Data were first collected as a function of temperature to validate the thermometry, then for longer times at 6.0\,K in the paramagnetic phase and at 5.3\,K where the (1$\pm\delta$\,0\,0) peaks reach their maximum intensity.  Figure~\ref{IC}(a) shows the baseline data from D20 in the paramagnetic phase at 6.0\,K, in Fig.~\ref{IC}(b) we present data at 5.3\,K, and the difference between these two scans is shown in Fig.~\ref{IC}(c).  Arrows mark artifacts from the imperfect subtraction of a few strong nuclear peaks, which is likely due to the peaks becoming sharper as phonons freeze out.  This difference was refined to determine the magnetic structure.  Since 5.3\,K is on the cusp of the multiple-$q$ phase, weak traces of the latter phase are visible, but the incommensurate peaks dominate.  A contribution from the multiple-$q$ phase was found to be present at the 1.3\%\ level and was refined as discussed below. 

Of the four possible magnetic irreducible representations, only $\Gamma_2$ gave an acceptable description of the magnetic intensity, but the unconstrained application of $\Gamma_2$ leads to physically unreasonable spin alignments on some ladders --- otherwise-equivalent ladders would have very different spin orientations, which would only be possible if the extremely weak interladder couplings overpowered the strong intraladder interactions.  If we constrain the Cu2 ordered moments to have equal magnitude on any given ladder, with spins on opposite legs exactly antialigned as in the lowest-temperature phase, and allow only one angle between consecutive spins on the central legs, a physically plausible magnetic model is obtained which does not lead to any reduction in the quality of the refinement.  As shown in Fig.~\ref{IC}(d), the incommensurate phase corresponds to elliptical-helical order of the antialigned ferromagnetic spins on the outer Cu2 legs of the ladder paired with cycloidal order on the inner Cu1 leg.  The helical and cycloidal order on selected Cu sites are also shown separately in Fig.~\ref{IC}(f), and a view of the elliptical order along the $a$ axis is shown in Fig.~\ref{IC}(e).  The propagation vector in this phase is ($\delta$\,0\,0), where $\delta\approx 0.1$ at 5.3\,K.  The elliptical helix on the outer legs has major and minor axes of 0.41(4) and 0.15(8)\,$\mu_\text{B}$, while the cycloidal ordered moment on the central leg is 0.09(7)\,$\mu_\text{B}$.  If we exclude magnetic peaks which overlap with nuclear peaks, the same order is obtained, including ellipticity, but with fewer peaks it is more difficult to quantify the ellipticity.  These are small fractions of the full 1\,$\mu_\text{B}$ moment, most likely because these refinements are performed on data collected at $\sim$95\% of \TN, where one would not ordinarily expect fully ordered moments.  There may also be a reduction in the ordered moments due to a competition between the inner and outer legs, for instance if these would prefer incompatible ordered states but are forced to compromise.  The angle between adjacent spins along the central leg refines to 108(40)$^\circ$, likely due to a competition between nearest-neighbor and next-nearest-neighbor interactions along the leg, combined with the exchanges to the outer legs. A magnetic crystallographic information file (mCIF) describing this refinement is available as an arXiv ancillary file as described in Appendix \ref{suppl}.  We note that several other forms of order were considered, for instance SDW+cycloidal, but these refinements produced visibly poorer fits, and in particular failed to adequately describe the peaks between 20 and 24$^\circ$.

Elliptical helical order has been reported previously in the Cu$^{2+}$ chain compounds NaCu$_2$O$_2$\,\cite{Capogna2010}, LiCu$_2$O$_2$\,\cite{Kobayashi2009}, and linarite PbCuSO$_4$(OH)$_2$\,\cite{Willenberg2012}, but these had smaller ellipticity.  The high ellipticity in antlerite suggests a strong proclivity of the Cu2 spins to point along $c$, as found in the antialigned ferromagnetic spin arrangement on this site at lower temperatures.  

\begin{figure}[t!]
    \includegraphics[width=\columnwidth]{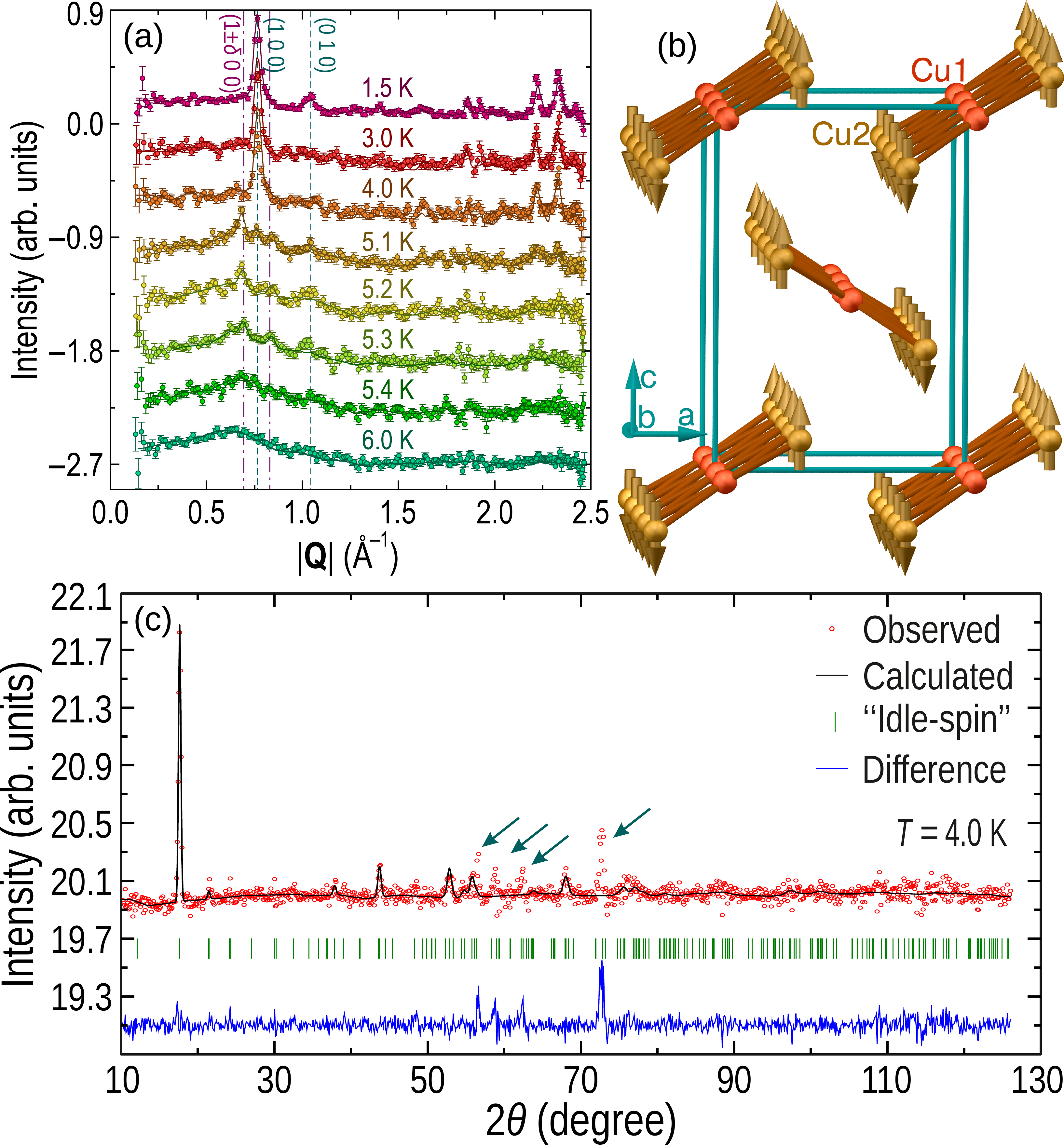}
    \caption{\label{fig:idle}Identifying the magnetic structure between 2.8 and 5.0\,K.  (a) Spin-flip (magnetic) scattering intensity as a function of temperature, collected using polarized neutrons on D7; data at higher temperatures have been shifted for clarity.  Solid curves are refinements according to the magnetic structures we identify.  Dashed lines mark the strong (100) magnetic peak and the weaker (010) peak at $|\bm{Q}|=1.0413$\,\AA$^{-1}$\ which is inconsistent with an idle-spin state, while purple dash-dotted lines indicate the incommensurate positions. (b) Refined magnetic structure at 4.0\,K, representative of the phase between 2.8 and 5.0\,K. (c)~Refinement of the magnetic intensity in the D1B data in the ``idle-spin'' state at 4.0\,K. Intensity at 6.0\,K has been subtracted; arrows mark the positions of strong nuclear peaks.}
\end{figure}

The phase between 2.8 and 5.0\,K, and in particular how it differs from the lowest-temperature magnetically ordered phase depicted in Fig.~\ref{fig:struct}(b), is more subtle.  As seen in Fig.~\ref{fig:Mickey}(a-c), the strongest magnetic peak does not appear to be affected by the transition at 2.8\,K.  Since most other magnetic Bragg peaks sit atop far stronger structural Bragg peaks, small changes there are more difficult to detect in conventional neutron diffraction.  We performed polarized-neutron diffuse scattering at the D7 diffractometer at the ILL in order to cleanly separate the magnetic intensity (in the spin-flip channel) from the structural contribution, using a six-point ($XYZ$) method\,\cite{Ehlers2013}.  Results are shown as a function of scattering vector $|\bm{Q}|$ in Fig.~\ref{fig:idle}(a) for several temperatures.  The behavior in the (100) and (1$\pm\delta$\,0\,0) magnetic reflections is consistent with that shown in Fig.~\ref{fig:Mickey}, but we observe nonmonotonic temperature dependence in the (010) peak at $|\bm{Q}|=1.0413$\,\AA$^{-1}$.  This peak is present in the low-temperature state at 1.5\,K, absent in the intermediate phase at 3.0 and 4.0\,K, and reappears in the multiple-$q$ and incommensurate phases above 5.0\,K.  

Our previous work on the lowest-temperature magnetically ordered state in antlerite identified magnetic intensity in a number of peaks that would not be present in the previously-proposed idle-spin state, including the (010) peak\,\cite{Kulbakov2022a}.  These peaks are associated both with antiferromagnetic order on the inner leg and strong canting on the outer legs, presumably due to strong coupling with the inner leg.  This contribution to the order is evidently reentrant.  The absence of these peaks between 2.8 and 5.0\,K indicates that the intermediate-temperature phase is essentially the idle-spin phase previously proposed to be the ground state\,\cite{Vilminot2003,Hara2011}.  The magnetic structure we refine for this phase based on data from D1B is shown in Fig.~\ref{fig:idle}(b), and the refinement is shown in Fig.~\ref{fig:idle}(c); an mCIF file is supplied as an ancillary file as described in Appendix \ref{suppl}.  This state features no long-range order on the central leg and no canting on the outer legs;   the refined magnetic moment on the outer legs is 0.802(8)\,$\mu_\text{B}$ at 4.0\,K, where the enhancement relative to the incommensurate phase is likely due to the lower temperature.  Refinements of D7 data at 3.0 and 4.0\,K found a similar state.  This is the same magnetic space group as at lower temperature, $Pn'm'a'$ (number 62.449).  We will refer to this phase as ``idle-spin'' as in the previous papers, and return to the question of what the Cu1 spins actually do in this phase.  

\begin{figure*}
  \includegraphics[width=\textwidth]{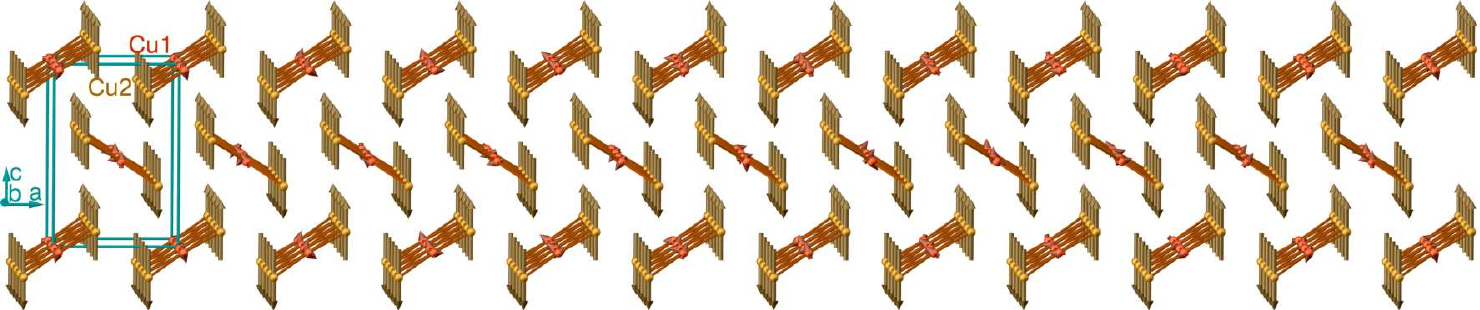}
  \caption{\label{multiQ}Multiple-$q$ phase, present in \antD\ between 5.0 and 5.3\,K, based on refinements at 5.3\,K.}
\end{figure*}

Since the previously proposed idle-spin phase is indeed present in the system but at a higher temperature, one may wonder whether the previous reports of an idle-spin state were due to a thermometry issue.  The neutron diffraction data that led to the original idle-spin conclusion were collected at 1.4\,K\,\cite{Vilminot2002,Vilminot2003}, which is half the temperature of the phase transition in question.  A thermometry issue of that magnitude is very unlikely.  The (010) peak is structurally forbidden, so interference from structural Bragg reflections is also not a satisfying explanation.  This peak may simply have been too weak to be clearly observed at the short wavelengths used.  We also note that antlerite is expected to be exquisitely sensitive to parameters such as strain, pressure, and chemical doping, so it is conceivable, although considerably less likely, that inadvertent doping or overly assertive packing of the powder sample could also have suppressed the 2.8-K transition.  

The collapse in diffuse magnetic background intensity below $\sim$3\,K in Fig.~\ref{fig:Mickey}(b) agrees with the significant reduction on cooling of the diffuse scattering hump centered around 0.7\,\AA$^{-1}$\ in Fig.~\ref{fig:idle}.  The differences between the lowest-temperature phase and the ``idle-spin'' state above 2.8\,K are the loss of antiferromagnetic order on the central Cu1 leg of the ladder and canting on the outer legs, so this large change in diffuse intensity is primarily associated with the Cu1 site.  

Having identified the phases below 5.0\,K and above 5.3\,K, we can now piece together the multiple-$q$ phase between 5.0 and 5.3\,K.  The outer legs of the ladder are presumably ordered ferromagnetically and antialigned, as they are in the lower-temperature phases, since the (100) magnetic peak is not sensitive to the transition at 5.0\,K.  Spins on the central leg evidently retain a strong tendency to stay in the plane of the cycloid, but fail to order below 5.0\,K.  This indicates that the multiple-$q$ phase is a combination of ferromagnetic outer legs with a cycloidal inner leg.  The elliptical helix on the outer legs of the ladder becomes constant-amplitude antialigned ferromagnetic order, which to some extent decouples the outer legs of the ladder from the central leg and allows the incommensurate wavevector to return to its higher-temperature value, again without any sign of locking into a commensurate value.  The multiple-$q$ order was refined as a minor component at 5.3\,K according to this model, producing an ordered moment on the outer legs of 0.87(28)\,$\mu_\text{B}$.  The resulting magnetic structure is depicted in Fig.~\ref{multiQ}, and an mCIF file is provided as an ancillary file as described in Appendix~\ref{suppl}.  The ordered moment and orientation on the central leg were constrained to be identical to that in the incommensurate phase.  

Within a ladder the spin order is commensurate in all four phases, although not collinear.  The relative spin orientation within every ladder is also the same, as would be expected because the interladder interactions are far weaker than the exchange coupling within a ladder, which we quantified in Ref.~\onlinecite{Kulbakov2022a}. The incommensurate modulation along $a$ is perpendicular to the ladders, implying that the interladder interactions play a key role in stabilizing the incommensurate order.  This indicates strong limitations to the picture of an isolated ladder, as has been used in calculating the phase diagrams for \antH\ as a function of the exchange parameters and its ground state\,\cite{Kulbakov2022a}.  The incommensurate order we have identified in antlerite at elevated temperatures arises from a relative rotation of the spins from one ladder to the next, and is not captured by the calculations, since they neglected interladder interactions.  We also note that although having an incommensurate modulation along $a$ is broadly reminiscent of the cycloidal state proposed for Se-substituted antlerite, Cu$_3$SeO$_4$(OH)$_4$\,\cite{Vilminot2007}, the ordered state is quite different.  In the Se analog similar order with a different pitch was reported ---  a commensurate $(\frac17\,0\,0)$ rather than a ($\sim0.1$\,0\,0) propagation vector; however, the spins on the outer legs were cycloidal rather than helical, while the spins on the inner leg did not participate. This was reported by the same group that found an idle-spin ground state in antlerite, so it is possible that nonzero ordered moments on the central leg in the Se analog are present but were missed for similar reasons, and it may be worth revisiting this material to look for signatures of ordering on the central leg. 

The spins on the outer legs in the incommensurate phase in antlerite prefer to lie parallel to ±c, as in the lowest-temperature state. Their spatial oscillation is presumably due to the interactions with the central leg destabilizing the preferred spin alignment.  The circular cycloid on the central leg suggests that this site is driving the incommensurate order.  Indeed, our DFT calculations\,\cite{Kulbakov2022a} found that the strongest exchange interaction in antlerite is the antiferromagnetic $J_5$ along the central leg of the ladder.

While strong antiferromagnetic interactions on the central leg evidently drive the transition to magnetic order at 5.55\,K, frustration results in noncollinear order along each central leg, and we additionally find a cycloidal state along $a$. The noncollinear order along the central leg of the ladder may arise from bond frustration --- DFT indicated that the second-neighbor exchange along this leg, $J_6$, is roughly half of its nearest-neighbor $J_5$ and is the third-strongest interaction in the system.  Alternating interactions with the outer legs may also play a role.

The second- and fourth-strongest interactions in antlerite are ferromagnetic exchanges on the outer legs.  With twice as many spins on the outer legs as on the central leg, these interactions may begin to overpower those on the central leg as they strengthen on cooling.  It appears that first the outer legs partially decouple and become constant-amplitude ferromagnetic, rather than helical, then their ferromagnetism becomes strong enough to destroy the cycloidal order on the central leg.  Finally, the central leg becomes antiferromagnetic, forming the lowest-temperature state. 

While incommensurate order as seen here is already relatively uncommon, multiple-$q$ phases are rarer still.  Such phases are especially uncommon in centrosymmetric materials, with the few prominent examples including USb\,\cite{Halg1986} and phase III in CeB$_6$\,\cite{Burlet1982,Effantin1985}.  However, these examples are rare-earth compounds.  Multiple-$q$ magnetic order is extremely uncommon in compounds based on $3d$ transition metals, with the best-known examples being $\gamma$-Mn alloys\,\cite{Long1993}; the recently-reported hedgehog- and skyrmion-lattice phases in SrFeO$_3$\,\cite{Ishiwata2020}; and the suggested triple-$q$ order in Na$_2$Co$_2$TeO$_6$\,\cite{Chen2021}. Multiple-$q$ order most commonly arises from Dzyaloshinskii-Moriya interactions, which sum to zero by symmetry over a unit cell in centrosymmetric crystal structures like that of antlerite and result from spin-orbit coupling which is generally weak in the $3d$ block; from an interplay between Kondo physics and exchange interactions, which is not expected in transition metal compounds; from bond-dependent Kitaev interactions as proposed in Na$_2$Co$_2$TeO$_6$\,\cite{Chen2021}; or from degeneracy among possible ordering directions in highly symmetric lattices as found in face-centered-cubic $\gamma$-Mn, which is unlikely to be the case in orthorhombic antlerite.  Here, the fine balance required for multiple-$q$ order evidently comes from competing exchange interactions.  

We also note that multiple-$q$ phases ordinarily exist only in very small regions of a material's magnetic phase diagram, usually forming as a bubble within other magnetically ordered phases.  This also seems to be the case in antlerite.  However, while these multiple-$q$ bubbles seldom extend over broad temperature ranges, the published magnetic phase diagrams of antlerite show an intriguing tail to zero temperature for fields around 1--1.8\,T along $c$ and around 3--4\,T along $a$\,\cite{Hara2011,Fujii2013}, apparently connected to the multiple-$q$ phase, suggesting that this phase may occupy a significant swath of the phase diagram.  This should be investigated further once sufficiently large single crystals are available to enable neutron diffraction in magnetic field.

An even rarer aspect of the multiple-$q$ phase in antlerite is that it combines commensurate antiferromagnetic order on one magnetic site with incommensurate order on another.  The only other reported example of such a situation is Co$_3$TeO$_6$\,\cite{Ivanov2012,Wang2013,Lee2017}, whose magnetic sublattice includes five magnetic Co sites (including both octahedral and tetrahedral coordination) and has an additional antiferroelectric transition within one of its magnetically ordered phases.  This structural complexity has greatly complicated the identification and investigation of its magnetic order, and led to the papers cited here to very different determinations of several of its phases.

\section{Diffuse Scattering in the Paramagnetic Phase}

\begin{figure*}
  \includegraphics[width=\textwidth]{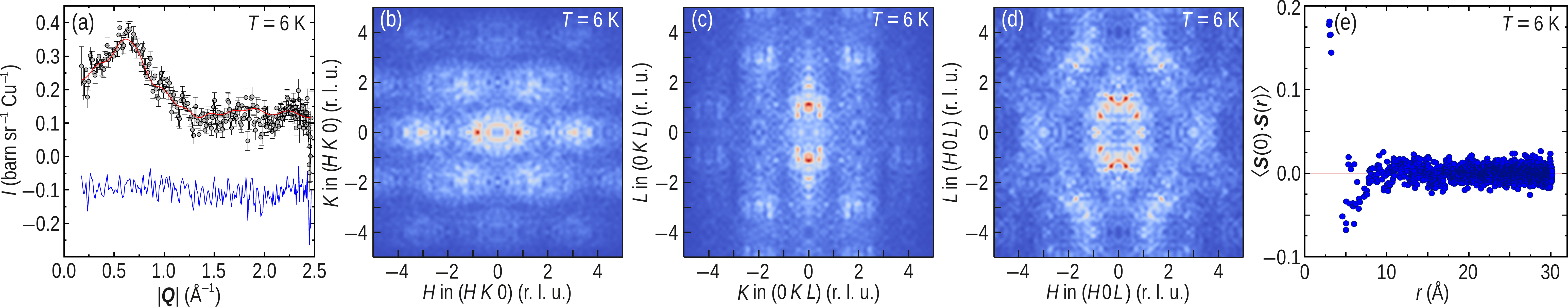}
  \caption{\label{diffuse}Diffuse scattering in \antD\ in the paramagnetic phase at 6.0\,K.  (a) Fit (red) of the magnetic diffuse scattering intensity (gray), and its residual (blue). (b-d) Reverse-Monte-Carlo reconstructed scattering intensity for $(HK0)$, $(0KL)$ and $(H0L)$ cuts through reciprocal space, respectively. (e) Extracted spin-spin correlation functions as a function of distance.  The ferromagnetic nearest neighbor correlations along the outer legs are visible at the lowest distances.}
\end{figure*}

We now return to the diffuse scattering presented in Fig.~\ref{fig:idle}. We used the {\sc Spinvert} software package\,\cite{Spinvert} to perform a reverse Monte Carlo refinement of the spin-flip neutron powder diffraction data, to recover the three-dimensional magnetic diffuse scattering pattern and the spin-pair correlation function. Reconstructed diffuse scattering intensity for antlerite in the paramagnetic phase at 6.0\,K based on a 5$\times$5$\times$5 supercell is shown in Fig.~\ref{diffuse}(b-d) for different cuts through reciprocal space. This is what we predict would be observed on a single crystal, if a sufficiently large crystal were available. Figure~\ref{diffuse}(a) shows the resulting fit to the powder data, and Fig.~\ref{diffuse}(e) shows the extracted spin-spin correlation function.  The shortest Cu--Cu distance in antlerite is a ferromagnetic exchange on the outer leg, followed by the antiferromagnetic link on the central leg and the second ferromagnetic exchange on the outer leg. Accordingly, the correlations in Fig.~\ref{diffuse}(e) are ferromagnetic at the shortest distance, then become mainly antiferromagnetic on average.  Beyond $\sim$8\,\AA\ the correlations average out to zero due to powder averaging --- the sphere corresponding to this radius begins to include a large number of spins.  

Spin-flip scattering was also performed to higher temperatures, as shown in Fig.~\ref{diffuse200K} in the Appendix, where the diffuse peak is seen to shift to lower $|\bm{Q}|$ on warming but does not vanish up to 200\,K, indicating persistent short-range correlations. We note that this measurement method integrates over the inelastic scattering from the sample, and hence probes the time-equal correlations rather than time-averaged correlations which give no energy change to the neutrons.  Correlations in paramagnetic fluctuations, which are all dynamic, will give a $|\bm{Q}|$ dependence in the time-equal correlations.

\section{Muon Spin Rotation and Relaxation}

We performed muon spin rotation and relaxation ($\mu$SR) experiments to obtain additional information about the magnetic phases in antlerite, and to distinguish whether the diffuse contributions are dynamic or static on the timescale of $\mu$SR. $\mu$SR probes local fields and their fluctuations through their effect on implanted positive muons $\mu^+$ (hereafter referred to as ``muons'' for simplicity), which stop at specific sites in the crystal structure, ordinarily near anions. Muons have a lifetime of 2.2\,$\mu$s and decay into a positron and neutrinos.  The decay positrons are emitted preferentially along the muon spin direction, leading to a decay asymmetry which rotates as the muon spins precess about local magnetic fields.  Positron detectors positioned along and opposite the direction of the incoming polarized muon beam yield a time-dependent asymmetry $A(t)$, the primary signal in $\mu$SR.

In a wTF experiment with a weak external magnetic field $B_\text{ext}$ applied perpendicular to the initial muon polarization, muon spins in {\slshape non-magnetic} environments will precess about this field, leading to oscillations in the asymmetry with a frequency $\nu_\mu = \gamma_\mu B_\text{ext} / 2\pi$, where the muon gyromagnetic ratio $\gamma_\mu$ = 135.54\,$\mu$s$^{-1}$T$^{-1}$. The asymmetry signal will be an oscillation within a decaying envelope: 
\begin{equation}
  A(t) = A(0)e^{-\lambda_\text{para}t} \cos(2\pi\nu_\mu t)\label{eq5} ,
\end{equation}
with $\lambda_\text{para}$ being the paramagnetic relaxation rate.

In \antD\ at higher temperatures $A(t)$ is consistent with a paramagnetic phase, but we find an initial asymmetry $A(0)$ of 0.23 in the paramagnetic regime.  This is significantly lower than the 0.28 expected for a fully polarized muon beam in the setup used (as also directly verified with a different sample). The asymmetries recorded in our ZF and wLF experiments are similarly reduced to $\sim$0.22, as seen in the paramagnetic state at 6.5\,K in Fig.~\ref{LFZF6K}, and this reduction is seen at all temperatures. A reduced initial asymmetry in a weak transverse field is an indication that some of the implanted muons form muonium, as is often found in insulating and semiconducting materials.  Muonium signals are determined by large hyperfine and dipolar interactions and are beyond our time resolution\,\cite{Cox1987}, but details of the mechanism for muonium formation in antlerite may be of interest for future investigation.

ZF and wLF experiments provide information about the temperature-dependent development, distributions, and spin dynamics of local magnetic fields $B_\text{loc}$ in the long-range ordered regime and on the approach to it.
The depolarization seen in zero field in the paramagnetic regime is dominated by strong damping with a shape that is well described by a static Gaussian Kubo-Toyabe function (Eq.~\ref{eq6})\,\cite{Uemura1985,Kubo1981}.  The application of a weak longitudinal field of 2\,mT suppresses this damping (see Fig.~\ref{LFZF6K}). This means that the field distribution in the paramagnetic state in \antD\ is due to small, randomly directed fields from static spins that can be overcome by even a small applied field.  The origin of this field distribution can be attributed to the nuclear spins, most likely of $^{17}$O when the muon is attached closely to oxygen.

Analysis of the 2-mT LF spectra reveals a much weaker damping which is now exponential due to fields from rapidly fluctuating electronic moments as expected in the paramagnetic state. The damping increases upon cooling toward the ordering temperature as the fluctuations slow down. Below 8\,K, however, we observe that an increasing part of the signal becomes much more rapidly damped. As addressed below, we find a reduction of wTF asymmetries (see Fig.~\ref{TFasy}) in the same temperature range. Both observations likely indicate an onset of short-range order above \TN, as is often found in low-dimensional magnetic systems.

When $\mu$SR is performed in a magnetically ordered state, additional local fields are present which are typically much larger than the applied field and are randomly oriented with respect to it.  Precession about these randomly oriented fields leads to depolarization of the muon spins.  This is reflected in a rapid loss of the asymmetry signal after implantation.  This clear difference in behavior enables the detection of ordered phases, including short-range order and fluctuations slower than the $\sim$10\,$\mu$s timescale to which muons are sensitive.  For antlerite in a transverse field of 5\,mT, the asymmetry starts to decrease below 8\,K, followed by a rapid drop below 6\,K, and saturates by 5.3\,K (see Fig.~\ref{TFasy}), demonstrating that the entire sample volume hosts long-range magnetic order. Within our temperature uncertainty of $\pm$0.3\,K, this finding is consistent with the first ordering temperature of \TN\,=\,5.55\,K derived from specific heat and magnetization data (Figs.~\ref{fig:M}, \ref{fig:cP}). In the ordered state we still find some minor residual asymmetry oscillating due to only the applied field, presumably from muons which stop in the sample holder; this temperature-independent background asymmetry $a_\text{BG}$\,=\,0.03 is taken into account in the analysis of the ZF and wLF experiments. 

\begin{figure}[tb]
  \includegraphics[width=\columnwidth]{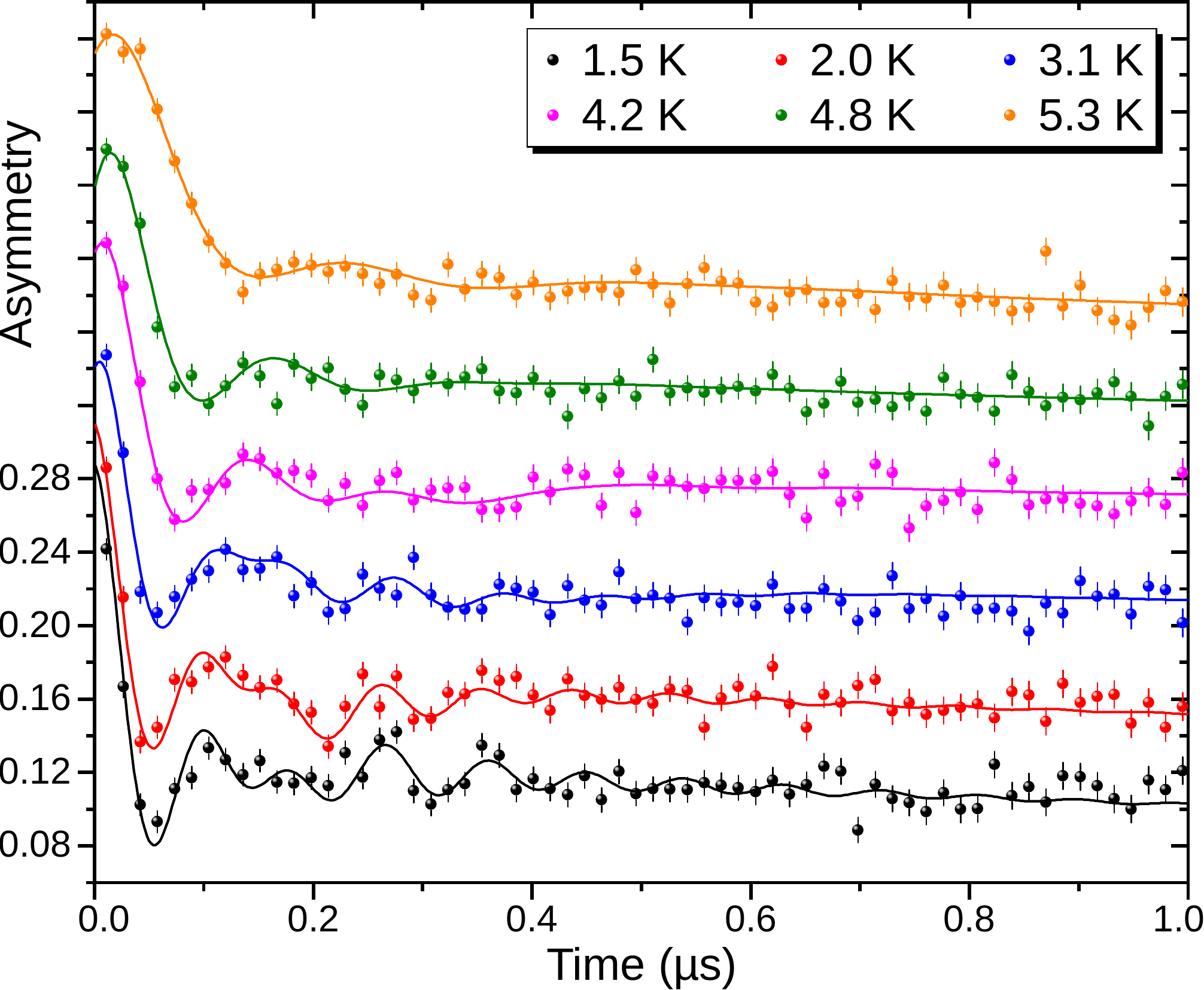}
 \caption{\label{spectraLF}Early-time $\mu^+$ spin-polarization asymmetry in \antD\ at temperatures below \TN\ from wLF measurements. Solid curves are fits to Eq.~\ref{eq7}.  The spectra at higher temperatures are shifted vertically for clarity; the common asymmetry scale is given only for the 1.5-K spectrum. Spectra up to 6\,$\mu$s are presented in Fig.~\ref{long6mis} in Appendix \ref{AppmuSR}.}
\end{figure}

Spectra below \TN\ in weak longitudinal field are shown in  Fig.~\ref{spectraLF} and to longer decay times in Fig.~\ref{long6mis}. Up to 4\,K they reveal at least two spontaneous rotation signals in addition to a broad, damped pattern that can be best reproduced again with a static Gaussian Kubo-Toyabe function. The different signals are associated with different muon stopping sites near oxygen sensing different local fields due to neighboring Cu ions. Unlike the Gaussian-damped signal found above \TN, the signal at low temperatures is not affected by an applied longitudinal field of 2\,mT, i.e., the field distribution comes from static fields much larger than 2\,mT and must be related to static electronic moments. For parametrizing these complex spectra we use a phenomenological approach described in Appendix \ref{AppmuSR}.  For each site $i$\,=\,1,2 we have a contribution $a_{\text{rot},i}$ to the asymmetry, a precession frequency $\nu_{\mu,i}$ about the local field $B_i$, a transverse relaxation rate $\lambda_{\text{t},i}$, and a longitudinal relaxation rate $\lambda_\text{l}$ which is directly proportional to the spin fluctuation rate. In addition we introduce a static Gaussian Kubo-Toyabe component with asymmetry contribution $a_\text{GKT}$ and damping parameter $\sigma$.

The longitudinal damping $\lambda_\text{l}$ (extracted from the behavior at long times once the oscillating signal is already damped out) is nearly zero up to 3.1\,K . This indicates that the electronic spin system is static on the timescale of $\mu$SR in all magnetically ordered phases, notably including the idle-spin state. Starting at 4.2\,K we see an increase of longitudinal damping that, as will be discussed below, can be related with the increased Kubo-Toyabe contribution. 

\begin{figure}[t!]
  \includegraphics[width=0.79\columnwidth]{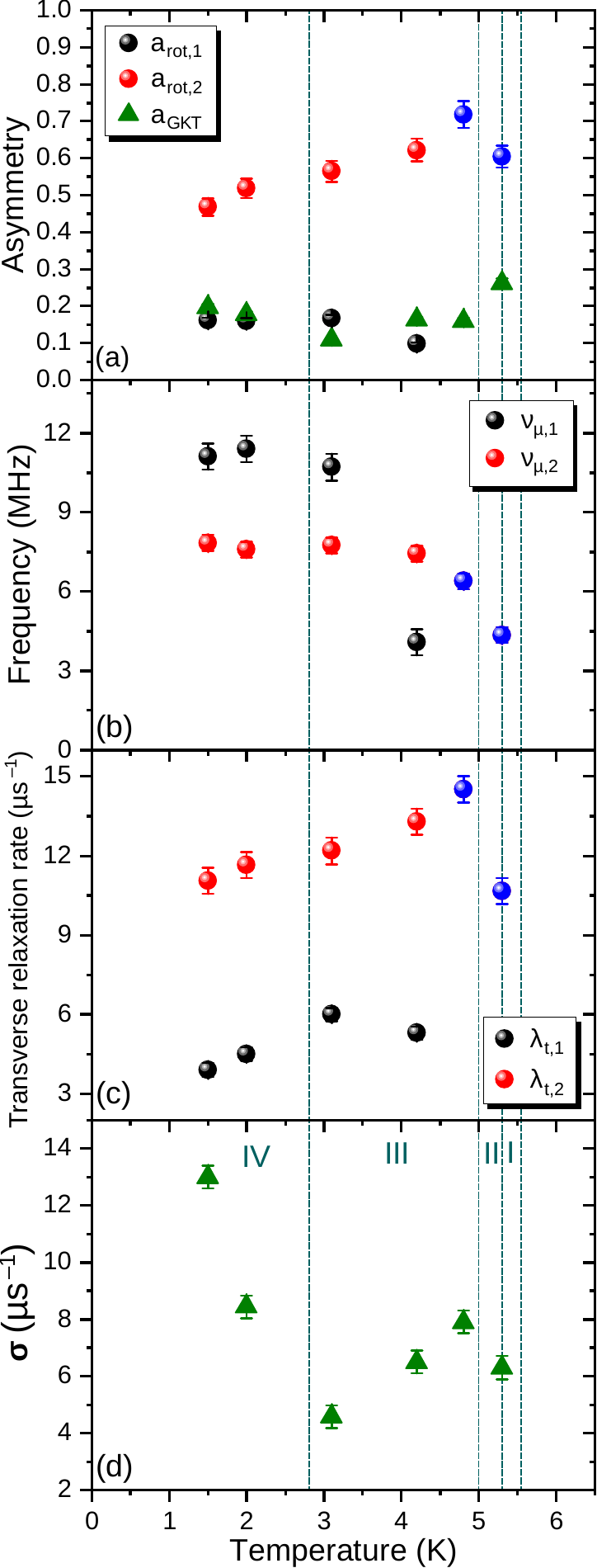}
  \caption{\label{MUSRparameter}Temperature dependence of weak-LF-$\mu$SR parameters in \antD: (a) normalized asymmetry, (b) spontaneous  $\mu^+$ spin precession frequency $\nu_{\mu,i}$, (c) transverse relaxation rate $\lambda_{\text{t},i}$, and (d) damping parameter $\sigma$ of the Gaussian component. Blue circles indicate the regime in which only one frequency is observed. Dashed lines indicate the phase transitions identified in magnetization and specific heat --- phase I is incommensurate, II is multiple-$q$, III is idle-spin, and IV is the fully-ordered low-temperature state.}
\end{figure}

Figure~\ref{MUSRparameter}(a) plots the temperature dependence of the fractional contributions $a_i$ to the background-corrected asymmetry.  The asymmetries of the rotating signals stay unchanged below 4.2\,K within the uncertainty. Starting from 4.2\,K the asymmetries of the Gaussian signal and one of the rotating signals increase at the expense of the other rotating signal. This means that at some muon sites there is a gradual change in spin order combined with an increase of static disorder. Finally, only a single frequency is observable at 4.8 and 5.3\,K. The latter two temperatures correspond approximately to the temperature range where incommensurate order was observed, and will be discussed separately.

The rotation frequencies $\nu_{\mu,i}$ vary only weakly up to 3\,K and then decrease with increasing temperature, as seen in Fig.~\ref{MUSRparameter}(b).  This represents the strength of the magnetic field on the muon site due to the ordered moments, so such a temperature dependence is expected.  Due to the limited number of temperatures, it is not possible to resolve the shape clearly enough to determine at which transition each frequency extrapolates to zero.

We present the transverse relaxation rates $\lambda_{\text{t},i}$ in Fig.~\ref{MUSRparameter}(c). The apparent strong damping of the rotation signals indicates a distribution of frequencies due to inhomogeneous broadening. For the first rotation signal we find a strong increase in $\lambda_\text{t}$ when warming from 2 to 3\,K. This means that the local field distribution at this muon site broadens around the lowest temperature transition and suggests that Cu1 is more strongly coupled to the first site than the second site. The second frequency has a much higher damping at low temperature, which is reduced somewhat on warming toward the second transition at 5.0\,K. There is clearly a much greater variation in local fields at this site than at site 1.  

The damping $\sigma$ of the Gaussian signal decreases significantly on warming from 1.5 to 3\,K, i.e., the static local field distribution narrows and the fields become more homogeneous. This may be related to a spin rearrangement in the neighborhood of the muon site. For higher temperatures the damping stays essentially unchanged within the uncertainty. 

We now return to the data at 4.8 and 5.3\,K.  The depolarization patterns at these temperatures are different in shape from all the others. Two distinct rotating signals can no longer be resolved, and at the earliest times (0--0.1\,$\mu$s) a finite phase shift $\varphi$ of $-$50(5)$^\circ$ is visible (see Fig.~\ref{spectraLF}).  Fits to the spectra at 3.1 and 4.2\,K give a reduced phase shift of $-$25(5)$^\circ$ for the rotating signal ($i=2$) with greater asymmetry. At 1.5 and 2\,K this phase shift is consistent with zero. Phase shifts of the order observed at 4.8 and 5.3\,K are fingerprints of the complex field distributions which arise in incommensurately modulated spin structures\,\cite{Yaouanc2011,Franke2018}, in agreement with the incommensurate or multiple-$q$ phases refined from neutron diffraction data in a similar temperature range.

Starting at 4.2\,K  the asymmetry contribution from the Gaussian signal has increased at the expense of the rotating signal [see Fig.~\ref{MUSRparameter}(a)].  In contrast to lower temperatures, there is a finite depolarization for times $t>1$\,$\mu$s (see Fig.~\ref{long6mis}) showing that the local fields are now fluctuating. For the data analysis we thus use a so-called dynamic Gaussian Kubo-Toyabe function\,\cite{Hayano1979,Reotier1992} instead of the static Kubo-Toyabe profile used at lower temperatures. We extract fluctuation rates that are relatively slow: 1.3(2) and 2.3(5)\,$\mu$s$^{-1}$ at 4.8 and 5.3\,K, respectively. A dynamic fit of the Gaussian signal at 4.2 K results in a considerably lower fluctuation rate of only 0.15(5)\,$\mu$s$^{-1}$.  At both temperatures the dynamic fluctuations are visible in the changes of the Gaussian damped signal, while the rotating signal indicates incommensurate order of the Cu spins. Note that at lower temperatures where we see two rotating signals, the fluctuations are slower, and we see a weaker, static Gaussian damped signal. Since the most probable muon sites lie close to oxygen atoms, which bridge the two Cu sites, the muons may be expected to sense fields originating from both Cu1 and Cu2. Without knowing the locations of the muon stopping sites it is unfortunately not possible to separate the roles of these sites.  

In summary, our $\mu$SR data indicate an onset of short-range order below 8\,K followed by full long-range order at 5.3\,K. Signals at 5.3 and 4.8\,K support the presence of incommensurate order and indicate slowly fluctuating local fields. At the lowest temperatures the two oscillating signals with their temperature-dependent frequencies are clearly associated with static, commensurate long-range order. A reduced phase shift in the idle-spin phase at 3.1 and 4.2\,K indicates that vestiges of the incommensurate modulation persist in this phase.  The spins are static on muon timescales here, which may indicate that the idle-spin phase is either glassy or short-range. The splitting of ZFC and FC magnetization below 5.0\,K argues against slow fluctuations, and there cannot be long-range order within a ladder and disorder from one ladder to the next, as this would produce clear signatures in the spin-flip neutron scattering data.  The splitting between ZFC and FC data is small, possibly because the idle spins remain constrained in the plane of the cycloid, or perhaps because the spins are trained into particular orientations when the system passes through the phases at higher and lower temperature.  The persistence of a minor Gaussian-damped signal to low temperatures indicates a random component in the local spin arrangement which remains disordered to the lowest temperatures, perhaps an additional canting angle.  This is most likely associated with the Cu1 site, where we refine a somewhat reduced ordered moment.

\section{Summary and Outlook}

We have confirmed that \antH\ undergoes three magnetic ordering transitions between 5.00 and 5.55\,K in zero field, followed by a fourth transition at 2.8\,K, as was suggested by the more-complicated magnetic phase diagram for antlerite proposed in Ref.~\onlinecite{Fujii2013}.  Our magnetization, $\mu$SR, and neutron scattering data have allowed us to identify the spin structures of these low-field phases.

The picture that emerges here is of a magnetic subsystem which first enters an incommensurate elliptical helical+cycloidal state, most likely driven by interladder interactions connecting the strongly antiferromagnetic central legs of the ladders.  While all spins participate, the refined moments are not the full moment of Cu$^{2+}$, and $\mu$SR indicates that this state still has a dynamic component.  The outer legs of the ladder have a strong tendency toward ferromagnetic order along $c$ which fights the incommensurate order, leading to the helical order on these sites and likely also to the rather small ordered moment on the central leg.  The incommensurate wavevector is perpendicular to the ladders, so although all sites are incommensurately ordered, the order along a ladder is commensurate, and this is neither the ``IC-IC-IC'' nor ``FM-IC-FM'' state proposed previously by density-matrix renormalization group (DMRG) based on DFT parameters\,\cite{Kulbakov2022a}, since in these phases the incommensurate (IC) order runs along the ladder. The DFT/DMRG results neglected interladder coupling and would not be able to generate the observed incommensurate state.  The observed incommensurate phase is also distinct from the cycloidal state proposed for Cu$_3$SeO$_4$(OH)$_4$\,\cite{Vilminot2007}.  The latter state has a similar but commensurate $(\frac17\,0\,0)$ propagation vector, but the spins on the outer legs were cycloidal rather than helical, while the spins on the inner leg did not participate.

As the incommensurate phase is cooled, the outer legs evidently push the incommensurability parameter $\delta$ toward zero.  They then order ferromagnetically, antialigned, partially decoupling from the central leg and leaving it to revert toward its original incommensurate wavevector in a multiple-$q$ structure.  As with the incommensurate phase, since the incommensurate wavevector presumably relies upon interladder interactions, the multiple-$q$ state is not the FM-IC-FM state proposed by DMRG to lie nearby in parameter space.  The discovery of multiple-$q$ magnetic order in a frustrated quantum magnet is likely to be of considerable interest, as the quantum nature of the spins may lead to entirely new and unexpected physics.

This is actually a particularly unique multiple-$q$ structure, insofar as one magnetic site hosts commensurate order while the other orders incommensurately.  This is the magnetic analog of misfit crystal structures, in which separate layers of a compound fail to line up due to a large lattice mismatch\,\cite{Wiegers1996,Ng2022}. The misfit materials only rarely host magnetic ions in both types of layers. In fact, to the authors' knowledge, there are only three misfit-layered compounds with magnetic ions in both layers on which magnetic properties have been investigated: (CeS)$_{1.20}$CrS$_2$ exhibits two transitions in the magnetization, which were attributed to the two separate layers ordering independently\,\cite{Suzuki1993}, while magnetization data on (GdS)$_{1.27}$CrS$_2$ suggested a transition at low temperature\,\cite{Lafond1993} which was not investigated further.  In neither of these materials has magnetic diffraction been applied to determine the spin arrangement in any layer.  Finally, in [Ca$_2$CoO$_3$]$_{0.62}$[CoO$_2$], spin-density-wave order in the CoO$_2$ layer\,\cite{Sugiyama2003} was recently found to compete with ferromagnetic clusters in the Ca$_2$CoO$_3$ layer to produce a glassy state\,\cite{Ahad2020}.  Due largely to a scarcity of suitable materials platforms, no long-range magnetic-misfit order has been identified in the misfit-layered materials.  The magnetic-misfit phase in antlerite, on the other hand, involves interpenetrating magnetic lattices that are structurally commensurate, and the magnetic misfit structure emerges spontaneously because of the different temperatures of the incommensurate-commensurate lock-in phase transitions in the ferro- and antiferromagnetic legs of the spin ladder.  As noted above, the results in Refs.~\onlinecite{Hara2011,Fujii2013} suggest that this novel phase occupies a large swath of the $H$--$T$ phase diagram.  The as-yet-unidentified ``$\gamma$1'' and ``$\gamma$2'' phases may also emerge from it at higher $H\parallel c$, which could indicate an ability to field-tune this magnetic-misfit phase.

Upon further cooling, the ferromagnetic order evidently strengthens on the outer legs, which appears to temporarily destabilize the order on the central leg.  We detect this as an ``idle-spin'' state, closely similar to that proposed previously as the ground state\,\cite{Vilminot2003,Hara2011}. However, the idle spins are likely static, at least on the timescale of $\mu$SR, possibly frozen in a glassy state with only short-range correlations, similar to the one found in the misfit cobaltate\,\cite{Ahad2020}. Finally, at 2.8\,K the central leg orders antiferromagnetically and the fully-ordered low-temperature state is obtained.

The long-range order on the central Cu1 leg is evidently reentrant.  The question remains what its spins are doing in the ``idle-spin'' phase between 2.8 and 5.0\,K.  Our magnetization data find an additional contribution to the magnetization for fields along $b$ when cooling into this phase.  This suggests that the ``idle'' Cu1 spins remain locked in the plane of the cycloid.  This would make the spins more readily polarized once their long-range order is extinguished, but only for fields perpendicular to the $ac$ plane of the failed cycloid.  When the Cu1 spins order antiferromagnetically below 2.8\,K, this enhanced magnetization falls away again.

DFT and DMRG calculations proposed that the inner and outer legs in antlerite are independently on the cusp of a phase transition, leading to a complex phase diagram with exchange parameters and $U$\,\cite{Kulbakov2022a}.  However, the complexity of these calculations required neglecting minor contributions such as anisotropy, Dzyaloshinskii-Moriya (DM) interactions, and interladder exchanges.  The phases we identify at zero field indicate that these neglected interactions contribute an additional rich complexity to the system.  The outer legs of the ladder clearly exhibit strong anisotropy, with the moments directed preferentially along $c$.  Strong canting in the previously reported lowest-temperature phase and within the central leg in the incommensurate phase may suggest a role for DM interactions.  Meanwhile, the incommensurate order presumably arises from relative rotations of the Cu1 moments between the ladders, which can only arise from interladder exchange interactions.  That the phases newly identified here were absent from the calculated phase diagram is most likely a result of having to exclude weaker interactions, making these phases inaccessible to DMRG.  The high tunability suggested by DMRG is clearly only the tip of the iceberg, and temperature is only one tuning knob of many.  More as-yet-unidentified phases have been seen in antlerite in applied magnetic fields\,\cite{Hara2011,Fujii2013}, while to our knowledge tuning by pressure and strain have not yet been reported.  Se substitution should be revisited\,\cite{Vilminot2007}, and a very different alternation of exchanges leads to a completely different ground state in szenicsite {Cu}$_3$({MoO}$_4$)({OH})$_4$\,\cite{Lebernegg2017}, which to our knowledge has never been synthesized, suggesting another potential substitution series.  We anticipate that antlerite and its derivatives will host a very rich collection of magnetic phases in an accessible range of parameters.  

\begin{acknowledgments}
The authors are grateful for the assistance of Philipp Schlender and Ellen H\"au\ss ler in initial sample characterization and Hubert Luetkens with $\mu$SR measurements.  This project was funded by the German Research Foundation (DFG) {\itshape via} the projects B01, B03, C03, and C06 of the Collaborative Research Center SFB 1143 (project-id 247310070); Research Training Group GRK 1621 (project-id 129760637); the W\"urzburg-Dresden Cluster of Excellence on Complexity and Topology in Quantum Matter\,--\,{\slshape ct.qmat} (EXC~2147, project-id 390858490); and through grants No.\ IN~209/9-1, PE~3318/2-1, and LI~244/12-2, and Project No.\ 422219907. This work is based in part on experiments performed at the Swiss Muon Source S$\mu$S, Paul Scherrer Institut, Villigen, Switzerland.  The authors acknowledge the support of the Australian Centre for Neutron Scattering, Australian Nuclear Science and Technology Organisation, in providing neutron research facilities used in this work.  The Institut Laue-Langevin, Grenoble (France) is acknowledged for providing neutron beam time; data underpinning the ILL diffraction results may be accessed {\itshape via} Refs.~\onlinecite{ILL-data1_D1B-2020,ILL-data_D20-2021,ILL-data_D7-2021}.
\end{acknowledgments}

\appendix

\section{UV-vis spectroscopy}

\begin{figure}[t!]
  \includegraphics[width=\columnwidth]{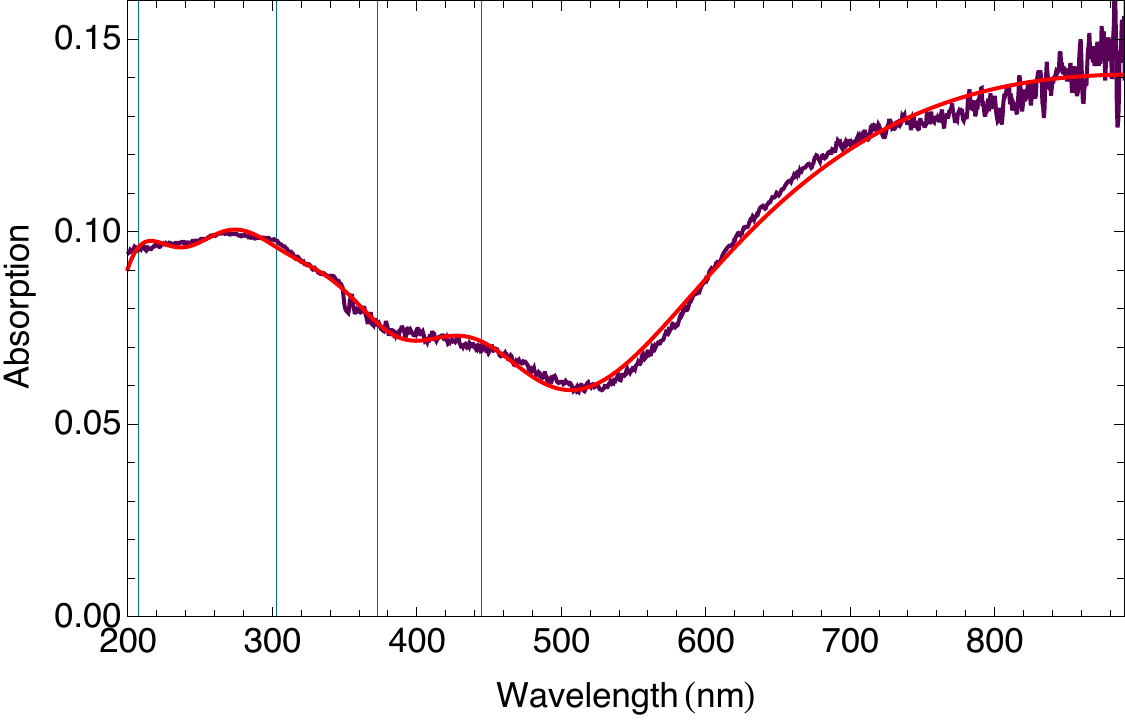}
  \caption{\label{UVvis}Optical absorption spectrum of \antH.  Data are in purple, while the fit described in the text is in red.}
\end{figure}

The infrared spectrum of antlerite has been reported several times\,\cite{Vilminot2003,Bissengaliyeva2012,Bissengaliyeva2013,Zittlau2013}, but we are only aware of one report of its optical response in the ultraviolet, visible, and near-infrared regimes\,\cite{Reddy2002b}, and this has significant gaps.  We show the absorption spectrum of \antH\ in these frequency regimes in Fig.~\ref{UVvis}.  This spectrum was fit to five Gaussians in the frequency domain, yielding peaks at 207, 303, 373, 444, and 909\,nm (48236, 32996, 26845, 22500, and 11007\,cm$^{-1}$), which are marked on the figure. The previous work found peaks in this frequency range at 49260, 44445, 16390, and 10990\,cm$^{-1}$, corresponding to 203, 225, 610, and 910\,nm\,\cite{Reddy2002b}. These were attributed to $^2\text{A}_\text{1g}(d_{x^2-y^2})\rightarrow {}^2\text{B}_\text{2g}(d_{xy})$ (10990\,cm$^{-1}$), $^2\text{A}_\text{1g}(d_{x^2-y^2})\rightarrow {}^2\text{B}_\text{3g}(d_{yz})$ (16390\,cm$^{-1}$), and charge transfer excitations at higher wavenumber. Our data would not allow us to clearly distinguish a shoulder at 225\,nm, and we are not able to resolve the weak hump reported at 610\,nm.  The previous report had large gaps in the spectrum, most notably between $\sim$250 and 595\,nm, which would have prevented the observation of additional features at intermediate wavelengths. 

We consider it plausible that the peaks at shorter wavelength are indeed charge transfer excitations as previously reported, while the longer-wavelength peaks are $d$--$d$ excitations.  A reliable assignment of all observed modes would require detailed modeling and calculation of excited states beyond the scope of this work --- antlerite has two Cu sites with different coordination environment\,\cite{Hawthorne1989,Vilminot2003}, neither site has a highly symmetric coordination sphere, there are multiple types of ligands, and electron-electron interactions are thought to be very strong in this system, all of which add considerable complexity to such a calculation.  Spectra in the ultraviolet, visible, and infrared frequency ranges can serve as a fingerprint of a substance, so having a broader spectrum available (when combined with the previously available infrared results) is likely to prove useful for chemical identification purposes.


\section{Magnetic Refinements}\label{suppl}

As arXiv ancillary files, we provide magnetic crystallographic information files (mCIF) describing our magnetic refinements:

\begin{tabular}{lcr}\\ \toprule\toprule
  Phase & Temperature & \multicolumn{1}{c}{Filename} \\ \midrule
  Idle Spin & 4.0\,K & {\tt \href{https://arxiv.org/src/2207.05606v2/anc/IdleSpin.mcif}{IdleSpin.mcif}}\\
  Multiple-$q$ & 5.3\,K & {\tt \href{https://arxiv.org/src/2207.05606v2/anc/MultiQ.mcif}{MultiQ.mcif}}\\
  Incommensurate & 5.3\,K & {\tt \href{https://arxiv.org/src/2207.05606v2/anc/Icom.mcif}{Icom.mcif}}\\ \bottomrule\bottomrule
\end{tabular}

\section{Diffuse Magnetic Intensity to Higher Temperatures}

\begin{figure}[t!]
    \includegraphics[width=\columnwidth]{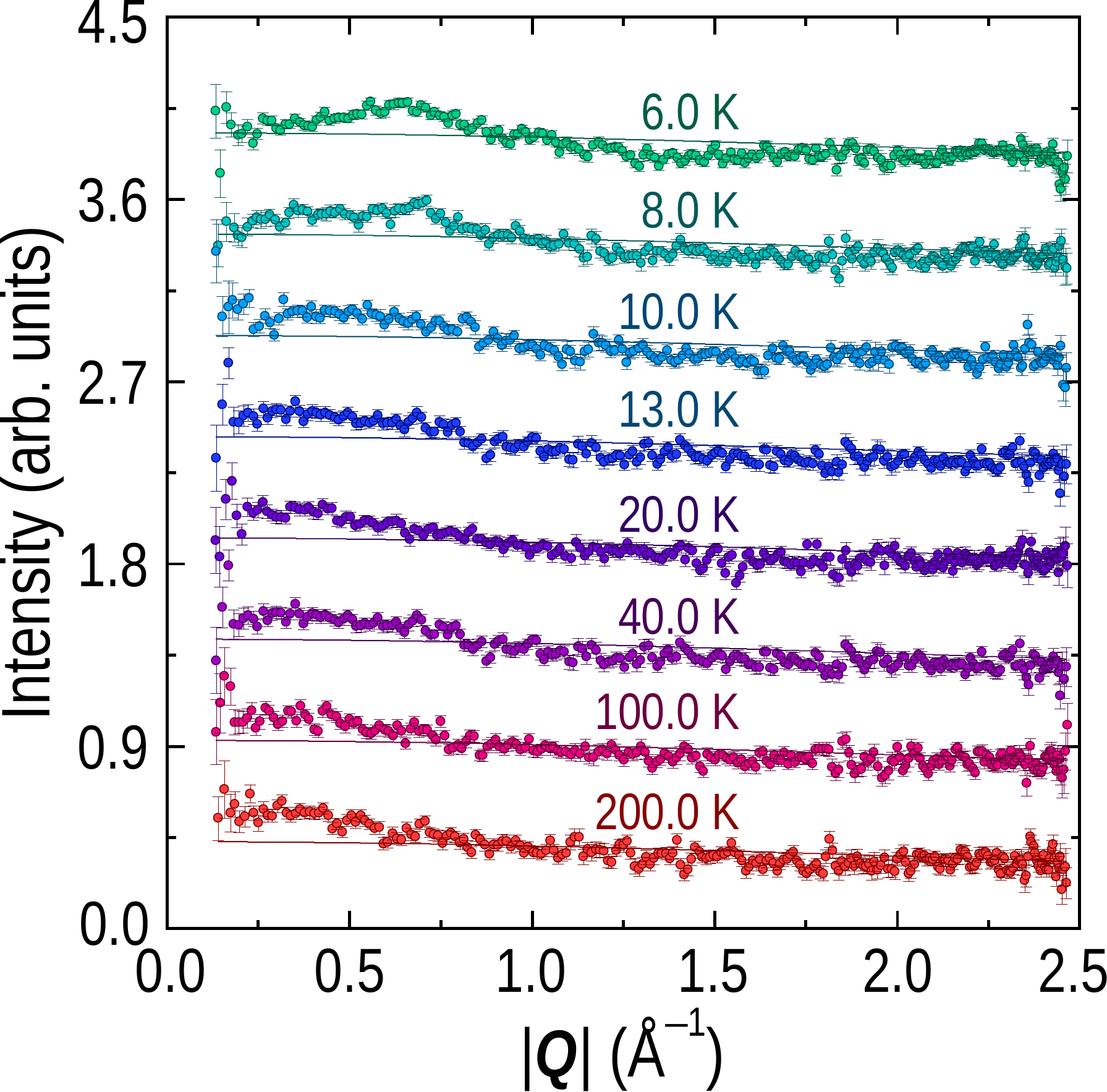}
    \caption{\label{diffuse200K}Diffuse magnetic intensity at higher temperatures.  Solid lines represent the magnetic form factor of Cu$^{2+}$. The 6.0-K dataset is identical to that in Fig.~\ref{diffuse}(a).}
\end{figure}

Figure \ref{diffuse200K} shows the diffuse magnetic intensity collected on the D7 diffractometer to higher temperatures. The data at all temperatures are fit to $I = AF^2$, where $I$ is the intensity, $F$ is the magnetic form factor for Cu$^{2+}$, and $A$ is 0.2290(19) for all datasets in the units used. The diffuse peak shifts to lower $|\bm{Q}|$, but a significant diffuse contribution persists to at least 200\,K.

\section{Additional Muon Spin Rotation and Relaxation ($\mu$SR) Results\label{AppmuSR}}

\begin{figure}[t!]
  \includegraphics[width=\columnwidth]{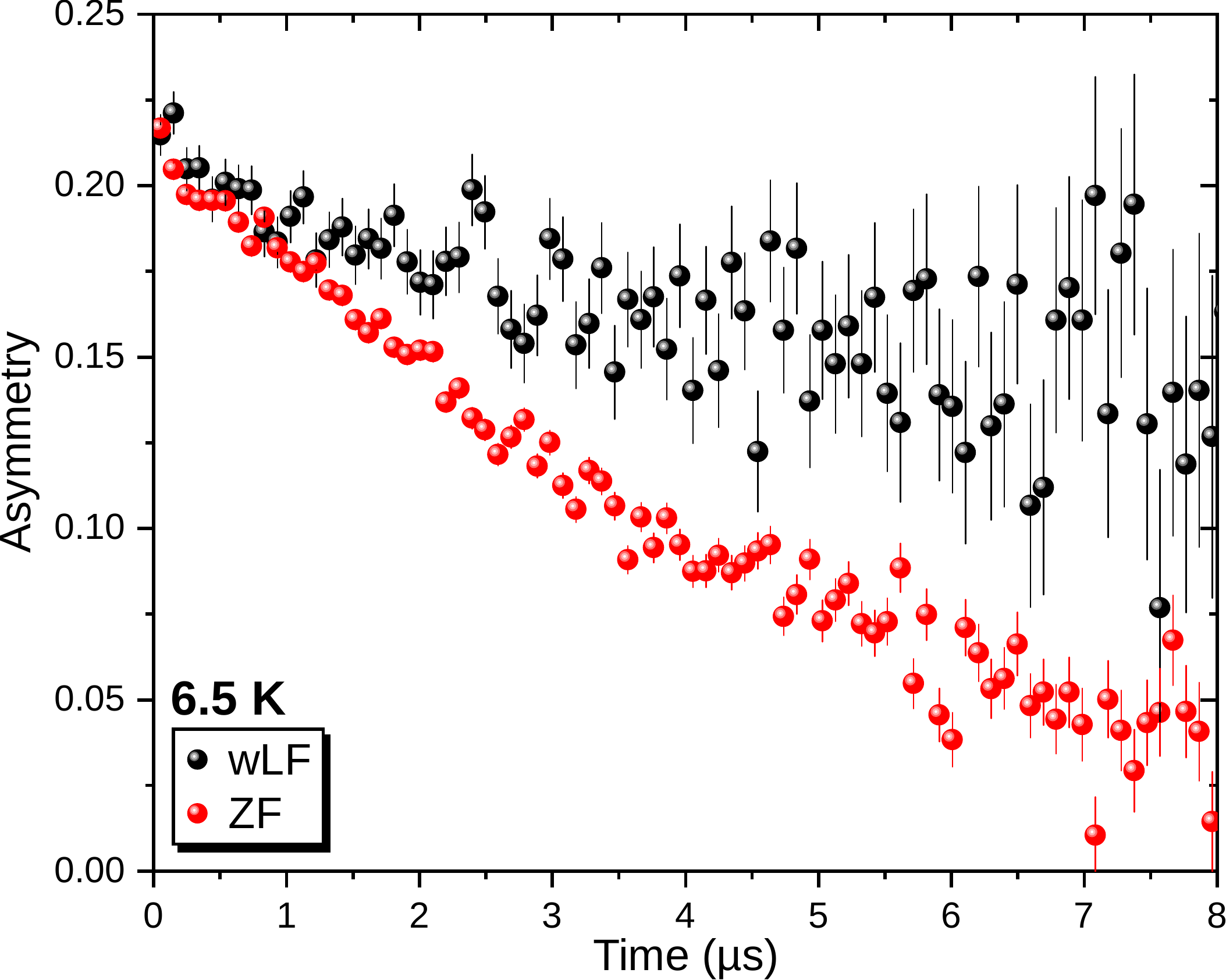}
 \caption{\label{LFZF6K}ZF and wLF spectra in the paramagnetic state at 6.5\,K. The damping is suppressed significantly by longitudinal field.}
\end{figure}

\begin{figure}[b!]
  \includegraphics[width=\columnwidth]{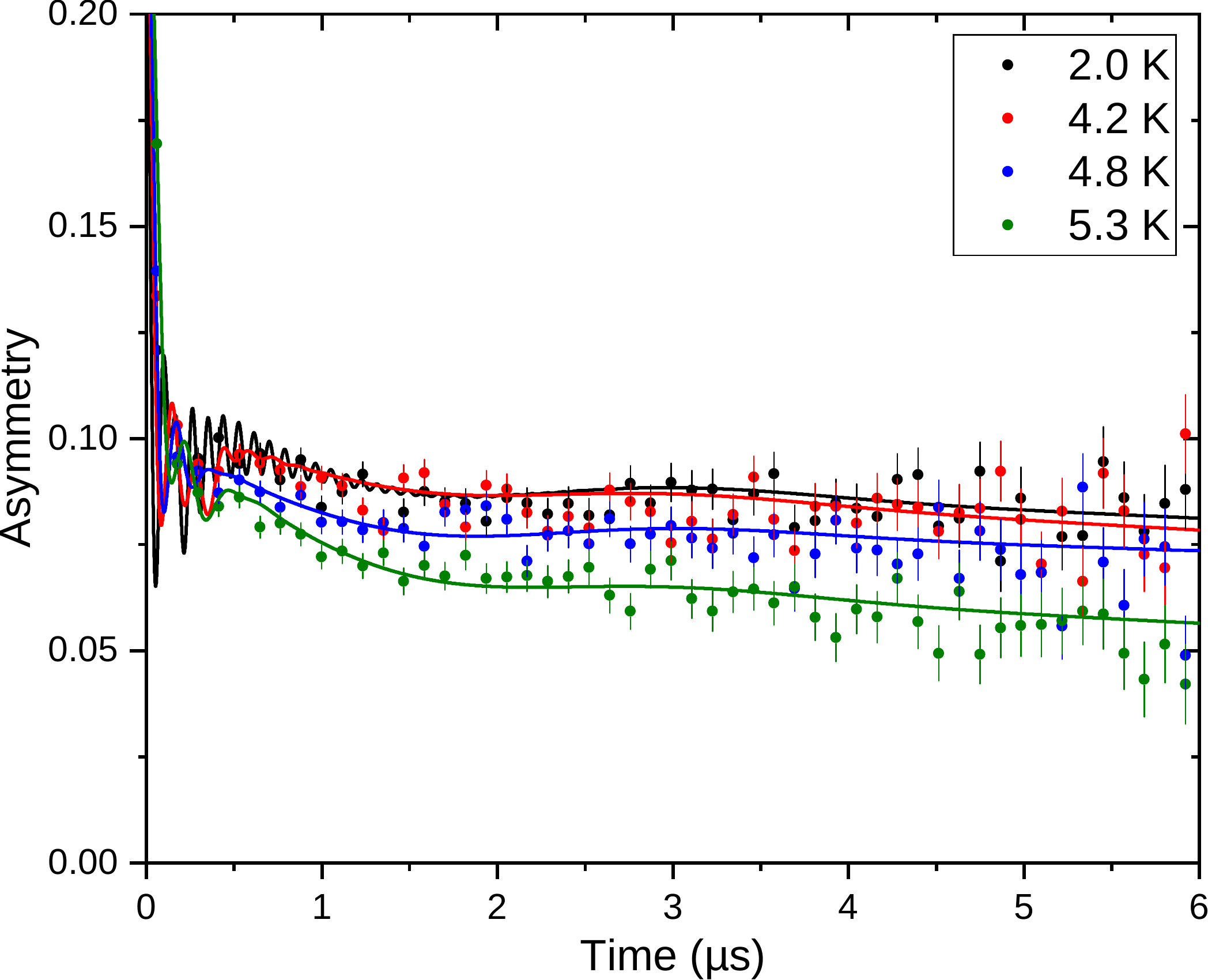}
 \caption{\label{long6mis} wLF spectra in the long-range magnetically ordered states.}
\end{figure}

The depolarization in ZF spectra in the paramagnetic regime is dominated by strong damping described by a static Gaussian Kubo-Toyabe function
\begin{equation}
 G_{z}(t)^\text{GKT} = \frac23 [1- (\sigma t)^2]e^{-\frac12(\sigma t)^2}  + \frac13 . \label{eq6}
\end{equation}
Here the Gaussian damping width $\sigma$ is $\gamma_\mu\Delta$, with $\Delta$ being the width of the Gaussian distribution of local field components $B_{x,y}$ around zero. $G_z$ is the component of the asymmetry along the direction of the incoming muon beam.

\begin{figure}[t!]
  \includegraphics[width=\columnwidth]{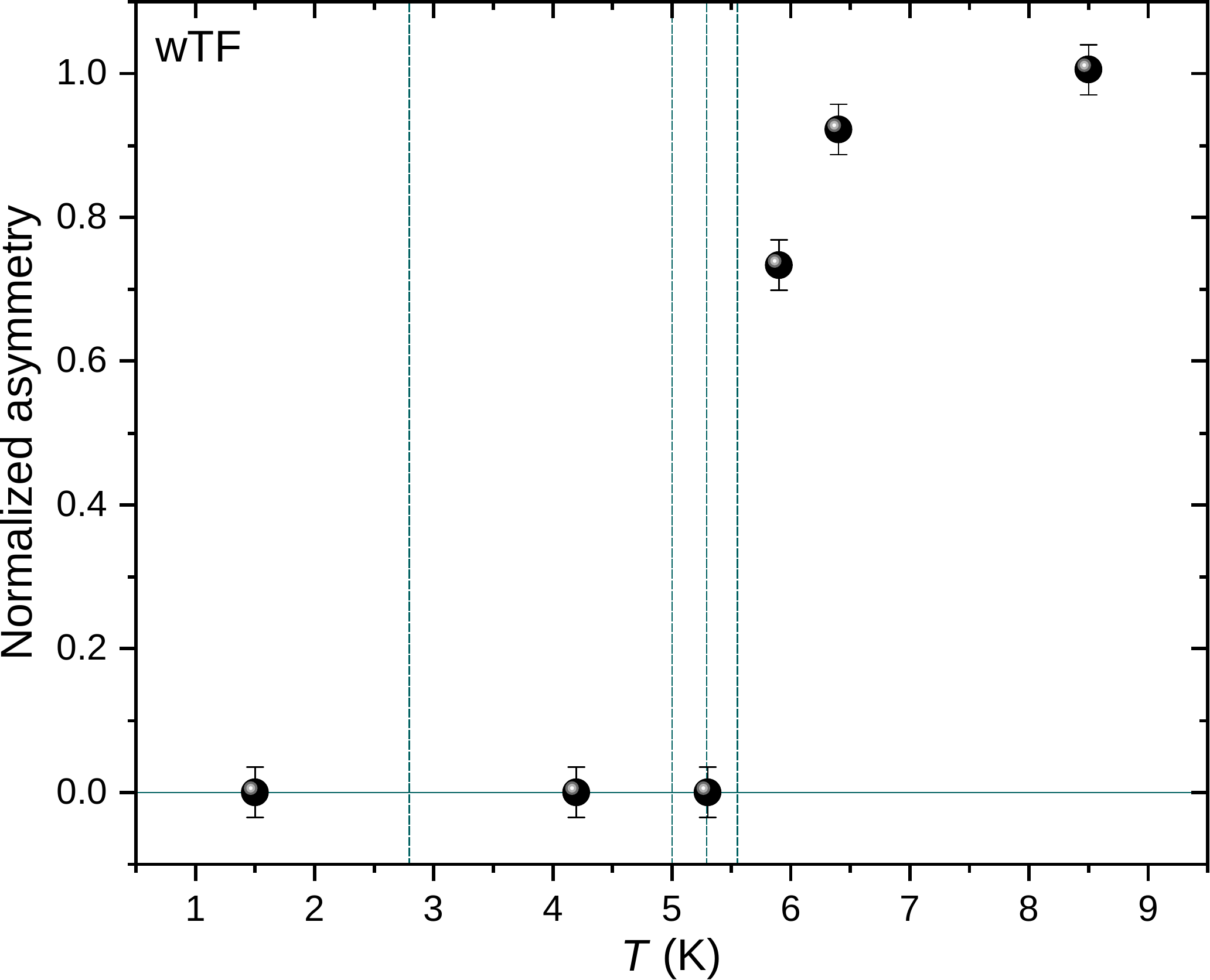}
 \caption{\label{TFasy}Normalized asymmetry from the fitted wTF spectra versus temperature.  Vertical dashed lines indicate the magnetic phase transitions extracted from specific heat data.}
\end{figure}

For parametrizing our asymmetry spectra below \TN\ we use a phenomenological approach:
\begin{equation}
  A(t) = \sum_{i=1}^2 a_{\text{rot,}i} G_{z,i}(t)^\text{rot}+a_\text{GKT} G_{z}(t)^\text{GKT}+\text{BG}\label{eq7}
\end{equation}
with
\begin{equation}
  G_{z,i}(t)^\text{rot} = \frac23 e^{-\lambda_{\text{t},i}t} \cos(2\pi\nu_{\mu,i}t +\varphi_i) + \frac13 e^{-\lambda_\text{l}t} ,\label{eq8}
\end{equation}
where $\nu_{\mu,i}$ is the rotation frequency in local internal fields caused by the ordered electronic moments, while $\lambda_\text{t}$ is the transverse damping parameter of the oscillations, mainly caused by field inhomogeneities, and can be seen as a direct measure of the width of the distribution of $B_i$ arising from local disorder in the long-range-ordered spin array. Here $a_\text{rot}$ is the contribution to the asymmetry from the rotating signal.  The two terms in $G_{z,i}(t)^\text{rot}$, transverse and longitudinal, arise from the usual approach to describe the isotropic average by assuming that $\frac23$ of the muons experience an internal field perpendicular to the initial orientation of their spins, resulting in precession. For the remaining $\frac13$ of the muons, the internal field is oriented parallel to the muon spins and no precession takes place. However, dynamic muon spin polarization, caused by fluctuations of the magnetic moments surrounding the muon, introduces a nonzero longitudinal term with a relaxation rate $\lambda_\text{l}$ proportional to the magnetic spin fluctuation rate.

Figure \ref{LFZF6K} shows the asymmetry in the paramagnetic state at 6.5\,K, where a longitudinal field of 2\,mT suppresses the damping. Figure \ref{long6mis} shows wLF spectra to longer decay times in the magnetically ordered state.  Figure \ref{TFasy} shows the temperature dependence of the normalized asymmetry of the wTF spectra.  A reduction is seen in points just above the highest magnetic transition, indicating an onset of short-range order, while points below the transition indicate ordering throughout the sample. 

\bibliography{antlerite,DFG}

\begin{thebibliography}{69}%
\makeatletter
\providecommand \@ifxundefined [1]{%
 \@ifx{#1\undefined}
}%
\providecommand \@ifnum [1]{%
 \ifnum #1\expandafter \@firstoftwo
 \else \expandafter \@secondoftwo
 \fi
}%
\providecommand \@ifx [1]{%
 \ifx #1\expandafter \@firstoftwo
 \else \expandafter \@secondoftwo
 \fi
}%
\providecommand \natexlab [1]{#1}%
\providecommand \enquote  [1]{``#1''}%
\providecommand \bibnamefont  [1]{#1}%
\providecommand \bibfnamefont [1]{#1}%
\providecommand \citenamefont [1]{#1}%
\providecommand \href@noop [0]{\@secondoftwo}%
\providecommand \href [0]{\begingroup \@sanitize@url \@href}%
\providecommand \@href[1]{\@@startlink{#1}\@@href}%
\providecommand \@@href[1]{\endgroup#1\@@endlink}%
\providecommand \@sanitize@url [0]{\catcode `\\12\catcode `\$12\catcode
  `\&12\catcode `\#12\catcode `\^12\catcode `\_12\catcode `\%12\relax}%
\providecommand \@@startlink[1]{}%
\providecommand \@@endlink[0]{}%
\providecommand \url  [0]{\begingroup\@sanitize@url \@url }%
\providecommand \@url [1]{\endgroup\@href {#1}{\urlprefix }}%
\providecommand \urlprefix  [0]{URL }%
\providecommand \Eprint [0]{\href }%
\providecommand \doibase [0]{https://doi.org/}%
\providecommand \selectlanguage [0]{\@gobble}%
\providecommand \bibinfo  [0]{\@secondoftwo}%
\providecommand \bibfield  [0]{\@secondoftwo}%
\providecommand \translation [1]{[#1]}%
\providecommand \BibitemOpen [0]{}%
\providecommand \bibitemStop [0]{}%
\providecommand \bibitemNoStop [0]{.\EOS\space}%
\providecommand \EOS [0]{\spacefactor3000\relax}%
\providecommand \BibitemShut  [1]{\csname bibitem#1\endcsname}%
\let\auto@bib@innerbib\@empty
\bibitem [{\citenamefont {Ramirez}(1994)}]{Ramirez1994}%
  \BibitemOpen
  \bibfield  {author} {\bibinfo {author} {\bibfnamefont {A.~P.}\ \bibnamefont
  {Ramirez}},\ }\bibfield  {title} {\bibinfo {title} {Strongly geometrically
  frustrated magnets},\ }\href
  {https://doi.org/10.1146/annurev.ms.24.080194.002321} {\bibfield  {journal}
  {\bibinfo  {journal} {Annu.\ Rev.\ Mater.\ Sci.}\ }\textbf {\bibinfo {volume}
  {24}},\ \bibinfo {pages} {453} (\bibinfo {year} {1994})}\BibitemShut
  {NoStop}%
\bibitem [{\citenamefont {Batista}\ \emph {et~al.}(2016)\citenamefont
  {Batista}, \citenamefont {Lin}, \citenamefont {Hayami},\ and\ \citenamefont
  {Kamiya}}]{Batista2016}%
  \BibitemOpen
  \bibfield  {author} {\bibinfo {author} {\bibfnamefont {C.~D.}\ \bibnamefont
  {Batista}}, \bibinfo {author} {\bibfnamefont {S.-Z.}\ \bibnamefont {Lin}},
  \bibinfo {author} {\bibfnamefont {S.}~\bibnamefont {Hayami}},\ and\ \bibinfo
  {author} {\bibfnamefont {Y.}~\bibnamefont {Kamiya}},\ }\bibfield  {title}
  {\bibinfo {title} {Frustration and chiral orderings in correlated electron
  systems},\ }\href {https://doi.org/10.1088/0034-4885/79/8/084504} {\bibfield
  {journal} {\bibinfo  {journal} {Rep.\ Prog.\ Phys.}\ }\textbf {\bibinfo
  {volume} {79}},\ \bibinfo {pages} {084504} (\bibinfo {year}
  {2016})}\BibitemShut {NoStop}%
\bibitem [{\citenamefont {Schmidt}\ and\ \citenamefont
  {Thalmeier}(2017)}]{Schmidt2017}%
  \BibitemOpen
  \bibfield  {author} {\bibinfo {author} {\bibfnamefont {B.}~\bibnamefont
  {Schmidt}}\ and\ \bibinfo {author} {\bibfnamefont {P.}~\bibnamefont
  {Thalmeier}},\ }\bibfield  {title} {\bibinfo {title} {Frustrated two
  dimensional quantum magnets},\ }\href
  {https://doi.org/10.1016/j.physrep.2017.06.004} {\bibfield  {journal}
  {\bibinfo  {journal} {Phys.\ Rep.}\ }\textbf {\bibinfo {volume} {703}},\
  \bibinfo {pages} {1} (\bibinfo {year} {2017})}\BibitemShut {NoStop}%
\bibitem [{\citenamefont {Lacroix}\ \emph {et~al.}(2011)\citenamefont
  {Lacroix}, \citenamefont {Mendels},\ and\ \citenamefont
  {Mila}}]{Lacroix2011}%
  \BibitemOpen
  \bibinfo {editor} {\bibfnamefont {C.}~\bibnamefont {Lacroix}}, \bibinfo
  {editor} {\bibfnamefont {P.}~\bibnamefont {Mendels}},\ and\ \bibinfo {editor}
  {\bibfnamefont {F.}~\bibnamefont {Mila}},\ eds.,\ \href
  {https://doi.org/10.1007/978-3-642-10589-0} {\emph {\bibinfo {title}
  {Introduction to Frustrated Magnetism}}},\ \bibinfo {series} {Springer Series
  in Solid-State Sciences}, Vol.\ \bibinfo {volume} {164}\ (\bibinfo
  {publisher} {Springer},\ \bibinfo {address} {Berlin},\ \bibinfo {year}
  {2011})\BibitemShut {NoStop}%
\bibitem [{\citenamefont {Park}\ and\ \citenamefont {Snyder}(1995)}]{Park1995}%
  \BibitemOpen
  \bibfield  {author} {\bibinfo {author} {\bibfnamefont {C.}~\bibnamefont
  {Park}}\ and\ \bibinfo {author} {\bibfnamefont {R.~L.}\ \bibnamefont
  {Snyder}},\ }\bibfield  {title} {\bibinfo {title} {Structures of
  high-temperature cuprate superconductors},\ }\href
  {https://doi.org/10.1111/j.1151-2916.1995.tb07953.x} {\bibfield  {journal}
  {\bibinfo  {journal} {J.\ Am.\ Ceram.\ Soc.}\ }\textbf {\bibinfo {volume}
  {78}},\ \bibinfo {pages} {3171} (\bibinfo {year} {1995})}\BibitemShut
  {NoStop}%
\bibitem [{\citenamefont {van Smaalen}(1999)}]{Smaalen1999}%
  \BibitemOpen
  \bibfield  {author} {\bibinfo {author} {\bibfnamefont {S.}~\bibnamefont {van
  Smaalen}},\ }\bibfield  {title} {\bibinfo {title} {Structural aspects of
  spin-chain and spin-ladder oxides},\ }\href
  {https://doi.org/10.1524/zkri.1999.214.12.786} {\bibfield  {journal}
  {\bibinfo  {journal} {Z.\ Kristallogr.\ Cryst.\ Mater.}\ }\textbf {\bibinfo
  {volume} {214}},\ \bibinfo {pages} {786} (\bibinfo {year}
  {1999})}\BibitemShut {NoStop}%
\bibitem [{\citenamefont {Kulbakov}\ \emph {et~al.}(2022)\citenamefont
  {Kulbakov}, \citenamefont {Kononenko}, \citenamefont {Nishimoto},
  \citenamefont {Stahl}, \citenamefont {Mannathanath~Chakkingal}, \citenamefont
  {Feig}, \citenamefont {Gumeniuk}, \citenamefont {Skourski}, \citenamefont
  {Bhaskaran}, \citenamefont {Zvyagin}, \citenamefont {Embs}, \citenamefont
  {Puente-Orench}, \citenamefont {Wildes}, \citenamefont {Geck}, \citenamefont
  {Janson}, \citenamefont {Inosov},\ and\ \citenamefont
  {Peets}}]{Kulbakov2022a}%
  \BibitemOpen
  \bibfield  {author} {\bibinfo {author} {\bibfnamefont {A.~A.}\ \bibnamefont
  {Kulbakov}}, \bibinfo {author} {\bibfnamefont {D.~Y.}\ \bibnamefont
  {Kononenko}}, \bibinfo {author} {\bibfnamefont {S.}~\bibnamefont
  {Nishimoto}}, \bibinfo {author} {\bibfnamefont {Q.}~\bibnamefont {Stahl}},
  \bibinfo {author} {\bibfnamefont {A.}~\bibnamefont
  {Mannathanath~Chakkingal}}, \bibinfo {author} {\bibfnamefont
  {M.}~\bibnamefont {Feig}}, \bibinfo {author} {\bibfnamefont {R.}~\bibnamefont
  {Gumeniuk}}, \bibinfo {author} {\bibfnamefont {Y.}~\bibnamefont {Skourski}},
  \bibinfo {author} {\bibfnamefont {L.}~\bibnamefont {Bhaskaran}}, \bibinfo
  {author} {\bibfnamefont {S.~A.}\ \bibnamefont {Zvyagin}}, \bibinfo {author}
  {\bibfnamefont {J.~P.}\ \bibnamefont {Embs}}, \bibinfo {author}
  {\bibfnamefont {I.}~\bibnamefont {Puente-Orench}}, \bibinfo {author}
  {\bibfnamefont {A.}~\bibnamefont {Wildes}}, \bibinfo {author} {\bibfnamefont
  {J.}~\bibnamefont {Geck}}, \bibinfo {author} {\bibfnamefont {O.}~\bibnamefont
  {Janson}}, \bibinfo {author} {\bibfnamefont {D.~S.}\ \bibnamefont {Inosov}},\
  and\ \bibinfo {author} {\bibfnamefont {D.~C.}\ \bibnamefont {Peets}},\
  }\bibfield  {title} {\bibinfo {title} {Coupled frustrated ferromagnetic and
  antiferromagnetic quantum spin chains in the quasi-one-dimensional mineral
  antlerite, {Cu$_3$SO$_4$(OH)$_4$}},\ }\href
  {https://doi.org/10.1103/PhysRevB.106.L020405} {\bibfield  {journal}
  {\bibinfo  {journal} {Phys.\ Rev.\ B}\ }\textbf {\bibinfo {volume} {106}},\
  \bibinfo {pages} {L020405} (\bibinfo {year} {2022})}\BibitemShut {NoStop}%
\bibitem [{\citenamefont {Inosov}(2018)}]{Inosov2018}%
  \BibitemOpen
  \bibfield  {author} {\bibinfo {author} {\bibfnamefont {D.~S.}\ \bibnamefont
  {Inosov}},\ }\bibfield  {title} {\bibinfo {title} {Quantum magnetism in
  minerals},\ }\href {https://doi.org/10.1080/00018732.2018.1571986} {\bibfield
   {journal} {\bibinfo  {journal} {Adv.\ Phys.}\ }\textbf {\bibinfo {volume}
  {67}},\ \bibinfo {pages} {149} (\bibinfo {year} {2018})}\BibitemShut
  {NoStop}%
\bibitem [{\citenamefont {Shores}\ \emph {et~al.}(2005)\citenamefont {Shores},
  \citenamefont {Nytko}, \citenamefont {Bartlett},\ and\ \citenamefont
  {Nocera}}]{Shores2005}%
  \BibitemOpen
  \bibfield  {author} {\bibinfo {author} {\bibfnamefont {M.~P.}\ \bibnamefont
  {Shores}}, \bibinfo {author} {\bibfnamefont {E.~A.}\ \bibnamefont {Nytko}},
  \bibinfo {author} {\bibfnamefont {B.~M.}\ \bibnamefont {Bartlett}},\ and\
  \bibinfo {author} {\bibfnamefont {D.~G.}\ \bibnamefont {Nocera}},\ }\bibfield
   {title} {\bibinfo {title} {A structurally perfect ${S} = \frac12$ kagom{\'e}
  antiferromagnet},\ }\href {https://doi.org/10.1021/ja053891p} {\bibfield
  {journal} {\bibinfo  {journal} {J.\ Am.\ Chem.\ Soc.}\ }\textbf {\bibinfo
  {volume} {127}},\ \bibinfo {pages} {13462} (\bibinfo {year}
  {2005})}\BibitemShut {NoStop}%
\bibitem [{\citenamefont {Norman}(2016)}]{Norman2016}%
  \BibitemOpen
  \bibfield  {author} {\bibinfo {author} {\bibfnamefont {M.~R.}\ \bibnamefont
  {Norman}},\ }\bibfield  {title} {\bibinfo {title} {Colloquium:
  Herbertsmithite and the search for the quantum spin liquid},\ }\href
  {https://doi.org/10.1103/RevModPhys.88.041002} {\bibfield  {journal}
  {\bibinfo  {journal} {Rev.\ Mod.\ Phys.}\ }\textbf {\bibinfo {volume} {88}},\
  \bibinfo {pages} {041002} (\bibinfo {year} {2016})}\BibitemShut {NoStop}%
\bibitem [{\citenamefont {Heinze}\ \emph {et~al.}(2021)\citenamefont {Heinze},
  \citenamefont {Jeschke}, \citenamefont {Mazin}, \citenamefont
  {Metavitsiadis}, \citenamefont {Reehuis}, \citenamefont {Feyerherm},
  \citenamefont {Hoffmann}, \citenamefont {Bartkowiak}, \citenamefont
  {Prokhnenko}, \citenamefont {Wolter}, \citenamefont {Ding}, \citenamefont
  {Zapf}, \citenamefont {Corval{\'a}n~Moya}, \citenamefont {Weickert},
  \citenamefont {Jaime}, \citenamefont {Rule}, \citenamefont {Menzel},
  \citenamefont {Valent{\'i}}, \citenamefont {Brenig},\ and\ \citenamefont
  {S{\"u}llow}}]{Heinze2021}%
  \BibitemOpen
  \bibfield  {author} {\bibinfo {author} {\bibfnamefont {L.}~\bibnamefont
  {Heinze}}, \bibinfo {author} {\bibfnamefont {H.~O.}\ \bibnamefont {Jeschke}},
  \bibinfo {author} {\bibfnamefont {I.~I.}\ \bibnamefont {Mazin}}, \bibinfo
  {author} {\bibfnamefont {A.}~\bibnamefont {Metavitsiadis}}, \bibinfo {author}
  {\bibfnamefont {M.}~\bibnamefont {Reehuis}}, \bibinfo {author} {\bibfnamefont
  {R.}~\bibnamefont {Feyerherm}}, \bibinfo {author} {\bibfnamefont {J.-U.}\
  \bibnamefont {Hoffmann}}, \bibinfo {author} {\bibfnamefont {M.}~\bibnamefont
  {Bartkowiak}}, \bibinfo {author} {\bibfnamefont {O.}~\bibnamefont
  {Prokhnenko}}, \bibinfo {author} {\bibfnamefont {A.~U.~B.}\ \bibnamefont
  {Wolter}}, \bibinfo {author} {\bibfnamefont {X.}~\bibnamefont {Ding}},
  \bibinfo {author} {\bibfnamefont {V.~S.}\ \bibnamefont {Zapf}}, \bibinfo
  {author} {\bibfnamefont {C.}~\bibnamefont {Corval{\'a}n~Moya}}, \bibinfo
  {author} {\bibfnamefont {F.}~\bibnamefont {Weickert}}, \bibinfo {author}
  {\bibfnamefont {M.}~\bibnamefont {Jaime}}, \bibinfo {author} {\bibfnamefont
  {K.~C.}\ \bibnamefont {Rule}}, \bibinfo {author} {\bibfnamefont
  {D.}~\bibnamefont {Menzel}}, \bibinfo {author} {\bibfnamefont
  {R.}~\bibnamefont {Valent{\'i}}}, \bibinfo {author} {\bibfnamefont
  {W.}~\bibnamefont {Brenig}},\ and\ \bibinfo {author} {\bibfnamefont
  {S.}~\bibnamefont {S{\"u}llow}},\ }\bibfield  {title} {\bibinfo {title}
  {Magnetization process of atacamite: A case of weakly coupled ${S}=1/2$
  sawtooth chains},\ }\href {https://doi.org/10.1103/PhysRevLett.126.207201}
  {\bibfield  {journal} {\bibinfo  {journal} {Phys.\ Rev.\ Lett.}\ }\textbf
  {\bibinfo {volume} {126}},\ \bibinfo {pages} {207201} (\bibinfo {year}
  {2021})}\BibitemShut {NoStop}%
\bibitem [{\citenamefont {Zhang}\ \emph {et~al.}(2020)\citenamefont {Zhang},
  \citenamefont {Zhao}, \citenamefont {Gautreau}, \citenamefont {Raczkowski},
  \citenamefont {Saha}, \citenamefont {Garlea}, \citenamefont {Cao},
  \citenamefont {Hong}, \citenamefont {Jeschke}, \citenamefont {Mahanti},
  \citenamefont {Birol}, \citenamefont {Assaad},\ and\ \citenamefont
  {Ke}}]{Zhang2020bot}%
  \BibitemOpen
  \bibfield  {author} {\bibinfo {author} {\bibfnamefont {H.}~\bibnamefont
  {Zhang}}, \bibinfo {author} {\bibfnamefont {Z.}~\bibnamefont {Zhao}},
  \bibinfo {author} {\bibfnamefont {D.}~\bibnamefont {Gautreau}}, \bibinfo
  {author} {\bibfnamefont {M.}~\bibnamefont {Raczkowski}}, \bibinfo {author}
  {\bibfnamefont {A.}~\bibnamefont {Saha}}, \bibinfo {author} {\bibfnamefont
  {V.~O.}\ \bibnamefont {Garlea}}, \bibinfo {author} {\bibfnamefont
  {H.}~\bibnamefont {Cao}}, \bibinfo {author} {\bibfnamefont {T.}~\bibnamefont
  {Hong}}, \bibinfo {author} {\bibfnamefont {H.~O.}\ \bibnamefont {Jeschke}},
  \bibinfo {author} {\bibfnamefont {S.~D.}\ \bibnamefont {Mahanti}}, \bibinfo
  {author} {\bibfnamefont {T.}~\bibnamefont {Birol}}, \bibinfo {author}
  {\bibfnamefont {F.~F.}\ \bibnamefont {Assaad}},\ and\ \bibinfo {author}
  {\bibfnamefont {X.}~\bibnamefont {Ke}},\ }\bibfield  {title} {\bibinfo
  {title} {Coexistence and interaction of spinons and magnons in an
  antiferromagnet with alternating antiferromagnetic and ferromagnetic quantum
  spin chains},\ }\href {https://doi.org/10.1103/PhysRevLett.125.037204}
  {\bibfield  {journal} {\bibinfo  {journal} {Phys. Rev. Lett.}\ }\textbf
  {\bibinfo {volume} {125}},\ \bibinfo {pages} {037204} (\bibinfo {year}
  {2020})}\BibitemShut {NoStop}%
\bibitem [{\citenamefont {Hawthorne}\ \emph {et~al.}(1989)\citenamefont
  {Hawthorne}, \citenamefont {Groat},\ and\ \citenamefont
  {Eby}}]{Hawthorne1989}%
  \BibitemOpen
  \bibfield  {author} {\bibinfo {author} {\bibfnamefont {F.~C.}\ \bibnamefont
  {Hawthorne}}, \bibinfo {author} {\bibfnamefont {L.~A.}\ \bibnamefont
  {Groat}},\ and\ \bibinfo {author} {\bibfnamefont {R.~R.}\ \bibnamefont
  {Eby}},\ }\bibfield  {title} {\bibinfo {title} {Antlerite,
  {Cu$_3$SO$_4$(OH)$_4$}, a heteropolyhedral wallpaper structure},\ }\href
  {https://pubs.geoscienceworld.org/canmin/article-abstract/27/2/205/12141/antlerite-cu-3-so-4-oh-4-a-heteropolyhedral}
  {\bibfield  {journal} {\bibinfo  {journal} {Can.\ Mineral.}\ }\textbf
  {\bibinfo {volume} {27}},\ \bibinfo {pages} {205} (\bibinfo {year}
  {1989})}\BibitemShut {NoStop}%
\bibitem [{\citenamefont {Vilminot}\ \emph {et~al.}(2003)\citenamefont
  {Vilminot}, \citenamefont {Richard-Plouet}, \citenamefont {Andr{\'e}},
  \citenamefont {Swierczynski}, \citenamefont {Guillot}, \citenamefont
  {Bour{\'e}e-Vigneron},\ and\ \citenamefont {Drillon}}]{Vilminot2003}%
  \BibitemOpen
  \bibfield  {author} {\bibinfo {author} {\bibfnamefont {S.}~\bibnamefont
  {Vilminot}}, \bibinfo {author} {\bibfnamefont {M.}~\bibnamefont
  {Richard-Plouet}}, \bibinfo {author} {\bibfnamefont {G.}~\bibnamefont
  {Andr{\'e}}}, \bibinfo {author} {\bibfnamefont {D.}~\bibnamefont
  {Swierczynski}}, \bibinfo {author} {\bibfnamefont {M.}~\bibnamefont
  {Guillot}}, \bibinfo {author} {\bibfnamefont {F.}~\bibnamefont
  {Bour{\'e}e-Vigneron}},\ and\ \bibinfo {author} {\bibfnamefont
  {M.}~\bibnamefont {Drillon}},\ }\bibfield  {title} {\bibinfo {title}
  {Magnetic structure and properties of {Cu$_3$(OH)$_4$SO$_4$} made of triple
  chains of spins $s$=1/2},\ }\href
  {https://doi.org/10.1016/S0022-4596(02)00081-6} {\bibfield  {journal}
  {\bibinfo  {journal} {J.\ Solid State Chem.}\ }\textbf {\bibinfo {volume}
  {170}},\ \bibinfo {pages} {255} (\bibinfo {year} {2003})}\BibitemShut
  {NoStop}%
\bibitem [{\citenamefont {Hara}\ \emph {et~al.}(2011)\citenamefont {Hara},
  \citenamefont {Kondo},\ and\ \citenamefont {Sato}}]{Hara2011}%
  \BibitemOpen
  \bibfield  {author} {\bibinfo {author} {\bibfnamefont {S.}~\bibnamefont
  {Hara}}, \bibinfo {author} {\bibfnamefont {H.}~\bibnamefont {Kondo}},\ and\
  \bibinfo {author} {\bibfnamefont {H.}~\bibnamefont {Sato}},\ }\bibfield
  {title} {\bibinfo {title} {Successive magnetic transitions in candidate
  ``idle-spin'' system, {Cu$_3$(OH)$_4$SO$_4$}},\ }\href
  {https://doi.org/10.1143/JPSJ.80.043701} {\bibfield  {journal} {\bibinfo
  {journal} {J.\ Phys.\ Soc.\ Jpn.}\ }\textbf {\bibinfo {volume} {80}},\
  \bibinfo {pages} {043701} (\bibinfo {year} {2011})}\BibitemShut {NoStop}%
\bibitem [{\citenamefont {Koo}\ \emph {et~al.}(2012)\citenamefont {Koo},
  \citenamefont {Kremer},\ and\ \citenamefont {Whangbo}}]{Koo2012}%
  \BibitemOpen
  \bibfield  {author} {\bibinfo {author} {\bibfnamefont {H.-J.}\ \bibnamefont
  {Koo}}, \bibinfo {author} {\bibfnamefont {R.~K.}\ \bibnamefont {Kremer}},\
  and\ \bibinfo {author} {\bibfnamefont {M.-H.}\ \bibnamefont {Whangbo}},\
  }\bibfield  {title} {\bibinfo {title} {Non-idle-spin behavior and
  field-induced magnetic transitions of the triple chain magnet
  {Cu$_3$(OH)$_4$SO$_4$}},\ }\href {https://doi.org/10.1143/JPSJ.81.063704}
  {\bibfield  {journal} {\bibinfo  {journal} {J.\ Phys.\ Soc.\ Jpn.}\ }\textbf
  {\bibinfo {volume} {81}},\ \bibinfo {pages} {063704} (\bibinfo {year}
  {2012})}\BibitemShut {NoStop}%
\bibitem [{\citenamefont {Fujii}\ \emph {et~al.}(2013)\citenamefont {Fujii},
  \citenamefont {Ishikawa}, \citenamefont {Kikuchi}, \citenamefont {Narumi},
  \citenamefont {Nojiri}, \citenamefont {Hara},\ and\ \citenamefont
  {Sato}}]{Fujii2013}%
  \BibitemOpen
  \bibfield  {author} {\bibinfo {author} {\bibfnamefont {Y.}~\bibnamefont
  {Fujii}}, \bibinfo {author} {\bibfnamefont {Y.}~\bibnamefont {Ishikawa}},
  \bibinfo {author} {\bibfnamefont {H.}~\bibnamefont {Kikuchi}}, \bibinfo
  {author} {\bibfnamefont {Y.}~\bibnamefont {Narumi}}, \bibinfo {author}
  {\bibfnamefont {H.}~\bibnamefont {Nojiri}}, \bibinfo {author} {\bibfnamefont
  {S.}~\bibnamefont {Hara}},\ and\ \bibinfo {author} {\bibfnamefont
  {H.}~\bibnamefont {Sato}},\ }\bibfield  {title} {\bibinfo {title} {Magnetic
  property of a single crystal of spin-1/2 triple-chain magnet
  {Cu$_3$(OH)$_4$SO$_4$}},\ }\href {https://doi.org/10.3938/jkps.62.2054}
  {\bibfield  {journal} {\bibinfo  {journal} {J.\ Korean Phys.\ Soc.}\ }\textbf
  {\bibinfo {volume} {62}},\ \bibinfo {pages} {2054} (\bibinfo {year}
  {2013})}\BibitemShut {NoStop}%
\bibitem [{\citenamefont {Studer}\ \emph {et~al.}(2006)\citenamefont {Studer},
  \citenamefont {Hagen},\ and\ \citenamefont {Noakes}}]{Wombat}%
  \BibitemOpen
  \bibfield  {author} {\bibinfo {author} {\bibfnamefont {A.~J.}\ \bibnamefont
  {Studer}}, \bibinfo {author} {\bibfnamefont {M.~E.}\ \bibnamefont {Hagen}},\
  and\ \bibinfo {author} {\bibfnamefont {T.~J.}\ \bibnamefont {Noakes}},\
  }\bibfield  {title} {\bibinfo {title} {Wombat: The high-intensity powder
  diffractometer at the {OPAL} reactor},\ }\href
  {https://doi.org/10.1016/j.physb.2006.05.323} {\bibfield  {journal} {\bibinfo
   {journal} {Physica B: Condens.\ Matter}\ }\textbf {\bibinfo {volume}
  {385--386}},\ \bibinfo {pages} {1013} (\bibinfo {year} {2006})}\BibitemShut
  {NoStop}%
\bibitem [{\citenamefont {Puente~Orench}\ \emph {et~al.}(2014)\citenamefont
  {Puente~Orench}, \citenamefont {Clergeau}, \citenamefont {Mart{\'i}nez},
  \citenamefont {Olmos}, \citenamefont {Fabelo},\ and\ \citenamefont
  {Campo}}]{D1B}%
  \BibitemOpen
  \bibfield  {author} {\bibinfo {author} {\bibfnamefont {I.}~\bibnamefont
  {Puente~Orench}}, \bibinfo {author} {\bibfnamefont {J.~F.}\ \bibnamefont
  {Clergeau}}, \bibinfo {author} {\bibfnamefont {S.}~\bibnamefont
  {Mart{\'i}nez}}, \bibinfo {author} {\bibfnamefont {M.}~\bibnamefont {Olmos}},
  \bibinfo {author} {\bibfnamefont {O.}~\bibnamefont {Fabelo}},\ and\ \bibinfo
  {author} {\bibfnamefont {J.}~\bibnamefont {Campo}},\ }\bibfield  {title}
  {\bibinfo {title} {The new powder diffractometer {D1B} of the {I}nstitut
  {L}aue {L}angevin},\ }\href {https://doi.org/10.1088/1742-6596/549/1/012003}
  {\bibfield  {journal} {\bibinfo  {journal} {J.\ Phys.: Conf.\ Ser.}\ }\textbf
  {\bibinfo {volume} {549}},\ \bibinfo {pages} {012003} (\bibinfo {year}
  {2014})}\BibitemShut {NoStop}%
\bibitem [{\citenamefont {Hansen}\ \emph {et~al.}(2008)\citenamefont {Hansen},
  \citenamefont {Henry}, \citenamefont {Fischer}, \citenamefont {Torregrossa},\
  and\ \citenamefont {Convert}}]{D20}%
  \BibitemOpen
  \bibfield  {author} {\bibinfo {author} {\bibfnamefont {T.~C.}\ \bibnamefont
  {Hansen}}, \bibinfo {author} {\bibfnamefont {P.~F.}\ \bibnamefont {Henry}},
  \bibinfo {author} {\bibfnamefont {H.~E.}\ \bibnamefont {Fischer}}, \bibinfo
  {author} {\bibfnamefont {J.}~\bibnamefont {Torregrossa}},\ and\ \bibinfo
  {author} {\bibfnamefont {P.}~\bibnamefont {Convert}},\ }\bibfield  {title}
  {\bibinfo {title} {The {D20} instrument at the {ILL}: a versatile
  high-intensity two-axis neutron diffractometer},\ }\href
  {https://doi.org/10.1088/0957-0233/19/3/034001} {\bibfield  {journal}
  {\bibinfo  {journal} {Meas.\ Sci.\ Technol.}\ }\textbf {\bibinfo {volume}
  {19}},\ \bibinfo {pages} {034001} (\bibinfo {year} {2008})}\BibitemShut
  {NoStop}%
\bibitem [{\citenamefont {Stewart}\ \emph {et~al.}(2009)\citenamefont
  {Stewart}, \citenamefont {Deen}, \citenamefont {Andersen}, \citenamefont
  {Schober}, \citenamefont {Barth{\'e}l{\'e}my}, \citenamefont {Hillier},
  \citenamefont {Murani}, \citenamefont {Hayes},\ and\ \citenamefont
  {Lindenau}}]{D7_2}%
  \BibitemOpen
  \bibfield  {author} {\bibinfo {author} {\bibfnamefont {J.~R.}\ \bibnamefont
  {Stewart}}, \bibinfo {author} {\bibfnamefont {P.~P.}\ \bibnamefont {Deen}},
  \bibinfo {author} {\bibfnamefont {K.~H.}\ \bibnamefont {Andersen}}, \bibinfo
  {author} {\bibfnamefont {H.}~\bibnamefont {Schober}}, \bibinfo {author}
  {\bibfnamefont {J.-F.}\ \bibnamefont {Barth{\'e}l{\'e}my}}, \bibinfo {author}
  {\bibfnamefont {J.~M.}\ \bibnamefont {Hillier}}, \bibinfo {author}
  {\bibfnamefont {A.~P.}\ \bibnamefont {Murani}}, \bibinfo {author}
  {\bibfnamefont {T.}~\bibnamefont {Hayes}},\ and\ \bibinfo {author}
  {\bibfnamefont {B.}~\bibnamefont {Lindenau}},\ }\bibfield  {title} {\bibinfo
  {title} {Disordered materials studied using neutron polarization analysis on
  the multi-detector spectrometer, {D7}},\ }\href
  {https://doi.org/10.1107/S0021889808039162} {\bibfield  {journal} {\bibinfo
  {journal} {J. Appl. Crystallogr.}\ }\textbf {\bibinfo {volume} {42}},\
  \bibinfo {pages} {69} (\bibinfo {year} {2009})}\BibitemShut {NoStop}%
\bibitem [{\citenamefont {Rodr{\'i}guez-{C}arvajal}(1993)}]{FullProf}%
  \BibitemOpen
  \bibfield  {author} {\bibinfo {author} {\bibfnamefont {J.}~\bibnamefont
  {Rodr{\'i}guez-{C}arvajal}},\ }\bibfield  {title} {\bibinfo {title} {Recent
  advances in magnetic structure determination by neutron powder diffraction},\
  }\href {https://doi.org/10.1016/0921-4526(93)90108-I} {\bibfield  {journal}
  {\bibinfo  {journal} {Physica B: Condens.\ Matter}\ }\textbf {\bibinfo
  {volume} {192}},\ \bibinfo {pages} {55} (\bibinfo {year} {1993})}\BibitemShut
  {NoStop}%
\bibitem [{\citenamefont {Sears}(1992)}]{Sears1992}%
  \BibitemOpen
  \bibfield  {author} {\bibinfo {author} {\bibfnamefont {V.~F.}\ \bibnamefont
  {Sears}},\ }\bibfield  {title} {\bibinfo {title} {Neutron scattering lengths
  and cross sections},\ }\href {https://doi.org/10.1080/10448639208218770}
  {\bibfield  {journal} {\bibinfo  {journal} {Neutron News}\ }\textbf {\bibinfo
  {volume} {3}},\ \bibinfo {pages} {26} (\bibinfo {year} {1992})}\BibitemShut
  {NoStop}%
\bibitem [{\citenamefont {Weschke}\ and\ \citenamefont
  {Schierle}(2018)}]{UE46}%
  \BibitemOpen
  \bibfield  {author} {\bibinfo {author} {\bibfnamefont {E.}~\bibnamefont
  {Weschke}}\ and\ \bibinfo {author} {\bibfnamefont {E.}~\bibnamefont
  {Schierle}},\ }\bibfield  {title} {\bibinfo {title} {The {UE46} {PGM}-1
  beamline at {BESSY II}},\ }\href {https://doi.org/10.17815/jlsrf-4-77}
  {\bibfield  {journal} {\bibinfo  {journal} {Journal of Large-Scale Research
  Facilities}\ }\textbf {\bibinfo {volume} {4}},\ \bibinfo {pages} {A127}
  (\bibinfo {year} {2018})}\BibitemShut {NoStop}%
\bibitem [{\citenamefont {Amato}\ \emph {et~al.}(2017)\citenamefont {Amato},
  \citenamefont {Luetkens}, \citenamefont {Sedlak}, \citenamefont {Stoykov},
  \citenamefont {Scheuermann}, \citenamefont {Elender}, \citenamefont
  {Raselli},\ and\ \citenamefont {Graf}}]{GPS}%
  \BibitemOpen
  \bibfield  {author} {\bibinfo {author} {\bibfnamefont {A.}~\bibnamefont
  {Amato}}, \bibinfo {author} {\bibfnamefont {H.}~\bibnamefont {Luetkens}},
  \bibinfo {author} {\bibfnamefont {K.}~\bibnamefont {Sedlak}}, \bibinfo
  {author} {\bibfnamefont {A.}~\bibnamefont {Stoykov}}, \bibinfo {author}
  {\bibfnamefont {R.}~\bibnamefont {Scheuermann}}, \bibinfo {author}
  {\bibfnamefont {M.}~\bibnamefont {Elender}}, \bibinfo {author} {\bibfnamefont
  {A.}~\bibnamefont {Raselli}},\ and\ \bibinfo {author} {\bibfnamefont
  {D.}~\bibnamefont {Graf}},\ }\bibfield  {title} {\bibinfo {title} {The new
  versatile general purpose surface-muon instrument {(GPS)} based on silicon
  photomultipliers for {$\mu$SR} measurements on a continuous-wave beam},\
  }\href {https://doi.org/10.1063/1.4986045} {\bibfield  {journal} {\bibinfo
  {journal} {Rev.\ Sci.\ Instrum.}\ }\textbf {\bibinfo {volume} {88}},\
  \bibinfo {pages} {093301} (\bibinfo {year} {2017})}\BibitemShut {NoStop}%
\bibitem [{\citenamefont {Suter}\ and\ \citenamefont {Wojek}(2012)}]{musrfit}%
  \BibitemOpen
  \bibfield  {author} {\bibinfo {author} {\bibfnamefont {A.}~\bibnamefont
  {Suter}}\ and\ \bibinfo {author} {\bibfnamefont {B.~M.}\ \bibnamefont
  {Wojek}},\ }\bibfield  {title} {\bibinfo {title} {{\sc Musrfit}: A free
  platform-independent framework for $\mu${SR} data analysis},\ }\href
  {https://doi.org/10.1016/j.phpro.2012.04.042} {\bibfield  {journal} {\bibinfo
   {journal} {Phys.\ Procedia}\ }\textbf {\bibinfo {volume} {30}},\ \bibinfo
  {pages} {69} (\bibinfo {year} {2012})}\BibitemShut {NoStop}%
\bibitem [{\citenamefont {Grioni}\ \emph {et~al.}(1992)\citenamefont {Grioni},
  \citenamefont {van Acker}, \citenamefont {Czy{\v{z}}yk},\ and\ \citenamefont
  {Fuggle}}]{Grioni1992}%
  \BibitemOpen
  \bibfield  {author} {\bibinfo {author} {\bibfnamefont {M.}~\bibnamefont
  {Grioni}}, \bibinfo {author} {\bibfnamefont {J.~F.}\ \bibnamefont {van
  Acker}}, \bibinfo {author} {\bibfnamefont {M.~T.}\ \bibnamefont
  {Czy{\v{z}}yk}},\ and\ \bibinfo {author} {\bibfnamefont {J.~C.}\ \bibnamefont
  {Fuggle}},\ }\bibfield  {title} {\bibinfo {title} {Unoccupied electronic
  structure and core-hole effects in the x-ray-absorption spectra of
  {Cu$_2$O}},\ }\href {https://doi.org/10.1103/PhysRevB.45.3309} {\bibfield
  {journal} {\bibinfo  {journal} {Phys.\ Rev.\ B}\ }\textbf {\bibinfo {volume}
  {45}},\ \bibinfo {pages} {3309} (\bibinfo {year} {1992})}\BibitemShut
  {NoStop}%
\bibitem [{\citenamefont {Grioni}\ \emph {et~al.}(1989)\citenamefont {Grioni},
  \citenamefont {Goedkoop}, \citenamefont {Schoorl}, \citenamefont {de~Groot},
  \citenamefont {Fuggle}, \citenamefont {Sch{\"a}fers}, \citenamefont {Koch},
  \citenamefont {Rossi}, \citenamefont {Esteva},\ and\ \citenamefont
  {Karnatak}}]{Grioni1989}%
  \BibitemOpen
  \bibfield  {author} {\bibinfo {author} {\bibfnamefont {M.}~\bibnamefont
  {Grioni}}, \bibinfo {author} {\bibfnamefont {J.~B.}\ \bibnamefont
  {Goedkoop}}, \bibinfo {author} {\bibfnamefont {R.}~\bibnamefont {Schoorl}},
  \bibinfo {author} {\bibfnamefont {F.~M.~F.}\ \bibnamefont {de~Groot}},
  \bibinfo {author} {\bibfnamefont {J.~C.}\ \bibnamefont {Fuggle}}, \bibinfo
  {author} {\bibfnamefont {F.}~\bibnamefont {Sch{\"a}fers}}, \bibinfo {author}
  {\bibfnamefont {E.~E.}\ \bibnamefont {Koch}}, \bibinfo {author}
  {\bibfnamefont {G.}~\bibnamefont {Rossi}}, \bibinfo {author} {\bibfnamefont
  {J.-M.}\ \bibnamefont {Esteva}},\ and\ \bibinfo {author} {\bibfnamefont
  {R.~C.}\ \bibnamefont {Karnatak}},\ }\bibfield  {title} {\bibinfo {title}
  {Studies of copper valence states with {C}u {${L}_3$} x-ray-absorption
  spectroscopy},\ }\href {https://doi.org/10.1103/PhysRevB.39.1541} {\bibfield
  {journal} {\bibinfo  {journal} {Phys.\ Rev.\ B}\ }\textbf {\bibinfo {volume}
  {39}},\ \bibinfo {pages} {1541} (\bibinfo {year} {1989})}\BibitemShut
  {NoStop}%
\bibitem [{\citenamefont {Jiang}\ \emph {et~al.}(2013)\citenamefont {Jiang},
  \citenamefont {Prendergast}, \citenamefont {Borondics}, \citenamefont
  {Porsgaard}, \citenamefont {Giovanetti}, \citenamefont {Pach}, \citenamefont
  {Newberg}, \citenamefont {Bluhm}, \citenamefont {Besenbacher},\ and\
  \citenamefont {Salmeron}}]{Jiang2013}%
  \BibitemOpen
  \bibfield  {author} {\bibinfo {author} {\bibfnamefont {P.}~\bibnamefont
  {Jiang}}, \bibinfo {author} {\bibfnamefont {D.}~\bibnamefont {Prendergast}},
  \bibinfo {author} {\bibfnamefont {F.}~\bibnamefont {Borondics}}, \bibinfo
  {author} {\bibfnamefont {S.}~\bibnamefont {Porsgaard}}, \bibinfo {author}
  {\bibfnamefont {L.}~\bibnamefont {Giovanetti}}, \bibinfo {author}
  {\bibfnamefont {E.}~\bibnamefont {Pach}}, \bibinfo {author} {\bibfnamefont
  {J.}~\bibnamefont {Newberg}}, \bibinfo {author} {\bibfnamefont
  {H.}~\bibnamefont {Bluhm}}, \bibinfo {author} {\bibfnamefont
  {F.}~\bibnamefont {Besenbacher}},\ and\ \bibinfo {author} {\bibfnamefont
  {M.}~\bibnamefont {Salmeron}},\ }\bibfield  {title} {\bibinfo {title}
  {Experimental and theoretical investigation of the electronic structure of
  {Cu$_2$O} and {CuO} thin films on {Cu(110)} using x-ray photoelectron and
  absorption spectroscopy},\ }\href {https://doi.org/10.1063/1.4773583}
  {\bibfield  {journal} {\bibinfo  {journal} {J.\ Chem.\ Phys.}\ }\textbf
  {\bibinfo {volume} {138}},\ \bibinfo {pages} {024704} (\bibinfo {year}
  {2013})}\BibitemShut {NoStop}%
\bibitem [{\citenamefont {Chen}(1997)}]{Chen1997}%
  \BibitemOpen
  \bibfield  {author} {\bibinfo {author} {\bibfnamefont {J.~G.}\ \bibnamefont
  {Chen}},\ }\bibfield  {title} {\bibinfo {title} {{NEXAFS} investigations of
  transition metal oxides, nitrides, carbides, sulfides and other interstitial
  compounds},\ }\href {https://doi.org/10.1016/S0167-5729(97)00011-3}
  {\bibfield  {journal} {\bibinfo  {journal} {Surf.\ Sci.\ Rep.}\ }\textbf
  {\bibinfo {volume} {30}},\ \bibinfo {pages} {1} (\bibinfo {year}
  {1997})}\BibitemShut {NoStop}%
\bibitem [{\citenamefont {Okubo}\ \emph {et~al.}(2009)\citenamefont {Okubo},
  \citenamefont {Yamamoto}, \citenamefont {Fujisawa}, \citenamefont {Ohta},
  \citenamefont {Nakamura},\ and\ \citenamefont {Kikuchi}}]{Okubo2009}%
  \BibitemOpen
  \bibfield  {author} {\bibinfo {author} {\bibfnamefont {S.}~\bibnamefont
  {Okubo}}, \bibinfo {author} {\bibfnamefont {H.}~\bibnamefont {Yamamoto}},
  \bibinfo {author} {\bibfnamefont {M.}~\bibnamefont {Fujisawa}}, \bibinfo
  {author} {\bibfnamefont {H.}~\bibnamefont {Ohta}}, \bibinfo {author}
  {\bibfnamefont {T.}~\bibnamefont {Nakamura}},\ and\ \bibinfo {author}
  {\bibfnamefont {H.}~\bibnamefont {Kikuchi}},\ }\bibfield  {title} {\bibinfo
  {title} {High field {ESR} measurements of quantum triple chain system
  {Cu$_3$(OH)$_4$SO$_4$}},\ }\href
  {https://doi.org/10.1088/1742-6596/150/4/042156} {\bibfield  {journal}
  {\bibinfo  {journal} {J.\ Phys.: Conf.\ Ser.}\ }\textbf {\bibinfo {volume}
  {150}},\ \bibinfo {pages} {042156} (\bibinfo {year} {2009})}\BibitemShut
  {NoStop}%
\bibitem [{\citenamefont {Leininger}\ \emph {et~al.}(2011)\citenamefont
  {Leininger}, \citenamefont {Chernyshov}, \citenamefont {Bosak}, \citenamefont
  {Berger},\ and\ \citenamefont {Inosov}}]{Leininger2011}%
  \BibitemOpen
  \bibfield  {author} {\bibinfo {author} {\bibfnamefont {P.}~\bibnamefont
  {Leininger}}, \bibinfo {author} {\bibfnamefont {D.}~\bibnamefont
  {Chernyshov}}, \bibinfo {author} {\bibfnamefont {A.}~\bibnamefont {Bosak}},
  \bibinfo {author} {\bibfnamefont {H.}~\bibnamefont {Berger}},\ and\ \bibinfo
  {author} {\bibfnamefont {D.~S.}\ \bibnamefont {Inosov}},\ }\bibfield  {title}
  {\bibinfo {title} {Competing charge density waves and temperature-dependent
  nesting in {$2H$-TaSe$_2$}},\ }\href
  {https://doi.org/10.1103/PhysRevB.83.233101} {\bibfield  {journal} {\bibinfo
  {journal} {Phys. Rev. B}\ }\textbf {\bibinfo {volume} {83}},\ \bibinfo
  {pages} {233101} (\bibinfo {year} {2011})}\BibitemShut {NoStop}%
\bibitem [{\citenamefont {Capogna}\ \emph {et~al.}(2010)\citenamefont
  {Capogna}, \citenamefont {Reehuis}, \citenamefont {Maljuk}, \citenamefont
  {Kremer}, \citenamefont {Ouladdiaf}, \citenamefont {Jansen},\ and\
  \citenamefont {Keimer}}]{Capogna2010}%
  \BibitemOpen
  \bibfield  {author} {\bibinfo {author} {\bibfnamefont {L.}~\bibnamefont
  {Capogna}}, \bibinfo {author} {\bibfnamefont {M.}~\bibnamefont {Reehuis}},
  \bibinfo {author} {\bibfnamefont {A.}~\bibnamefont {Maljuk}}, \bibinfo
  {author} {\bibfnamefont {R.~K.}\ \bibnamefont {Kremer}}, \bibinfo {author}
  {\bibfnamefont {B.}~\bibnamefont {Ouladdiaf}}, \bibinfo {author}
  {\bibfnamefont {M.}~\bibnamefont {Jansen}},\ and\ \bibinfo {author}
  {\bibfnamefont {B.}~\bibnamefont {Keimer}},\ }\bibfield  {title} {\bibinfo
  {title} {Magnetic structure of the edge-sharing copper oxide chain compound
  {NaCu}$_2${O}$_2$},\ }\href {https://doi.org/10.1103/PhysRevB.82.014407}
  {\bibfield  {journal} {\bibinfo  {journal} {Phys.\ Rev.\ B}\ }\textbf
  {\bibinfo {volume} {82}},\ \bibinfo {pages} {014407} (\bibinfo {year}
  {2010})}\BibitemShut {NoStop}%
\bibitem [{\citenamefont {Kobayashi}\ \emph {et~al.}(2009)\citenamefont
  {Kobayashi}, \citenamefont {Sato}, \citenamefont {Yasui}, \citenamefont
  {Moyoshi}, \citenamefont {Sato},\ and\ \citenamefont
  {Kakurai}}]{Kobayashi2009}%
  \BibitemOpen
  \bibfield  {author} {\bibinfo {author} {\bibfnamefont {Y.}~\bibnamefont
  {Kobayashi}}, \bibinfo {author} {\bibfnamefont {K.}~\bibnamefont {Sato}},
  \bibinfo {author} {\bibfnamefont {Y.}~\bibnamefont {Yasui}}, \bibinfo
  {author} {\bibfnamefont {T.}~\bibnamefont {Moyoshi}}, \bibinfo {author}
  {\bibfnamefont {M.}~\bibnamefont {Sato}},\ and\ \bibinfo {author}
  {\bibfnamefont {K.}~\bibnamefont {Kakurai}},\ }\bibfield  {title} {\bibinfo
  {title} {Studies of multiferroic system of {LiCu$_2$O$_2$}: {II}.\ magnetic
  structures of two ordered phases with incommensurate modulations},\ }\href
  {https://doi.org/10.1143/JPSJ.78.084721} {\bibfield  {journal} {\bibinfo
  {journal} {J.\ Phys.\ Soc.\ Jpn.}\ }\textbf {\bibinfo {volume} {78}},\
  \bibinfo {pages} {084721} (\bibinfo {year} {2009})}\BibitemShut {NoStop}%
\bibitem [{\citenamefont {Willenberg}\ \emph {et~al.}(2012)\citenamefont
  {Willenberg}, \citenamefont {Sch{\"a}pers}, \citenamefont {Rule},
  \citenamefont {S{\"u}llow}, \citenamefont {Reehuis}, \citenamefont {Ryll},
  \citenamefont {Klemke}, \citenamefont {Kiefer}, \citenamefont
  {Schottenhamel}, \citenamefont {B{\"u}chner}, \citenamefont {Ouladdiaf},
  \citenamefont {Uhlarz}, \citenamefont {Beyer}, \citenamefont {Wosnitza},\
  and\ \citenamefont {Wolter}}]{Willenberg2012}%
  \BibitemOpen
  \bibfield  {author} {\bibinfo {author} {\bibfnamefont {B.}~\bibnamefont
  {Willenberg}}, \bibinfo {author} {\bibfnamefont {M.}~\bibnamefont
  {Sch{\"a}pers}}, \bibinfo {author} {\bibfnamefont {K.~C.}\ \bibnamefont
  {Rule}}, \bibinfo {author} {\bibfnamefont {S.}~\bibnamefont {S{\"u}llow}},
  \bibinfo {author} {\bibfnamefont {M.}~\bibnamefont {Reehuis}}, \bibinfo
  {author} {\bibfnamefont {H.}~\bibnamefont {Ryll}}, \bibinfo {author}
  {\bibfnamefont {B.}~\bibnamefont {Klemke}}, \bibinfo {author} {\bibfnamefont
  {K.}~\bibnamefont {Kiefer}}, \bibinfo {author} {\bibfnamefont
  {W.}~\bibnamefont {Schottenhamel}}, \bibinfo {author} {\bibfnamefont
  {B.}~\bibnamefont {B{\"u}chner}}, \bibinfo {author} {\bibfnamefont
  {B.}~\bibnamefont {Ouladdiaf}}, \bibinfo {author} {\bibfnamefont
  {M.}~\bibnamefont {Uhlarz}}, \bibinfo {author} {\bibfnamefont
  {R.}~\bibnamefont {Beyer}}, \bibinfo {author} {\bibfnamefont
  {J.}~\bibnamefont {Wosnitza}},\ and\ \bibinfo {author} {\bibfnamefont
  {A.~U.~B.}\ \bibnamefont {Wolter}},\ }\bibfield  {title} {\bibinfo {title}
  {Magnetic frustration in a quantum spin chain: The case of linarite
  {PbCuSO}$_4${(OH)}$_2$},\ }\href
  {https://doi.org/10.1103/PhysRevLett.108.117202} {\bibfield  {journal}
  {\bibinfo  {journal} {Phys.\ Rev.\ Lett.}\ }\textbf {\bibinfo {volume}
  {108}},\ \bibinfo {pages} {117202} (\bibinfo {year} {2012})}\BibitemShut
  {NoStop}%
\bibitem [{\citenamefont {Ehlers}\ \emph {et~al.}(2013)\citenamefont {Ehlers},
  \citenamefont {Stewart}, \citenamefont {Wildes}, \citenamefont {Deen},\ and\
  \citenamefont {Andersen}}]{Ehlers2013}%
  \BibitemOpen
  \bibfield  {author} {\bibinfo {author} {\bibfnamefont {G.}~\bibnamefont
  {Ehlers}}, \bibinfo {author} {\bibfnamefont {J.~R.}\ \bibnamefont {Stewart}},
  \bibinfo {author} {\bibfnamefont {A.~R.}\ \bibnamefont {Wildes}}, \bibinfo
  {author} {\bibfnamefont {P.~P.}\ \bibnamefont {Deen}},\ and\ \bibinfo
  {author} {\bibfnamefont {K.~H.}\ \bibnamefont {Andersen}},\ }\bibfield
  {title} {\bibinfo {title} {Generalization of the classical {\itshape
  xyz}-polarization analysis technique to out-of-plane and inelastic
  scattering},\ }\href {https://doi.org/10.1063/1.4819739} {\bibfield
  {journal} {\bibinfo  {journal} {Rev.\ Sci.\ Instrum.}\ }\textbf {\bibinfo
  {volume} {84}},\ \bibinfo {pages} {093901} (\bibinfo {year}
  {2013})}\BibitemShut {NoStop}%
\bibitem [{\citenamefont {Vilminot}\ \emph {et~al.}(2002)\citenamefont
  {Vilminot}, \citenamefont {Richard-Plouet}, \citenamefont {Andr{\'e}},
  \citenamefont {Swierczynski}, \citenamefont {Bour{\'e}e-Vigneron},
  \citenamefont {Marino},\ and\ \citenamefont {Guillot}}]{Vilminot2002}%
  \BibitemOpen
  \bibfield  {author} {\bibinfo {author} {\bibfnamefont {S.}~\bibnamefont
  {Vilminot}}, \bibinfo {author} {\bibfnamefont {M.}~\bibnamefont
  {Richard-Plouet}}, \bibinfo {author} {\bibfnamefont {G.}~\bibnamefont
  {Andr{\'e}}}, \bibinfo {author} {\bibfnamefont {D.}~\bibnamefont
  {Swierczynski}}, \bibinfo {author} {\bibfnamefont {F.}~\bibnamefont
  {Bour{\'e}e-Vigneron}}, \bibinfo {author} {\bibfnamefont {E.}~\bibnamefont
  {Marino}},\ and\ \bibinfo {author} {\bibfnamefont {M.}~\bibnamefont
  {Guillot}},\ }\bibfield  {title} {\bibinfo {title} {Synthesis, structure and
  magnetic properties of copper hydroxysulfates},\ }\href
  {https://doi.org/10.1016/S1463-0184(02)00027-8} {\bibfield  {journal}
  {\bibinfo  {journal} {Cryst. Eng.}\ }\textbf {\bibinfo {volume} {5}},\
  \bibinfo {pages} {177} (\bibinfo {year} {2002})}\BibitemShut {NoStop}%
\bibitem [{\citenamefont {Vilminot}\ \emph {et~al.}(2007)\citenamefont
  {Vilminot}, \citenamefont {Andr{\'e}}, \citenamefont {Bour{\'e}e-Vigneron},
  \citenamefont {Richard-Plouet},\ and\ \citenamefont {Kurmoo}}]{Vilminot2007}%
  \BibitemOpen
  \bibfield  {author} {\bibinfo {author} {\bibfnamefont {S.}~\bibnamefont
  {Vilminot}}, \bibinfo {author} {\bibfnamefont {G.}~\bibnamefont {Andr{\'e}}},
  \bibinfo {author} {\bibfnamefont {F.}~\bibnamefont {Bour{\'e}e-Vigneron}},
  \bibinfo {author} {\bibfnamefont {M.}~\bibnamefont {Richard-Plouet}},\ and\
  \bibinfo {author} {\bibfnamefont {M.}~\bibnamefont {Kurmoo}},\ }\bibfield
  {title} {\bibinfo {title} {Magnetic properties and magnetic structures of
  {Cu$_3$(OD)$_4X$O$_4$}, {$X$} = {Se} or {S}: Cycloidal versus collinear
  antiferromagnetic structure},\ }\href {https://doi.org/10.1021/ic700900b}
  {\bibfield  {journal} {\bibinfo  {journal} {Inorg.\ Chem.}\ }\textbf
  {\bibinfo {volume} {46}},\ \bibinfo {pages} {10079} (\bibinfo {year}
  {2007})}\BibitemShut {NoStop}%
\bibitem [{\citenamefont {H{\"a}lg}\ and\ \citenamefont
  {Furrer}(1986)}]{Halg1986}%
  \BibitemOpen
  \bibfield  {author} {\bibinfo {author} {\bibfnamefont {B.}~\bibnamefont
  {H{\"a}lg}}\ and\ \bibinfo {author} {\bibfnamefont {A.}~\bibnamefont
  {Furrer}},\ }\bibfield  {title} {\bibinfo {title} {Anisotropic exchange and
  spin dynamics in the type-{I} (-{IA}) antiferromagnets {CeAs}, {CeSb}, and
  {USb}: {A} neutron study},\ }\href {https://doi.org/10.1103/PhysRevB.34.6258}
  {\bibfield  {journal} {\bibinfo  {journal} {Phys.\ Rev.\ B}\ }\textbf
  {\bibinfo {volume} {34}},\ \bibinfo {pages} {6258} (\bibinfo {year}
  {1986})}\BibitemShut {NoStop}%
\bibitem [{\citenamefont {Burlet}\ \emph {et~al.}(1982)\citenamefont {Burlet},
  \citenamefont {Rossat-Mignod}, \citenamefont {Effantin}, \citenamefont
  {Kasuya}, \citenamefont {Kunii},\ and\ \citenamefont
  {Komatsubara}}]{Burlet1982}%
  \BibitemOpen
  \bibfield  {author} {\bibinfo {author} {\bibfnamefont {P.}~\bibnamefont
  {Burlet}}, \bibinfo {author} {\bibfnamefont {J.}~\bibnamefont
  {Rossat-Mignod}}, \bibinfo {author} {\bibfnamefont {J.~M.}\ \bibnamefont
  {Effantin}}, \bibinfo {author} {\bibfnamefont {T.}~\bibnamefont {Kasuya}},
  \bibinfo {author} {\bibfnamefont {S.}~\bibnamefont {Kunii}},\ and\ \bibinfo
  {author} {\bibfnamefont {T.}~\bibnamefont {Komatsubara}},\ }\bibfield
  {title} {\bibinfo {title} {Magnetic ordering in cerium hexaboride
  {CeB$_6$}},\ }\href {https://doi.org/10.1063/1.330762} {\bibfield  {journal}
  {\bibinfo  {journal} {J.\ Appl.\ Phys.}\ }\textbf {\bibinfo {volume} {53}},\
  \bibinfo {pages} {2149} (\bibinfo {year} {1982})}\BibitemShut {NoStop}%
\bibitem [{\citenamefont {Effantin}\ \emph {et~al.}(1985)\citenamefont
  {Effantin}, \citenamefont {{Rossat-Mignod}}, \citenamefont {Burlet},
  \citenamefont {Bartholin}, \citenamefont {Kunii},\ and\ \citenamefont
  {Kasuya}}]{Effantin1985}%
  \BibitemOpen
  \bibfield  {author} {\bibinfo {author} {\bibfnamefont {J.~M.}\ \bibnamefont
  {Effantin}}, \bibinfo {author} {\bibfnamefont {J.}~\bibnamefont
  {{Rossat-Mignod}}}, \bibinfo {author} {\bibfnamefont {P.}~\bibnamefont
  {Burlet}}, \bibinfo {author} {\bibfnamefont {H.}~\bibnamefont {Bartholin}},
  \bibinfo {author} {\bibfnamefont {S.}~\bibnamefont {Kunii}},\ and\ \bibinfo
  {author} {\bibfnamefont {T.}~\bibnamefont {Kasuya}},\ }\bibfield  {title}
  {\bibinfo {title} {Magnetic phase diagram of {CeB$_6$}},\ }\href
  {https://doi.org/10.1016/0304-8853(85)90382-8} {\bibfield  {journal}
  {\bibinfo  {journal} {J.\ Magn.\ Magn.\ Mater.}\ }\textbf {\bibinfo {volume}
  {47--48}},\ \bibinfo {pages} {145} (\bibinfo {year} {1985})}\BibitemShut
  {NoStop}%
\bibitem [{\citenamefont {Long}(1993)}]{Long1993}%
  \BibitemOpen
  \bibfield  {author} {\bibinfo {author} {\bibfnamefont {M.~W.}\ \bibnamefont
  {Long}},\ }\bibfield  {title} {\bibinfo {title} {Multiple-{\itshape {q}}
  structures in frustrated antiferromagnets},\ }\href
  {https://doi.org/10.1142/S0217979293003127} {\bibfield  {journal} {\bibinfo
  {journal} {Int.\ J.\ Mod.\ Phys.\ B}\ }\textbf {\bibinfo {volume} {07}},\
  \bibinfo {pages} {2981} (\bibinfo {year} {1993})}\BibitemShut {NoStop}%
\bibitem [{\citenamefont {Ishiwata}\ \emph {et~al.}(2020)\citenamefont
  {Ishiwata}, \citenamefont {Nakajima}, \citenamefont {Kim}, \citenamefont
  {Inosov}, \citenamefont {Kanazawa}, \citenamefont {White}, \citenamefont
  {Gavilano}, \citenamefont {Georgii}, \citenamefont {Seemann}, \citenamefont
  {Brandl}, \citenamefont {Manuel}, \citenamefont {Khalyavin}, \citenamefont
  {Seki}, \citenamefont {Tokunaga}, \citenamefont {Kinoshita}, \citenamefont
  {Long}, \citenamefont {Kaneko}, \citenamefont {Taguchi}, \citenamefont
  {Arima}, \citenamefont {Keimer},\ and\ \citenamefont
  {Tokura}}]{Ishiwata2020}%
  \BibitemOpen
  \bibfield  {author} {\bibinfo {author} {\bibfnamefont {S.}~\bibnamefont
  {Ishiwata}}, \bibinfo {author} {\bibfnamefont {T.}~\bibnamefont {Nakajima}},
  \bibinfo {author} {\bibfnamefont {J.-H.}\ \bibnamefont {Kim}}, \bibinfo
  {author} {\bibfnamefont {D.~S.}\ \bibnamefont {Inosov}}, \bibinfo {author}
  {\bibfnamefont {N.}~\bibnamefont {Kanazawa}}, \bibinfo {author}
  {\bibfnamefont {J.~S.}\ \bibnamefont {White}}, \bibinfo {author}
  {\bibfnamefont {J.~L.}\ \bibnamefont {Gavilano}}, \bibinfo {author}
  {\bibfnamefont {R.}~\bibnamefont {Georgii}}, \bibinfo {author} {\bibfnamefont
  {K.~M.}\ \bibnamefont {Seemann}}, \bibinfo {author} {\bibfnamefont
  {G.}~\bibnamefont {Brandl}}, \bibinfo {author} {\bibfnamefont
  {P.}~\bibnamefont {Manuel}}, \bibinfo {author} {\bibfnamefont {D.~D.}\
  \bibnamefont {Khalyavin}}, \bibinfo {author} {\bibfnamefont {S.}~\bibnamefont
  {Seki}}, \bibinfo {author} {\bibfnamefont {Y.}~\bibnamefont {Tokunaga}},
  \bibinfo {author} {\bibfnamefont {M.}~\bibnamefont {Kinoshita}}, \bibinfo
  {author} {\bibfnamefont {Y.~W.}\ \bibnamefont {Long}}, \bibinfo {author}
  {\bibfnamefont {Y.}~\bibnamefont {Kaneko}}, \bibinfo {author} {\bibfnamefont
  {Y.}~\bibnamefont {Taguchi}}, \bibinfo {author} {\bibfnamefont
  {T.}~\bibnamefont {Arima}}, \bibinfo {author} {\bibfnamefont
  {B.}~\bibnamefont {Keimer}},\ and\ \bibinfo {author} {\bibfnamefont
  {Y.}~\bibnamefont {Tokura}},\ }\bibfield  {title} {\bibinfo {title} {Emergent
  topological spin structures in the centrosymmetric cubic perovskite
  {SrFeO}$_3$},\ }\href {https://doi.org/10.1103/PhysRevB.101.134406}
  {\bibfield  {journal} {\bibinfo  {journal} {Phys. Rev. B}\ }\textbf {\bibinfo
  {volume} {101}},\ \bibinfo {pages} {134406} (\bibinfo {year}
  {2020})}\BibitemShut {NoStop}%
\bibitem [{\citenamefont {Chen}\ \emph {et~al.}(2021)\citenamefont {Chen},
  \citenamefont {Li}, \citenamefont {Hu}, \citenamefont {Hu}, \citenamefont
  {Yue}, \citenamefont {Sutarto}, \citenamefont {He}, \citenamefont {Iida},
  \citenamefont {Kamazawa}, \citenamefont {Yu}, \citenamefont {Lin},\ and\
  \citenamefont {Li}}]{Chen2021}%
  \BibitemOpen
  \bibfield  {author} {\bibinfo {author} {\bibfnamefont {W.}~\bibnamefont
  {Chen}}, \bibinfo {author} {\bibfnamefont {X.}~\bibnamefont {Li}}, \bibinfo
  {author} {\bibfnamefont {Z.}~\bibnamefont {Hu}}, \bibinfo {author}
  {\bibfnamefont {Z.}~\bibnamefont {Hu}}, \bibinfo {author} {\bibfnamefont
  {L.}~\bibnamefont {Yue}}, \bibinfo {author} {\bibfnamefont {R.}~\bibnamefont
  {Sutarto}}, \bibinfo {author} {\bibfnamefont {F.}~\bibnamefont {He}},
  \bibinfo {author} {\bibfnamefont {K.}~\bibnamefont {Iida}}, \bibinfo {author}
  {\bibfnamefont {K.}~\bibnamefont {Kamazawa}}, \bibinfo {author}
  {\bibfnamefont {W.}~\bibnamefont {Yu}}, \bibinfo {author} {\bibfnamefont
  {X.}~\bibnamefont {Lin}},\ and\ \bibinfo {author} {\bibfnamefont
  {Y.}~\bibnamefont {Li}},\ }\bibfield  {title} {\bibinfo {title} {Spin-orbit
  phase behavior of {Na}$_2${Co}$_2${TeO}$_6$ at low temperatures},\ }\href
  {https://doi.org/10.1103/PhysRevB.103.L180404} {\bibfield  {journal}
  {\bibinfo  {journal} {Phys.\ Rev.\ B}\ }\textbf {\bibinfo {volume} {103}},\
  \bibinfo {pages} {L180404} (\bibinfo {year} {2021})}\BibitemShut {NoStop}%
\bibitem [{\citenamefont {Ivanov}\ \emph {et~al.}(2012)\citenamefont {Ivanov},
  \citenamefont {Tellgren}, \citenamefont {Ritter}, \citenamefont {Nordblad},
  \citenamefont {Mathieu}, \citenamefont {Andr{\'e}}, \citenamefont {Golubko},
  \citenamefont {Politova},\ and\ \citenamefont {Weil}}]{Ivanov2012}%
  \BibitemOpen
  \bibfield  {author} {\bibinfo {author} {\bibfnamefont {S.~A.}\ \bibnamefont
  {Ivanov}}, \bibinfo {author} {\bibfnamefont {R.}~\bibnamefont {Tellgren}},
  \bibinfo {author} {\bibfnamefont {C.}~\bibnamefont {Ritter}}, \bibinfo
  {author} {\bibfnamefont {P.}~\bibnamefont {Nordblad}}, \bibinfo {author}
  {\bibfnamefont {R.}~\bibnamefont {Mathieu}}, \bibinfo {author} {\bibfnamefont
  {G.}~\bibnamefont {Andr{\'e}}}, \bibinfo {author} {\bibfnamefont {N.~V.}\
  \bibnamefont {Golubko}}, \bibinfo {author} {\bibfnamefont {E.~D.}\
  \bibnamefont {Politova}},\ and\ \bibinfo {author} {\bibfnamefont
  {M.}~\bibnamefont {Weil}},\ }\bibfield  {title} {\bibinfo {title}
  {Temperature-dependent multi-k magnetic structure in multiferroic
  {Co$_3$TeO$_6$}},\ }\href
  {https://doi.org/10.1016/j.materresbull.2011.10.003} {\bibfield  {journal}
  {\bibinfo  {journal} {Mater.\ Res.\ Bull.}\ }\textbf {\bibinfo {volume}
  {47}},\ \bibinfo {pages} {63} (\bibinfo {year} {2012})}\BibitemShut {NoStop}%
\bibitem [{\citenamefont {Wang}\ \emph {et~al.}(2013)\citenamefont {Wang},
  \citenamefont {Lee}, \citenamefont {Li}, \citenamefont {Wu}, \citenamefont
  {Li}, \citenamefont {Chou}, \citenamefont {Yang}, \citenamefont {Lynn},
  \citenamefont {Huang}, \citenamefont {Harris},\ and\ \citenamefont
  {Berger}}]{Wang2013}%
  \BibitemOpen
  \bibfield  {author} {\bibinfo {author} {\bibfnamefont {C.-W.}\ \bibnamefont
  {Wang}}, \bibinfo {author} {\bibfnamefont {C.-H.}\ \bibnamefont {Lee}},
  \bibinfo {author} {\bibfnamefont {C.-Y.}\ \bibnamefont {Li}}, \bibinfo
  {author} {\bibfnamefont {C.-M.}\ \bibnamefont {Wu}}, \bibinfo {author}
  {\bibfnamefont {W.-H.}\ \bibnamefont {Li}}, \bibinfo {author} {\bibfnamefont
  {C.-C.}\ \bibnamefont {Chou}}, \bibinfo {author} {\bibfnamefont {H.-D.}\
  \bibnamefont {Yang}}, \bibinfo {author} {\bibfnamefont {J.~W.}\ \bibnamefont
  {Lynn}}, \bibinfo {author} {\bibfnamefont {Q.}~\bibnamefont {Huang}},
  \bibinfo {author} {\bibfnamefont {A.~B.}\ \bibnamefont {Harris}},\ and\
  \bibinfo {author} {\bibfnamefont {H.}~\bibnamefont {Berger}},\ }\bibfield
  {title} {\bibinfo {title} {Complex magnetic couplings in {Co$_3$TeO$_6$}},\
  }\href {https://doi.org/10.1103/PhysRevB.88.184427} {\bibfield  {journal}
  {\bibinfo  {journal} {Phys.\ Rev.\ B}\ }\textbf {\bibinfo {volume} {88}},\
  \bibinfo {pages} {184427} (\bibinfo {year} {2013})}\BibitemShut {NoStop}%
\bibitem [{\citenamefont {Lee}\ \emph {et~al.}(2017)\citenamefont {Lee},
  \citenamefont {Wang}, \citenamefont {Zhao}, \citenamefont {Li}, \citenamefont
  {Lynn}, \citenamefont {Harris}, \citenamefont {Rule}, \citenamefont {Yang},\
  and\ \citenamefont {Berger}}]{Lee2017}%
  \BibitemOpen
  \bibfield  {author} {\bibinfo {author} {\bibfnamefont {C.-H.}\ \bibnamefont
  {Lee}}, \bibinfo {author} {\bibfnamefont {C.-W.}\ \bibnamefont {Wang}},
  \bibinfo {author} {\bibfnamefont {Y.}~\bibnamefont {Zhao}}, \bibinfo {author}
  {\bibfnamefont {W.-H.}\ \bibnamefont {Li}}, \bibinfo {author} {\bibfnamefont
  {J.~W.}\ \bibnamefont {Lynn}}, \bibinfo {author} {\bibfnamefont {A.~B.}\
  \bibnamefont {Harris}}, \bibinfo {author} {\bibfnamefont {K.}~\bibnamefont
  {Rule}}, \bibinfo {author} {\bibfnamefont {H.-D.}\ \bibnamefont {Yang}},\
  and\ \bibinfo {author} {\bibfnamefont {H.}~\bibnamefont {Berger}},\
  }\bibfield  {title} {\bibinfo {title} {Complex magnetic incommensurability
  and electronic charge transfer through the ferroelectric transition in
  multiferroic {Co$_3$TeO$_6$}},\ }\href
  {https://doi.org/10.1038/s41598-017-06651-9} {\bibfield  {journal} {\bibinfo
  {journal} {Sci.\ Rep.}\ }\textbf {\bibinfo {volume} {7}},\ \bibinfo {pages}
  {6437} (\bibinfo {year} {2017})}\BibitemShut {NoStop}%
\bibitem [{\citenamefont {Paddison}\ \emph {et~al.}(2013)\citenamefont
  {Paddison}, \citenamefont {Stewart},\ and\ \citenamefont
  {Goodwin}}]{Spinvert}%
  \BibitemOpen
  \bibfield  {author} {\bibinfo {author} {\bibfnamefont {J.~A.~M.}\
  \bibnamefont {Paddison}}, \bibinfo {author} {\bibfnamefont {J.~R.}\
  \bibnamefont {Stewart}},\ and\ \bibinfo {author} {\bibfnamefont {A.~L.}\
  \bibnamefont {Goodwin}},\ }\bibfield  {title} {\bibinfo {title} {{\sc
  Spinvert}: a program for refinement of paramagnetic diffuse scattering
  data},\ }\href {https://doi.org/10.1088/0953-8984/25/45/454220} {\bibfield
  {journal} {\bibinfo  {journal} {J.\ Phys.: Condens.\ Matter}\ }\textbf
  {\bibinfo {volume} {25}},\ \bibinfo {pages} {454220} (\bibinfo {year}
  {2013})}\BibitemShut {NoStop}%
\bibitem [{\citenamefont {Cox}(1987)}]{Cox1987}%
  \BibitemOpen
  \bibfield  {author} {\bibinfo {author} {\bibfnamefont {S.~F.~J.}\
  \bibnamefont {Cox}},\ }\bibfield  {title} {\bibinfo {title} {Implanted muon
  studies in condensed matter science},\ }\href
  {https://doi.org/10.1088/0022-3719/20/22/005} {\bibfield  {journal} {\bibinfo
   {journal} {J.\ Phys.\ C: Solid State Phys.}\ }\textbf {\bibinfo {volume}
  {20}},\ \bibinfo {pages} {3187} (\bibinfo {year} {1987})}\BibitemShut
  {NoStop}%
\bibitem [{\citenamefont {Uemura}\ \emph {et~al.}(1985)\citenamefont {Uemura},
  \citenamefont {Yamazaki}, \citenamefont {Harshman}, \citenamefont {Senba},\
  and\ \citenamefont {Ansaldo}}]{Uemura1985}%
  \BibitemOpen
  \bibfield  {author} {\bibinfo {author} {\bibfnamefont {Y.~J.}\ \bibnamefont
  {Uemura}}, \bibinfo {author} {\bibfnamefont {T.}~\bibnamefont {Yamazaki}},
  \bibinfo {author} {\bibfnamefont {D.~R.}\ \bibnamefont {Harshman}}, \bibinfo
  {author} {\bibfnamefont {M.}~\bibnamefont {Senba}},\ and\ \bibinfo {author}
  {\bibfnamefont {E.~J.}\ \bibnamefont {Ansaldo}},\ }\bibfield  {title}
  {\bibinfo {title} {Muon-spin relaxation in {AuFe} and {CuMn} spin glasses},\
  }\href {https://doi.org/10.1103/PhysRevB.31.546} {\bibfield  {journal}
  {\bibinfo  {journal} {Phys.\ Rev.\ B}\ }\textbf {\bibinfo {volume} {31}},\
  \bibinfo {pages} {546} (\bibinfo {year} {1985})}\BibitemShut {NoStop}%
\bibitem [{\citenamefont {{Kubo}}(1981)}]{Kubo1981}%
  \BibitemOpen
  \bibfield  {author} {\bibinfo {author} {\bibfnamefont {R.}~\bibnamefont
  {{Kubo}}},\ }\bibfield  {title} {\bibinfo {title} {{A stochastic theory of
  spin relaxation}},\ }\href {https://doi.org/10.1007/BF01037553} {\bibfield
  {journal} {\bibinfo  {journal} {Hyperfine Interact.}\ }\textbf {\bibinfo
  {volume} {8}},\ \bibinfo {pages} {731} (\bibinfo {year} {1981})}\BibitemShut
  {NoStop}%
\bibitem [{\citenamefont {Yaouanc}\ and\ \citenamefont {{Dalmas de
  R{\'e}otier}}(2011)}]{Yaouanc2011}%
  \BibitemOpen
  \bibfield  {author} {\bibinfo {author} {\bibfnamefont {A.}~\bibnamefont
  {Yaouanc}}\ and\ \bibinfo {author} {\bibfnamefont {P.}~\bibnamefont {{Dalmas
  de R{\'e}otier}}},\ }\href@noop {} {\emph {\bibinfo {title} {Muon Spin
  Rotation, Relaxation, and Resonance: Applications to Condensed Matter}}}\
  (\bibinfo  {publisher} {Oxford University Press},\ \bibinfo {address}
  {Oxford, UK},\ \bibinfo {year} {2011})\BibitemShut {NoStop}%
\bibitem [{\citenamefont {Franke}\ \emph {et~al.}(2018)\citenamefont {Franke},
  \citenamefont {Huddart}, \citenamefont {Hicken}, \citenamefont {Xiao},
  \citenamefont {Blundell}, \citenamefont {Pratt}, \citenamefont {Crisanti},
  \citenamefont {Barker}, \citenamefont {Clark}, \citenamefont {\ifmmode
  \check{S}\else \v{S}\fi{}tefan\ifmmode \check{c}\else
  \v{c}\fi{}i\ifmmode~\check{c}\else \v{c}\fi{}}, \citenamefont {Hatnean},
  \citenamefont {Balakrishnan},\ and\ \citenamefont {Lancaster}}]{Franke2018}%
  \BibitemOpen
  \bibfield  {author} {\bibinfo {author} {\bibfnamefont {K.~J.~A.}\
  \bibnamefont {Franke}}, \bibinfo {author} {\bibfnamefont {B.~M.}\
  \bibnamefont {Huddart}}, \bibinfo {author} {\bibfnamefont {T.~J.}\
  \bibnamefont {Hicken}}, \bibinfo {author} {\bibfnamefont {F.}~\bibnamefont
  {Xiao}}, \bibinfo {author} {\bibfnamefont {S.~J.}\ \bibnamefont {Blundell}},
  \bibinfo {author} {\bibfnamefont {F.~L.}\ \bibnamefont {Pratt}}, \bibinfo
  {author} {\bibfnamefont {M.}~\bibnamefont {Crisanti}}, \bibinfo {author}
  {\bibfnamefont {J.~A.~T.}\ \bibnamefont {Barker}}, \bibinfo {author}
  {\bibfnamefont {S.~J.}\ \bibnamefont {Clark}}, \bibinfo {author}
  {\bibfnamefont {A.}~\bibnamefont {\ifmmode \check{S}\else
  \v{S}\fi{}tefan\ifmmode \check{c}\else \v{c}\fi{}i\ifmmode~\check{c}\else
  \v{c}\fi{}}}, \bibinfo {author} {\bibfnamefont {M.~C.}\ \bibnamefont
  {Hatnean}}, \bibinfo {author} {\bibfnamefont {G.}~\bibnamefont
  {Balakrishnan}},\ and\ \bibinfo {author} {\bibfnamefont {T.}~\bibnamefont
  {Lancaster}},\ }\bibfield  {title} {\bibinfo {title} {Magnetic phases of
  skyrmion-hosting {GaV}$_4${S}$_{8-y}${Se}$_y$ ($y$=0,2,4,8) probed with muon
  spectroscopy},\ }\href {https://doi.org/10.1103/PhysRevB.98.054428}
  {\bibfield  {journal} {\bibinfo  {journal} {Phys.\ Rev.\ B}\ }\textbf
  {\bibinfo {volume} {98}},\ \bibinfo {pages} {054428} (\bibinfo {year}
  {2018})}\BibitemShut {NoStop}%
\bibitem [{\citenamefont {Hayano}\ \emph {et~al.}(1979)\citenamefont {Hayano},
  \citenamefont {Uemura}, \citenamefont {Imazato}, \citenamefont {Nishida},
  \citenamefont {Yamazaki},\ and\ \citenamefont {Kubo}}]{Hayano1979}%
  \BibitemOpen
  \bibfield  {author} {\bibinfo {author} {\bibfnamefont {R.~S.}\ \bibnamefont
  {Hayano}}, \bibinfo {author} {\bibfnamefont {Y.~J.}\ \bibnamefont {Uemura}},
  \bibinfo {author} {\bibfnamefont {J.}~\bibnamefont {Imazato}}, \bibinfo
  {author} {\bibfnamefont {N.}~\bibnamefont {Nishida}}, \bibinfo {author}
  {\bibfnamefont {T.}~\bibnamefont {Yamazaki}},\ and\ \bibinfo {author}
  {\bibfnamefont {R.}~\bibnamefont {Kubo}},\ }\bibfield  {title} {\bibinfo
  {title} {Zero-and low-field spin relaxation studied by positive muons},\
  }\href {https://doi.org/10.1103/PhysRevB.20.850} {\bibfield  {journal}
  {\bibinfo  {journal} {Phys. Rev. B}\ }\textbf {\bibinfo {volume} {20}},\
  \bibinfo {pages} {850} (\bibinfo {year} {1979})}\BibitemShut {NoStop}%
\bibitem [{\citenamefont {de~Reotier}\ and\ \citenamefont
  {Yaouanc}(1992)}]{Reotier1992}%
  \BibitemOpen
  \bibfield  {author} {\bibinfo {author} {\bibfnamefont {P.~D.}\ \bibnamefont
  {de~Reotier}}\ and\ \bibinfo {author} {\bibfnamefont {A.}~\bibnamefont
  {Yaouanc}},\ }\bibfield  {title} {\bibinfo {title} {Quantum calculation of
  the muon depolarization function: effect of spin dynamics in nuclear dipole
  systems},\ }\href {https://doi.org/10.1088/0953-8984/4/18/020} {\bibfield
  {journal} {\bibinfo  {journal} {J.\ Phys.: Condens.\ Matter}\ }\textbf
  {\bibinfo {volume} {4}},\ \bibinfo {pages} {4533} (\bibinfo {year}
  {1992})}\BibitemShut {NoStop}%
\bibitem [{\citenamefont {Wiegers}(1996)}]{Wiegers1996}%
  \BibitemOpen
  \bibfield  {author} {\bibinfo {author} {\bibfnamefont {G.~A.}\ \bibnamefont
  {Wiegers}},\ }\bibfield  {title} {\bibinfo {title} {Misfit layer compounds:
  Structures and physical properties},\ }\href
  {https://doi.org/10.1016/0079-6786(95)00007-0} {\bibfield  {journal}
  {\bibinfo  {journal} {Prog.\ Solid State Chem.}\ }\textbf {\bibinfo {volume}
  {24}},\ \bibinfo {pages} {1} (\bibinfo {year} {1996})}\BibitemShut {NoStop}%
\bibitem [{\citenamefont {Ng}\ and\ \citenamefont {Mc{Q}ueen}(2022)}]{Ng2022}%
  \BibitemOpen
  \bibfield  {author} {\bibinfo {author} {\bibfnamefont {N.}~\bibnamefont
  {Ng}}\ and\ \bibinfo {author} {\bibfnamefont {T.~M.}\ \bibnamefont
  {Mc{Q}ueen}},\ }\bibfield  {title} {\bibinfo {title} {Misfit layered
  compounds: Unique, tunable heterostructured materials with untapped
  properties},\ }\href {https://doi.org/10.1063/5.0101429} {\bibfield
  {journal} {\bibinfo  {journal} {APL Mater.}\ }\textbf {\bibinfo {volume}
  {10}},\ \bibinfo {pages} {100901} (\bibinfo {year} {2022})}\BibitemShut
  {NoStop}%
\bibitem [{\citenamefont {Suzuki}\ \emph {et~al.}(1993)\citenamefont {Suzuki},
  \citenamefont {Kondo}, \citenamefont {Iwasaki},\ and\ \citenamefont
  {Enoki}}]{Suzuki1993}%
  \BibitemOpen
  \bibfield  {author} {\bibinfo {author} {\bibfnamefont {K.}~\bibnamefont
  {Suzuki}}, \bibinfo {author} {\bibfnamefont {T.}~\bibnamefont {Kondo}},
  \bibinfo {author} {\bibfnamefont {M.}~\bibnamefont {Iwasaki}},\ and\ \bibinfo
  {author} {\bibfnamefont {T.}~\bibnamefont {Enoki}},\ }\bibfield  {title}
  {\bibinfo {title} {Variety of magnetism in incommensurate misfit layer
  compounds ({{\slshape {RE}}}{S})$_x${{\slshape {T}}}{S}$_2$ ({{\slshape
  {RE}}}=rare earths, {{\slshape {T}}}={Ta}, {V}, {Ti}, {Cr})},\ }\href
  {https://doi.org/10.7567/jjaps.32s3.341} {\bibfield  {journal} {\bibinfo
  {journal} {Jpn.\ J.\ Appl.\ Phys.}\ }\textbf {\bibinfo {volume} {32}},\
  \bibinfo {pages} {341} (\bibinfo {year} {1993})}\BibitemShut {NoStop}%
\bibitem [{\citenamefont {Lafond}\ and\ \citenamefont
  {Meerschaut}(1993)}]{Lafond1993}%
  \BibitemOpen
  \bibfield  {author} {\bibinfo {author} {\bibfnamefont {A.}~\bibnamefont
  {Lafond}}\ and\ \bibinfo {author} {\bibfnamefont {A.}~\bibnamefont
  {Meerschaut}},\ }\bibfield  {title} {\bibinfo {title} {Pr{\'e}paration,
  d{\'e}termination structurale et propri{\'e}t{\'e}s magn{\'e}tiques d'un
  nouveau compos{\'e} lamellaire incommensurable: Le sulfure de gadolinium et
  de chrome},\ }\href {https://doi.org/10.1016/0025-5408(93)90135-Z} {\bibfield
   {journal} {\bibinfo  {journal} {Mater.\ Res.\ Bull.}\ }\textbf {\bibinfo
  {volume} {28}},\ \bibinfo {pages} {979} (\bibinfo {year} {1993})}\BibitemShut
  {NoStop}%
\bibitem [{\citenamefont {Sugiyama}\ \emph {et~al.}(2003)\citenamefont
  {Sugiyama}, \citenamefont {Brewer}, \citenamefont {Ansaldo}, \citenamefont
  {Itahara}, \citenamefont {Dohmae}, \citenamefont {Seno}, \citenamefont
  {Xia},\ and\ \citenamefont {Tani}}]{Sugiyama2003}%
  \BibitemOpen
  \bibfield  {author} {\bibinfo {author} {\bibfnamefont {J.}~\bibnamefont
  {Sugiyama}}, \bibinfo {author} {\bibfnamefont {J.~H.}\ \bibnamefont
  {Brewer}}, \bibinfo {author} {\bibfnamefont {E.~J.}\ \bibnamefont {Ansaldo}},
  \bibinfo {author} {\bibfnamefont {H.}~\bibnamefont {Itahara}}, \bibinfo
  {author} {\bibfnamefont {K.}~\bibnamefont {Dohmae}}, \bibinfo {author}
  {\bibfnamefont {Y.}~\bibnamefont {Seno}}, \bibinfo {author} {\bibfnamefont
  {C.}~\bibnamefont {Xia}},\ and\ \bibinfo {author} {\bibfnamefont
  {T.}~\bibnamefont {Tani}},\ }\bibfield  {title} {\bibinfo {title} {Hidden
  magnetic transitions in the thermoelectric layered cobaltite
  [{Ca}$_2${CoO}$_3$]$_{0.62}$[{CoO}$_2$]},\ }\href
  {https://doi.org/10.1103/PhysRevB.68.134423} {\bibfield  {journal} {\bibinfo
  {journal} {Phys.\ Rev.\ B}\ }\textbf {\bibinfo {volume} {68}},\ \bibinfo
  {pages} {134423} (\bibinfo {year} {2003})}\BibitemShut {NoStop}%
\bibitem [{\citenamefont {Ahad}\ \emph {et~al.}(2020)\citenamefont {Ahad},
  \citenamefont {Gautam}, \citenamefont {Dey}, \citenamefont {Majid},
  \citenamefont {Rahman}, \citenamefont {Sharma}, \citenamefont {Coaquira},
  \citenamefont {da~Silva}, \citenamefont {Welter},\ and\ \citenamefont
  {Shukla}}]{Ahad2020}%
  \BibitemOpen
  \bibfield  {author} {\bibinfo {author} {\bibfnamefont {A.}~\bibnamefont
  {Ahad}}, \bibinfo {author} {\bibfnamefont {K.}~\bibnamefont {Gautam}},
  \bibinfo {author} {\bibfnamefont {K.}~\bibnamefont {Dey}}, \bibinfo {author}
  {\bibfnamefont {S.~S.}\ \bibnamefont {Majid}}, \bibinfo {author}
  {\bibfnamefont {F.}~\bibnamefont {Rahman}}, \bibinfo {author} {\bibfnamefont
  {S.~K.}\ \bibnamefont {Sharma}}, \bibinfo {author} {\bibfnamefont {J.~A.~H.}\
  \bibnamefont {Coaquira}}, \bibinfo {author} {\bibfnamefont {I.}~\bibnamefont
  {da~Silva}}, \bibinfo {author} {\bibfnamefont {E.}~\bibnamefont {Welter}},\
  and\ \bibinfo {author} {\bibfnamefont {D.~K.}\ \bibnamefont {Shukla}},\
  }\bibfield  {title} {\bibinfo {title} {Magnetic correlations in subsystems of
  the misfit [{Ca}$_2${CoO}$_3$]$_{0.62}$[{CoO}$_2$] cobaltate},\ }\href
  {https://doi.org/10.1103/PhysRevB.102.094428} {\bibfield  {journal} {\bibinfo
   {journal} {Phys.\ Rev.\ B}\ }\textbf {\bibinfo {volume} {102}},\ \bibinfo
  {pages} {094428} (\bibinfo {year} {2020})}\BibitemShut {NoStop}%
\bibitem [{\citenamefont {Lebernegg}\ \emph {et~al.}(2017)\citenamefont
  {Lebernegg}, \citenamefont {Janson}, \citenamefont {Rousochatzakis},
  \citenamefont {Nishimoto}, \citenamefont {Rosner},\ and\ \citenamefont
  {Tsirlin}}]{Lebernegg2017}%
  \BibitemOpen
  \bibfield  {author} {\bibinfo {author} {\bibfnamefont {S.}~\bibnamefont
  {Lebernegg}}, \bibinfo {author} {\bibfnamefont {O.}~\bibnamefont {Janson}},
  \bibinfo {author} {\bibfnamefont {I.}~\bibnamefont {Rousochatzakis}},
  \bibinfo {author} {\bibfnamefont {S.}~\bibnamefont {Nishimoto}}, \bibinfo
  {author} {\bibfnamefont {H.}~\bibnamefont {Rosner}},\ and\ \bibinfo {author}
  {\bibfnamefont {A.~A.}\ \bibnamefont {Tsirlin}},\ }\bibfield  {title}
  {\bibinfo {title} {Frustrated spin chain physics near the {M}ajumdar-{G}hosh
  point in szenicsite {Cu}$_3$({MoO}$_4$)({OH})$_4$},\ }\href
  {https://doi.org/10.1103/PhysRevB.95.035145} {\bibfield  {journal} {\bibinfo
  {journal} {Phys.\ Rev.\ B}\ }\textbf {\bibinfo {volume} {95}},\ \bibinfo
  {pages} {035145} (\bibinfo {year} {2017})}\BibitemShut {NoStop}%
\bibitem [{\citenamefont {Inosov}\ \emph {et~al.}(2020)\citenamefont {Inosov},
  \citenamefont {Kulbakov}, \citenamefont {Peets},\ and\ \citenamefont
  {Puente-Orench}}]{ILL-data1_D1B-2020}%
  \BibitemOpen
  \bibfield  {author} {\bibinfo {author} {\bibfnamefont {D.}~\bibnamefont
  {Inosov}}, \bibinfo {author} {\bibfnamefont {A.~A.}\ \bibnamefont
  {Kulbakov}}, \bibinfo {author} {\bibfnamefont {D.~C.}\ \bibnamefont
  {Peets}},\ and\ \bibinfo {author} {\bibfnamefont {I.}~\bibnamefont
  {Puente-Orench}},\ }\href {https://doi.org/10.5291/ILL-DATA.5-31-2734}
  {\bibinfo {title} {Determining the zero-field magnetic structure of
  antlerite}} (\bibinfo {year} {2020}),\ \bibinfo {note} {{I}nstitut
  {L}aue-{L}angevin ({ILL}) data, Grenoble.
  \href{https://doi.org/10.5291/ILL-DATA.5-31-2734}{doi:10.5291/ILL-DATA.5-31-2734}}\BibitemShut
  {NoStop}%
\bibitem [{\citenamefont {Inosov}\ \emph
  {et~al.}(2021{\natexlab{a}})\citenamefont {Inosov}, \citenamefont {Kulbakov},
  \citenamefont {Mannathanath~Chakkingal}, \citenamefont {Peets}, \citenamefont
  {Puente-Orench},\ and\ \citenamefont {Ritter}}]{ILL-data_D20-2021}%
  \BibitemOpen
  \bibfield  {author} {\bibinfo {author} {\bibfnamefont {D.~S.}\ \bibnamefont
  {Inosov}}, \bibinfo {author} {\bibfnamefont {A.~A.}\ \bibnamefont
  {Kulbakov}}, \bibinfo {author} {\bibfnamefont {A.}~\bibnamefont
  {Mannathanath~Chakkingal}}, \bibinfo {author} {\bibfnamefont {D.~C.}\
  \bibnamefont {Peets}}, \bibinfo {author} {\bibfnamefont {I.}~\bibnamefont
  {Puente-Orench}},\ and\ \bibinfo {author} {\bibfnamefont {C.}~\bibnamefont
  {Ritter}},\ }\href {https://doi.org/10.5291/ILL-DATA.5-31-2864} {\bibinfo
  {title} {Determining the intermediate-temperature incommensurate magnetic
  structure of antlerite}} (\bibinfo {year} {2021}{\natexlab{a}}),\ \bibinfo
  {note} {{I}nstitut {L}aue-{L}angevin ({ILL}) data, Grenoble.
  \href{https://doi.org/10.5291/ILL-DATA.5-31-2864}{doi:10.5291/ILL-DATA.5-31-2864}}\BibitemShut
  {NoStop}%
\bibitem [{\citenamefont {Inosov}\ \emph
  {et~al.}(2021{\natexlab{b}})\citenamefont {Inosov}, \citenamefont {Kulbakov},
  \citenamefont {Mannathanath~Chakkingal}, \citenamefont {Peets},\ and\
  \citenamefont {Wildes}}]{ILL-data_D7-2021}%
  \BibitemOpen
  \bibfield  {author} {\bibinfo {author} {\bibfnamefont {D.~S.}\ \bibnamefont
  {Inosov}}, \bibinfo {author} {\bibfnamefont {A.~A.}\ \bibnamefont
  {Kulbakov}}, \bibinfo {author} {\bibfnamefont {A.}~\bibnamefont
  {Mannathanath~Chakkingal}}, \bibinfo {author} {\bibfnamefont {D.~C.}\
  \bibnamefont {Peets}},\ and\ \bibinfo {author} {\bibfnamefont
  {A.}~\bibnamefont {Wildes}},\ }\href
  {https://doi.org/10.5291/ILL-DATA.5-32-923} {\bibinfo {title} {Verifying the
  magnetic origin of the low-temperature diffuse signal in antlerite powder}}
  (\bibinfo {year} {2021}{\natexlab{b}}),\ \bibinfo {note} {{I}nstitut
  {L}aue-{L}angevin ({ILL}) data, Grenoble.
  \href{https://doi.org/10.5291/ILL-DATA.5-32-923}{doi:10.5291/ILL-DATA.5-32-923}}\BibitemShut
  {NoStop}%
\bibitem [{\citenamefont {Bissengaliyeva}\ \emph {et~al.}(2012)\citenamefont
  {Bissengaliyeva}, \citenamefont {Ogorodova}, \citenamefont {Mel'chakova},\
  and\ \citenamefont {Vigasina}}]{Bissengaliyeva2012}%
  \BibitemOpen
  \bibfield  {author} {\bibinfo {author} {\bibfnamefont {M.~R.}\ \bibnamefont
  {Bissengaliyeva}}, \bibinfo {author} {\bibfnamefont {L.~P.}\ \bibnamefont
  {Ogorodova}}, \bibinfo {author} {\bibfnamefont {L.~V.}\ \bibnamefont
  {Mel'chakova}},\ and\ \bibinfo {author} {\bibfnamefont {M.~F.}\ \bibnamefont
  {Vigasina}},\ }\bibfield  {title} {\bibinfo {title} {Thermochemical
  investigation of natural antlerite},\ }\href
  {https://doi.org/10.1007/s10973-011-1763-7} {\bibfield  {journal} {\bibinfo
  {journal} {J.\ Therm.\ Anal.\ Calorim.}\ }\textbf {\bibinfo {volume} {109}},\
  \bibinfo {pages} {467} (\bibinfo {year} {2012})}\BibitemShut {NoStop}%
\bibitem [{\citenamefont {Bissengaliyeva}\ \emph {et~al.}(2013)\citenamefont
  {Bissengaliyeva}, \citenamefont {Bekturganov}, \citenamefont {Gogol},
  \citenamefont {Taimassova}, \citenamefont {Koketai},\ and\ \citenamefont
  {Bespyatov}}]{Bissengaliyeva2013}%
  \BibitemOpen
  \bibfield  {author} {\bibinfo {author} {\bibfnamefont {M.~R.}\ \bibnamefont
  {Bissengaliyeva}}, \bibinfo {author} {\bibfnamefont {N.~S.}\ \bibnamefont
  {Bekturganov}}, \bibinfo {author} {\bibfnamefont {D.~B.}\ \bibnamefont
  {Gogol}}, \bibinfo {author} {\bibfnamefont {S.~T.}\ \bibnamefont
  {Taimassova}}, \bibinfo {author} {\bibfnamefont {T.~A.}\ \bibnamefont
  {Koketai}},\ and\ \bibinfo {author} {\bibfnamefont {M.~A.}\ \bibnamefont
  {Bespyatov}},\ }\bibfield  {title} {\bibinfo {title} {Heat capacities of
  natural antlerite and brochantite at low temperature},\ }\href
  {https://doi.org/10.1021/je400130b} {\bibfield  {journal} {\bibinfo
  {journal} {J.\ Chem.\ Eng.\ Data}\ }\textbf {\bibinfo {volume} {58}},\
  \bibinfo {pages} {2904} (\bibinfo {year} {2013})}\BibitemShut {NoStop}%
\bibitem [{\citenamefont {Zittlau}\ \emph {et~al.}(2013)\citenamefont
  {Zittlau}, \citenamefont {Shi}, \citenamefont {Boerio-Goates}, \citenamefont
  {Woodfield},\ and\ \citenamefont {Majzlan}}]{Zittlau2013}%
  \BibitemOpen
  \bibfield  {author} {\bibinfo {author} {\bibfnamefont {A.~H.}\ \bibnamefont
  {Zittlau}}, \bibinfo {author} {\bibfnamefont {Q.}~\bibnamefont {Shi}},
  \bibinfo {author} {\bibfnamefont {J.}~\bibnamefont {Boerio-Goates}}, \bibinfo
  {author} {\bibfnamefont {B.~F.}\ \bibnamefont {Woodfield}},\ and\ \bibinfo
  {author} {\bibfnamefont {J.}~\bibnamefont {Majzlan}},\ }\bibfield  {title}
  {\bibinfo {title} {Thermodynamics of the basic copper sulfates antlerite,
  posnjakite, and brochantite},\ }\href
  {https://doi.org/10.1016/j.chemer.2012.12.002} {\bibfield  {journal}
  {\bibinfo  {journal} {Geochemistry}\ }\textbf {\bibinfo {volume} {73}},\
  \bibinfo {pages} {39} (\bibinfo {year} {2013})}\BibitemShut {NoStop}%
\bibitem [{\citenamefont {Rama Subba~Reddy}\ \emph {et~al.}(2002)\citenamefont
  {Rama Subba~Reddy}, \citenamefont {Lakshmi~Reddy}, \citenamefont
  {Siva~Reddy},\ and\ \citenamefont {Reddy}}]{Reddy2002b}%
  \BibitemOpen
  \bibfield  {author} {\bibinfo {author} {\bibfnamefont {R.}~\bibnamefont {Rama
  Subba~Reddy}}, \bibinfo {author} {\bibfnamefont {S.}~\bibnamefont
  {Lakshmi~Reddy}}, \bibinfo {author} {\bibfnamefont {G.}~\bibnamefont
  {Siva~Reddy}},\ and\ \bibinfo {author} {\bibfnamefont {B.~J.}\ \bibnamefont
  {Reddy}},\ }\bibfield  {title} {\bibinfo {title} {Spectral studies of
  divalent copper in antlerite mineral},\ }\href
  {https://doi.org/10.1002/1521-4079(200205)37:5<485::AID-CRAT485>3.0.CO;2-E}
  {\bibfield  {journal} {\bibinfo  {journal} {Cryst.\ Res.\ Technol.}\ }\textbf
  {\bibinfo {volume} {37}},\ \bibinfo {pages} {485} (\bibinfo {year}
  {2002})}\BibitemShut {NoStop}%
\end{thebibliography}%

\end{document}